\documentclass[fleqn,usenatbib]{mnras}

\usepackage[T1]{fontenc}

\DeclareRobustCommand{\VAN}[3]{#2}
\let\VANthebibliography\thebibliography
\def\thebibliography{\DeclareRobustCommand{\VAN}[3]{##3}\VANthebibliography}

\usepackage{graphicx}	
\usepackage{amsmath}	
\usepackage{amssymb}	

\usepackage[nopatch]{microtype}
\usepackage{booktabs}

\usepackage{tabularx}
\usepackage[skip=2pt]{caption}
\usepackage{txfonts}
\usepackage{orcidlink}

\usepackage{bm}
\usepackage{array}
\usepackage{relsize}
\usepackage{acro}
\usepackage{multicol}
\usepackage{multirow}
\usepackage{xfrac}
\usepackage{lipsum}
\usepackage{cleveref}
\usepackage{hyperref}

\DeclareAcronym{agn}{
  short = AGN ,
  short-plural-form = AGN ,
  long  = active galactic nucleus ,
  long-plural-form = active galactic nuclei ,
  tag = abbrev
}

\DeclareAcronym{smbh}{
  short = SMBH ,
  long  = supermassive black hole ,
  tag = abbrev
}

\DeclareAcronym{mq}{
  short = MQ ,
  long  = massive quiescent ,
  tag = abbrev
}

\DeclareAcronym{mqg}{
  short = MQG ,
  long  = massive quiescent galaxy ,
  long-plural-form = massive quiescent galaxies ,
  tag = abbrev
}

\DeclareAcronym{rq}{
  short = RQ ,
  long  = recent quiescent ,
  tag = abbrev
}

\DeclareAcronym{rqg}{
  short = RQG ,
  long  = recently quenched galaxy ,
  long-plural-form = recently quenched galaxies ,
  tag = abbrev
}

\DeclareAcronym{lq}{
  short = LQ ,
  long  = long quiescent ,
  tag = abbrev
}

\DeclareAcronym{lqg}{
  short = LQG ,
  long  = long quiescent galaxy ,
  long-plural-form = long quiescent galaxies ,
  tag = abbrev
}

\DeclareAcronym{coda}{
  short = CoDaIII ,
  long  = Cosmic Dawn III ,
  tag = abbrev
}

\DeclareAcronym{illtng}{
  short = TNG ,
  long  = Illustris: The Next Generation ,
  tag = abbrev
}

\DeclareAcronym{cpc}{
  short = cpc ,
  short-plural =  ,
  long  = comoving parsec ,
  tag = abbrev
}

\DeclareAcronym{ckpc}{
  short = ckpc ,
  short-plural =  ,
  long  = comoving kiloparsec ,
  tag = abbrev
}

\DeclareAcronym{cMpc}{
  short = cMpc ,
  short-plural =  ,
  long  = comoving megaparsec ,
  tag = abbrev
}

\DeclareAcronym{cGpc}{
  short = cGpc ,
  short-plural =  ,
  long  = comoving gigaparsec ,
  tag = abbrev
}

\DeclareAcronym{ppc}{
  short = ppc ,
  short-plural =  ,
  long  = physical parsec ,
  tag = abbrev
}

\DeclareAcronym{pkpc}{
  short = pkpc ,
  short-plural =  ,
  long  = physical kiloparsec ,
  tag = abbrev
}

\DeclareAcronym{pMpc}{
  short = pMpc ,
  short-plural =  ,
  long  = physical megaparsec ,
  tag = abbrev
}

\DeclareAcronym{pGpc}{
  short = pGpc ,
  short-plural =  ,
  long  = physical gigaparsec ,
  tag = abbrev
}

\DeclareAcronym{lt}{
  short = LT ,
  long  = logarithmic transformation ,
  tag = abbrev
}

\DeclareAcronym{gqt}{
  short = GQT ,
  long  = Gaussian quantile transformation ,
  tag = abbrev
}

\DeclareAcronym{cdf}{
  short = CDF ,
  long  = cumulative distribution function ,
  tag = abbrev
}

\DeclareAcronym{imf}{
  short = IMF ,
  long  = initial mass function ,
  tag = abbrev
}

\DeclareAcronym{sfh}{
  short = SFH ,
  long  = star formation history ,
  long-plural-form = star formation histories ,
  tag = abbrev
}

\DeclareAcronym{zh}{
  short = ZH ,
  long  = metallicity history ,
  long-plural-form = metallicity histories ,
  tag = abbrev
}

\DeclareAcronym{bgs}{
  short = BGS ,
  long  = Bright Galaxy Survey ,
  tag = abbrev
}

\DeclareAcronym{grispy}{
  short = GriSPy ,
  long  = Grid Search In Python ,
  tag = abbrev
}

\DeclareAcronym{disperse}{
  short = DisPerSE ,
  long  = Discrete Persistent Structure Extractor ,
  tag = abbrev
}

\DeclareAcronym{pca}{
  short = PCA ,
  long  = principal component analysis ,
  long-plural-form = principal component analyses ,
  tag = abbrev
}

\DeclareAcronym{pc}{
  short = PC ,
  long  = principal component ,
  tag = abbrev
}

\DeclareAcronym{lrd}{
  short = LRD ,
  long  = Little Red Dot ,
  tag = abbrev
}

\DeclareAcronym{gmm}{
  short = GMM ,
  long  = Gaussian mixture model ,
  tag = abbrev
}

\DeclareAcronym{em}{
  short = EM ,
  long  = expectation-maximisation ,
  tag = abbrev
}

\DeclareAcronym{ghc}{
  short = GHC ,
  long  = galaxy-halo connection ,
  tag = abbrev
}

\DeclareAcronym{shmr}{
  short = SHMR ,
  long  = stellar-halo mass relation ,
  tag = abbrev
}

\DeclareAcronym{mzr}{
  short = MZR ,
  long  = mass-metallicity relation ,
  tag = abbrev
}

\DeclareAcronym{mhzr}{
  short = HMZR ,
  long  = halo mass-metallicity relation ,
  tag = abbrev
}

\DeclareAcronym{ssp}{
  short = SSP ,
  long  = simple stellar population ,
  tag = abbrev
}

\DeclareAcronym{srcc}{
  short = $r_S$ ,
  long  = Spearman's rank correlation coefficient ,
  tag = abbrev
}

\DeclareAcronym{mwa}{
  short = MWA ,
  long  = mass-weighted age ,
  tag = abbrev
}

\DeclareAcronym{zwa}{
  short = ZWA ,
  long  = metallicity-weighted age ,
  tag = abbrev
}

\DeclareAcronym{mpb}{
  short = MPB ,
  long  = main progenitor branch ,
  long-plural-form = main progenitor branches ,
  tag = abbrev
}

\DeclareAcronym{spl}{
  short = SPL ,
  long  = secondary progenitor list ,
  tag = abbrev
}

\DeclareAcronym{rfr}{
  short = RFR ,
  long  = random forest regressor ,
  tag = abbrev
}

\DeclareAcronym{ert}{
  short = ERT ,
  long  = extremely randomised tree ,
  tag = abbrev
}

\DeclareAcronym{fof}{
  short = FoF ,
  long  = Friends-of-Friends ,
  tag = abbrev
}

\DeclareAcronym{relu}{
  short = ReLU ,
  long  = rectified linear unit ,
  tag = abbrev
}

\DeclareAcronym{lrelu}{
  short = L-ReLU ,
  long  = leaky rectified linear unit ,
  tag = abbrev
}

\DeclareAcronym{elu}{
  short = ELU ,
  long  = exponential linear unit ,
  tag = abbrev
}

\DeclareAcronym{fsps}{
  short = FSPS ,
  long  = Flexible Stellar Population Synthesis ,
  tag = abbrev
}

\DeclareAcronym{UM}{
  short = UM ,
  long  = UniverseMachine ,
  tag = abbrev
}

\DeclareAcronym{desi}{
  short = DESI ,
  long  = Dark Energy Spectroscopic Instrument ,
  tag = abbrev
}

\defcitealias{Chittenden}{CT22}
\defcitealias{Behera}{BCT24}
\defcitealias{Chittenden2}{CTB24}

\title[The physics of massive quiescent galaxies]{On the unique evolutionary mechanisms of massive quiescent galaxies in the epoch of reionisation}

\author[H. G. Chittenden et al. 2026]{
Harry George Chittenden$^{1,2}$\thanks{E-mail: hchittenden@swin.edu.au}\textsuperscript{\orcidlink{https://orcid.org/0000-0001-5856-8713}}, Karl Glazebrook$^{1,2}$\textsuperscript{\orcidlink{https://orcid.org/0000-0002-3254-9044}}, Themiya Nanayakkara$^{1,2}$\textsuperscript{\orcidlink{https://orcid.org/0000-0003-2804-0648}}, \newauthor \ Lalitwadee Kawinwanichakij$^{1,2}$\textsuperscript{\orcidlink{https://orcid.org/0000-0003-4032-2445}}, Claudia Lagos$^{3,4,5}$\textsuperscript{\orcidlink{https://orcid.org/0000-0003-3021-8564}}, Lucas Kimmig$^{6}$\textsuperscript{\orcidlink{https://orcid.org/0009-0006-8337-8712}} and Rhea-Silvia Remus$^{6}$\textsuperscript{\orcidlink{https://orcid.org/0009-0008-9260-7278}}
\\
$^{1}$Centre for Astrophysics and Supercomputing [CAS], Swinburne University of Technology, P.O. Box 218, Hawthorn VIC 3122, Melbourne, Australia\\
$^{2}$JWST Australia Data Centre [JADC], Swinburne Advanced Manufacturing and Design Centre [AMDC], John Street, Hawthorn VIC 3122, Australia\\
$^{3}$International Centre for Radio Astronomy Research [ICRAR], M468, University of Western Australia, 35 Stirling Hwy, Crawley WA 6009, Perth, Australia\\
$^{4}$ARC Centre of Excellence for All Sky Astrophysics in 3 Dimensions [ASTRO 3D]\\
$^{5}$Cosmic Dawn Center [DAWN], Niels Bohr Institute, University of Copenhagen, Rådmandsgade 64, 2200 København N, Denmark\\
$^{6}$Universitäts-Sternwarte München, Fakultät für Physik, Ludwig-Maximilians-Universität, Scheinerstr. 1, D-81679 München, Germany
}
\date{Accepted XXX. Received YYY; in original form ZZZ}

\pubyear{2026}

\begin{document}
\label{firstpage}
\pagerange{\pageref{firstpage}--\pageref{lastpage}}
\maketitle

\begin{abstract}
We investigate the evolutionary histories of a population of high mass, high redshift, quiescent galaxies in the cosmohydrodynamical simulation \textsc{Thesan}, studying the characteristic properties of their haloes and environments over the epoch of reionisation. \textsc{Thesan} employs a modified version of the \textsc{Arepo} moving-mesh code utilised in \textsc{IllustrisTNG}, which incorporates on-the-fly radiative transfer to couple haloes and galaxies with the evolving radiation field. \textsc{Thesan} exhibits nine massive quiescent galaxies at $z=5.5$, in a $(95.5 \text{cMpc})^3$ volume, with no counterpart in \textsc{IllustrisTNG}. A numerical issue in the simulation reduces AGN feedback efficiency by a factor of 25 while enhancing accretion rates, creating a regime of suppressed feedback. We find their stellar mass assembles rapidly through smooth halo accretion in dense environments, particularly from massive neighbouring structures, while their early-forming haloes develop fast-growing potential wells hosting massive black holes. This suppressed feedback allows prolonged black hole growth before eventual kinetic-mode quenching, providing insight into galaxy evolution under weakened AGN regulation. We find that megaparsec-scale overdensities and halo masses continue growing after quenching, suggesting these galaxies will reside in some of the largest haloes and densest regions of space by $z=6$. With massive quiescent galaxies found in \textit{JWST} data, the identification of such galaxies in \textsc{Thesan} enables isolation of halo and environmental conditions most conducive to their evolution under this suppressed feedback regime, guiding future deep surveys and N-body simulation studies of analogous systems.
\end{abstract}




\section{Introduction}
\label{sec:intro}

The advent of high-redshift galaxy surveys, with modern state-of-the-art telescopes such as \textit{JWST}, has culminated in a new suite of spectroscopic data for some of the earliest galaxies, which will continue to grow with the advent of future facilities such as the ELT. This data challenges a number of consensuses on galaxy evolution and post-reionisation cosmology, such as the establishment of a larger-than-expected UV luminosity function \citep{Xu, XShen} warranting new theories of the rapid assembly of massive and luminous galaxies, during and following cosmic reionisation \citep[for review see][]{Adamo}.

Of particular interest, there are exotic galaxies in the \textit{JWST} data which were found to have grown to Milky Way masses, compacted and quenched during cosmic reionisation \citep{Carnallmnras, Kakimoto}, some even having quenched by $z\sim 11$ \citep{Glazebrook2024}; which we refer to as \acp{mqg}. Similar observations have confirmed that these supposedly rare distant galaxies are more abundant at these redshifts than expected \citep{Carnallnature,Carnallmnras2,Nanayakkara,Nanayakkara2,Park}. These surveys effectively predict a greater number density of such galaxies and their potential submillimetre progenitors than what is found in standard numerical models \citep{Valentino,Carnallmnras,Russell,Labbe,Lagos}, and hypothesise an enhancement of star formation efficiency in the reionisation epoch. However, matching the abundance of high-redshift quiescent galaxies seen in such observations remains difficult for most simulations, especially when also aiming for consistency with galaxy evolution at lower redshifts.

Cosmic simulations have long served as a valuable tool for modelling the evolution of galaxies over time. While most standard $\Lambda$CDM simulations invoke an approximate solution to cosmic reionisation, specialised reionisation simulations instead invoke explicit, time-dependent treatment of photon synthesis and radiative transfer; and as a result, they produce a catalogue of high-redshift galaxies which are physically linked to the epoch of reionisation.

The \textsc{Thesan} simulation suite \citep{Thesan,Garaldi,Smith} is based on the widely used \textsc{IllustrisTNG} simulations \citep{Nelson2018, IllustrisTNG, Pillepich2017, Springel, Marinacci, Naiman}. \textsc{Thesan} runs \textsc{Arepo-RT} \citep{ArepoRT}: a modified version of \textsc{IllustrisTNG}'s magnetohydrodynamical model \textsc{Arepo} \citep{Arepo}. The key difference lies in its treatment of radiative hydrodynamics, allowing it to model radiation-gas interactions directly. This enables explicit calculations of radiative feedback from young stars and \acp{agn}.

At \textsc{Thesan}’s final redshift of $z=5.5$ (approximately 1.04 Gyr after the Big Bang), the simulation includes nine central galaxies with stellar masses above $10^{10}M_\odot$ and instantaneous specific star formation rates below $1\text{Gyr}^{-1}$. Four of these galaxies have sustained low star formation over the past 100 Myr, show stellar mass growth closely linked to their \acp{agn}, and have stellar half-mass radii under 4 \acp{ckpc} \citep{XShen2}, consistent with the five quenched massive galaxies reported by \citet{Carnallmnras2} at the same epoch.

The \textsc{Thesan} collaboration has revealed that the current public release of the simulation contains a numerical issue affecting the implementation of AGN feedback \citep{ThesanAddendum}. The reduced speed of light used in the simulation ($\tilde{c} = 0.2c$) inadvertently overwrote the physical value in black hole routines, reducing energy injection by both quasar and radio mode feedback by a factor of $(c/\tilde{c})^2 = 25$, and allowing accretion up to quintuple the Eddington rate limit; compounded by equation of state modifications which further boosted Bondi accretion rates. Consequently, black holes in \textsc{Thesan} can grow to unusually high masses at early times, with the subsequent transition to kinetic feedback quenching galaxies despite the weakened coupling. The physical picture traced by \textsc{Thesan} therefore does not represent the originally anticipated balance between radiation-gas coupling and standard AGN self-regulation, and cannot be expected to reproduce the galaxy demographics of \textsc{IllustrisTNG} at low redshift; but rather, \textsc{Thesan} represents a scenario in which galaxies evolve under suppressed feedback efficiency in the first billion years of the universe.

This realisation effectively reframes the interpretation of the physics of massive galaxies in \textsc{Thesan} as a unique exploration of galaxy and black hole coevolution in a regime of suppressed feedback efficiency: an important limiting case for understanding the onset of quenching during the epoch of reionisation. Observations of high-redshift galaxies do suggest the existence of weakly coupled \ac{agn} feedback in at least some systems: \citet{WWang} identify a compact $10^{11} M_\odot$ galaxy at $z = 3.5892$ where radiative feedback couples at efficiencies below $10^{-4}$, and rapidly growing, overmassive black holes are now frequently identified in \textit{JWST} data \citep{Maiolino}. In this context, \textsc{Thesan} offers a numerical experiment into early supermassive black hole growth and delayed kinetic-mode quenching, highlighting how feedback strength modulates both star formation and black hole-galaxy scaling relations in the first billion years of the universe.

\begin{table*}
\centering
\caption{A summary of the attributes of the nine massive quiescent galaxies investigated in this work. All snapshot-specific fields are quoted at $z=5.5$ unless explicitly stated otherwise. From left to right, this table shows the indices of each halo in the \textsc{Thesan-1} FoF table at $z=5.5$, the equivalent central subhalo index in the Subfind table, the halo's final dark matter component mass, the galaxy's stellar mass, the galaxy's stellar half-mass radius, the specific star formation rate at $z=6$, the specific star formation rate at $z=5.5$, the halo's half-mass formation redshift, the galaxy's stellar half-mass formation redshift, and finally, the redshift at which the galaxy is quenched. Galaxies in this table are defined as long quiescents if the specific star formation rate at $z=6$ is below the threshold value of $1 \text{Gyr}^{-1}$, or equivalently, if the quenching redshift is greater than $z=6$.}
\label{tab:MQs}
\begin{tabular}{| c | c | c | c | c | c | c | c | c | c |}
\hline
FoF ID & Subfind ID & $\log_{10} M_h / M_\odot$ & $\log_{10} M_s / M_\odot$ & $R_s / \text{ckpc}$ & $\log_{10} \text{sSFR} / \text{Gyr}^{-1} (z=6)$ & $\log_{10} \text{sSFR} / \text{Gyr}^{-1} (z=5.5)$ & $z_\frac{1}{2}^h$ & $z_\frac{1}{2}^s$ & $z_Q$ \\
\hline
2 & 806 & 11.866 & 10.350 & 3.614 & -1.306 & -2.729 & 6.59 & 6.52 & 6.10 \\
5 & 1393 & 11.975 & 10.449 & 2.735 & -3.438 & -2.312 & 6.52 & 6.94 & 6.28 \\
7 & 1752 & 12.044 & 10.480 & 3.201 & -1.862 & -2.948 & 6.94 & 7.25 & 6.72 \\
12 & 2362 & 11.682 & 10.356 & 3.921 & 0.631 & -0.951 & 7.17 & 6.40 & 5.55 \\
20 & 3228 & 11.775 & 10.422 & 2.676 & 0.739 & -1.492 & 6.87 & 6.34 & 5.69 \\
62 & 7299 & 11.674 & 10.246 & 6.096 & 0.731 & -0.043 & 7.01 & 6.40 & 5.59 \\
98 & 9717 & 11.586 & 10.270 & 2.668 & -1.387 & -3.690 & 7.50 & 6.87 & 6.10 \\
103 & 10073 & 11.582 & 10.289 & 3.162 & 0.458 & -4.319 & 7.68 & 6.79 & 5.99 \\
112 & 10689 & 11.523 & 10.147 & 3.246 & 0.779 & -1.821 & 7.09 & 6.28 & 5.59 \\
\hline
\end{tabular}
\end{table*}

Such galaxies do not appear at similar redshifts in \textsc{IllustrisTNG}. In contrast, their number density in \textsc{Thesan} reaches $1.033 \times 10^{-5}~\text{cMpc}^{-3}$ by $z=5.5$, more than an order of magnitude higher than in other recent simulations. These include the reionisation-focused zoom-in \textsc{Flares} simulations \citep{FlaresVIII} and the chemically detailed \textsc{Simba-C} simulation \citep{SimbaC}, neither of which were specifically calibrated to produce quenched galaxies at high redshift. Despite this, \textsc{Simba-C} yields a notable quenched population between $z=2$ and $z=4$ \citep{Szpila}.

\begin{figure}
\includegraphics[width=\linewidth]{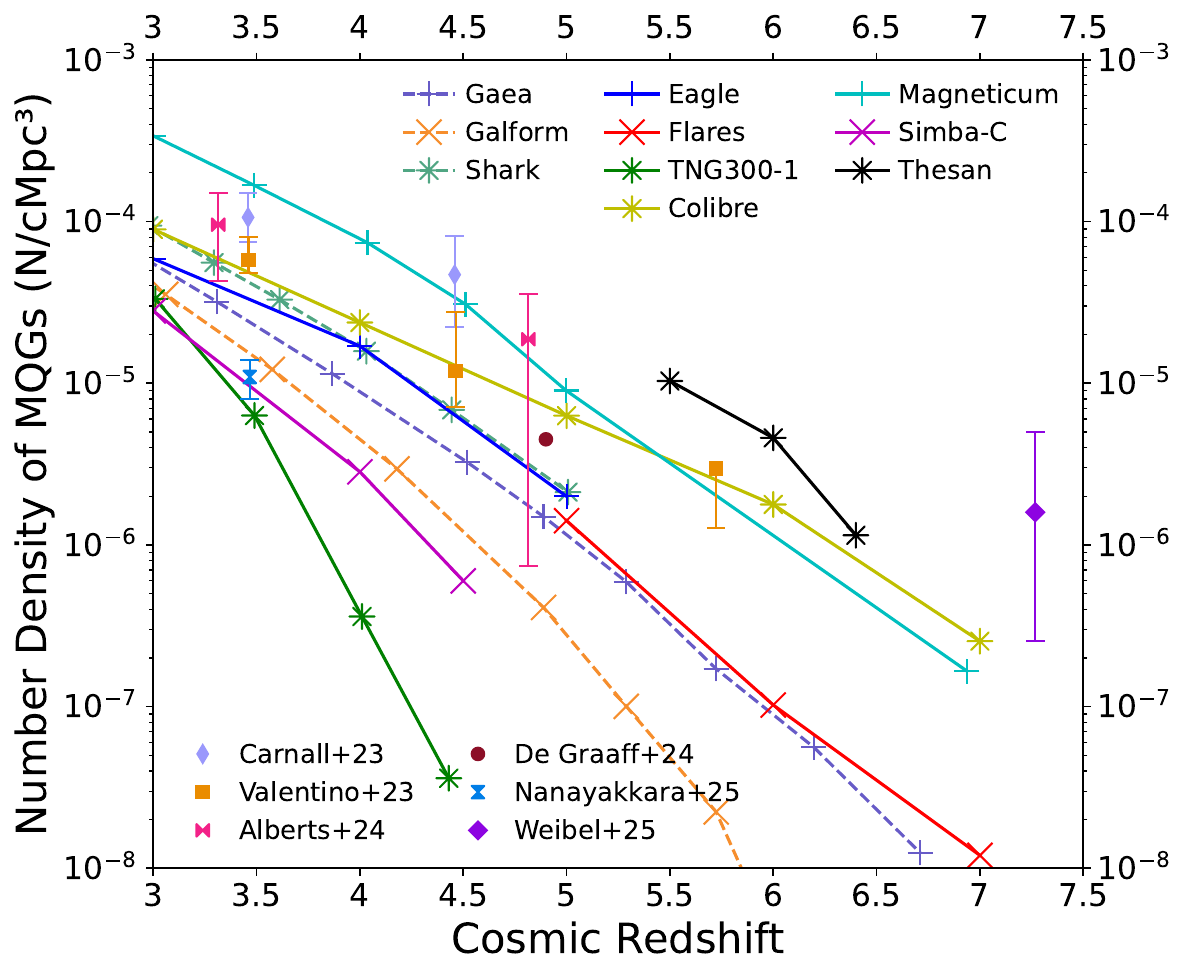}
\vspace*{-20pt}
\caption{A comparison of the redshift-dependent, volume-normalised number densities of quiescent galaxies above $10^{10} M_\odot$ in stellar mass. The abundance of \acp{mqg} over time in \textsc{Thesan} (black) is shown in relation to conventional hydrodynamical simulation models such as \textsc{Eagle} (blue), \textsc{IllustrisTNG} (green) and \textsc{Magneticum} (cyan); as well as the reionisation-era zoom simulation \textsc{Flares} (red), the chemical enrichment adapted simulation \textsc{Simba-C} (magenta), and the \textsc{Colibre L200m6} simulation (yellow), which invokes explicit multiphase modelling of the cold ISM. With dashed lines, we show predictions from the semi-analytic models \textsc{GAEA} (purple), \textsc{Galform} (orange) and \textsc{Shark} (green). This simulation data is accompanied by observational estimates of \ac{mqg} number densities from \textit{JWST}-era studies, denoted with unconnected points and error bars, showing an abundance of \acp{mqg} exceeding most simulations at low redshift. It should be noted, however, that logistical differences, such as photometric and spectroscopic selection criteria, and different stellar mass and star formation rate cuts, exist between these papers. This figure shows that \textsc{Thesan} entails a much higher quantity of \acp{mqg} at high redshift, aligning with high redshift observational estimates, indicating the suite's analogue for the abundant quiescent galaxy population seen in deep surveys.}
\label{fig:nQcomp}
\end{figure}

\Cref{fig:nQcomp} compares the number densities of \acp{mqg} in \textsc{Thesan} to those in other hydrodynamical simulations, as well as to predictions from semi-analytic models \citep{Galform,Shark,Shark2,Gaea} and observational estimates from several recent \textit{JWST}-era studies \citep{Carnallmnras,Valentino2023,Alberts,deGraaff,Nanayakkara2,Weibel}. Unlike most simulations, observations up to $z \sim 7$ report a significantly higher abundance of these rare galaxies than predicted by current models \citep{Valentino, Carnallmnras, Baker, Weibel}, highlighting the need to revise standard approaches to modelling galaxy evolution.

One exception is the \textsc{Magneticum Pathfinder} simulation\footnote{Throughout this work, we refer to the \textsc{Magneticum Pathfinder} suite simply as “\textsc{Magneticum},” following common usage in the literature. All mentions of \textsc{Magneticum} \acp{mqg} refer to the Box3/uhr run used in \citet{Kimmig}, unless stated otherwise.}. \citet{Kimmig, RemusKimmig} show that \textsc{Magneticum} produces a higher number of \acp{mqg} from $z = 3$ to $z = 4$, matching observed abundances at those redshifts, though it overpredicts their abundance by $z = 2$. While this suggests that \textsc{Magneticum} is well-suited for studying high-redshift quiescent systems, it also predicts quenching that occurs earlier than observed between $z = 3$ and $z = 4$ \citep{Nanayakkara2}, and overproduces intermediate-mass quiescent galaxies from $z = 2$ to $z = 3$. These discrepancies likely stem from its strong \ac{agn} feedback implementation \citep{Steinborn} and relatively coarse resolution \citep[][see Appendix B]{Lagos}.

A further point of comparison is the \textsc{Colibre} simulation suite, the successor to \textsc{Eagle} with explicit multiphase ISM modelling \citep{Colibre2, Colibre}. \citet{ChandroGomez} demonstrate that \textsc{Colibre} reproduces the observed number densities of \acp{mqg} at $z \sim 2-4$, yielding abundances comparable to those found in both \textsc{Magneticum} and \textsc{Thesan} at $z > 5$. As with \textsc{Magneticum}, however, agreement in number densities does not uniquely constrain the underlying quenching pathways. The \textsc{Colibre} feedback model incorporates calibrated stellar and \ac{agn} prescriptions tuned to match low-redshift galaxy scaling relations, which may influence the timing and efficiency of quenching at earlier epochs.

Taken together, \textsc{Magneticum}, \textsc{Colibre} and \textsc{Thesan} span a range of feedback implementations, from strong coupling to calibrated multiphase prescriptions to globally suppressed coupling. Their differing quenching histories, despite broadly comparable high-redshift \ac{mqg} abundances, suggest that current simulations bracket a physically plausible range of \ac{agn} self-regulation scenarios rather than converging on a single, uniquely constrained pathway.

\Aclp{mqg} may provide insight into galaxy evolution during the epoch of reionisation, offering constraints on the environmental conditions and feedback processes operating at these early times. Given the similar increase in \ac{mqg} abundance over time in \textsc{Colibre}, \textsc{Magneticum} and \textsc{Thesan}, we compare our results with those from \textsc{Magneticum} and \textsc{Colibre}. Differences in the properties of \acp{mqg} between the two simulations may help identify the physical processes that future deep observations will probe.

Several studies have developed machine learning models that predict galaxy properties using only the dark matter component of simulations \citep{Agarwal,JoKim,Lovell,McGibbonKhochfar,McGibbonKhochfar2,Chittenden,Cuevas,Wu}. These approaches allow large-volume N-body simulations without baryonic physics to be analysed effectively \citep{Chadayammuri,Chittenden2}. However, the rarity of \acp{mqg} makes them difficult to predict using general-purpose models, and presents significant challenges for developing dedicated machine learning frameworks.

In this work, we study the properties of \aclp{mqg} in the \textsc{Thesan} simulation and compare their galaxy and halo formation histories to those of high-mass star-forming counterparts under this suppressed feedback regime. This comparison sheds light on the evolutionary pathways that lead to early quenching under weakened \ac{agn} regulation.

By identifying the halo and environmental features most predictive of \ac{mqg} formation, whether conventional or novel, we lay the groundwork for future models tailored to rare, early-forming galaxies and their progenitors. Unlike previous studies which focus mainly on galaxy properties, we apply principal component analysis and probabilistic parameter space clustering to isolate a statistically robust subset of haloes with the physical conditions necessary for early quenching.

This method formalises the selection of \acp{mqg} within physical parameter space, and reveals the combinations of halo assembly history, environmental overdensity, and growth symmetry which best distinguish them from other high-mass systems. These findings provide a physically grounded interpretation of massive galaxy growth and quenching in the early universe.

Although current \ac{mqg} samples are too small to support generalised evolutionary machine learning models, the features identified in this study provide a physically motivated framework for locating analogues in larger N-body simulations. Our results indicate that \ac{mqg} formation in this regime is not driven solely by internal processes such as \ac{agn} feedback, but is strongly influenced by early cosmic environment and anisotropic gas accretion. These findings underscore the importance of including such variables in future simulations and predictive models. They also offer testable predictions for \textit{JWST} and other deep surveys aiming to characterise the quenching of massive galaxies at high redshift.

This paper is organised as follows. In \cref{sec:sims}, we outline the properties of the \textsc{Thesan} simulation data and the data which was used to investigate \acp{mqg}. In \cref{sec:selection}, we outline the method of selecting \ac{mqg} candidates from the simulation dataset, and detail the distinguishing properties of quenched samples in \cref{sec:comp}. We then explain how these \acp{mqg} evolved according to the implemented physics and reconcile this with observationally supported expectations in \cref{sec:disc}, and summarise our findings in \cref{sec:conc}.

\section{Simulation Overview and Physical Modelling}
\label{sec:sims}

\subsection{The \textsc{Thesan} Simulation Suite}
\label{sec:Thesan}

\subsubsection{Public data release}
\label{sec:ThesanData}

The \textsc{Thesan} simulation suite \citep{Thesan,Garaldi,Smith} consists of a set of cubic simulation boxes of side length 95.5 \acp{cMpc}, evaluated down to a redshift of $z=5.5$. Each simulation is run with the same initial conditions, creating a near identical matter distribution with cross-matched subhalo catalogs between simulations of different resolutions and different physical models.

The fiducial \textsc{Thesan} model is run in the \textsc{Thesan-1} simulation: the source of all data discussed in this paper unless stated otherwise, with $2100^3$ dark matter particles of mass $3.12\times10^6M_\odot$ and $2100^3$ gas cells of mass $5.82\times10^5M_\odot$. The fiducial model is also run in the lower-resolution \textsc{Thesan-2} simulation, with $1050^3$ units of each type, and respective masses of $2.49\times10^7M_\odot$ and $4.66\times10^6M_\odot$. The suite also contains a pure dark matter run for each of the aforementioned simulations, and a series of alternative models such as one including strong dark acoustic oscillations, and another including high ionising photon escape fractions.

The \textsc{Thesan} public data release\footnote{\hyperlink{https://thesan-project.com/data.html}{https://www.thesan-project.com/data.html}} consists of a series of snapshot particle data, halo group catalogs per snapshot, halo merger trees and supplementary data. The time domain for these simulations comprises 81 snapshots from $z=20$ to $z=5.5$, with a mean and a median separation in time of approximately 10Myr. All \textsc{Thesan} simulation runs assume the same \citet{Planck} $\Lambda$CDM cosmological parameters as \textsc{IllustrisTNG}: $\Omega_m=0.3089$, $\Omega_\Lambda=0.6911$, $\Omega_b=0.0486$, $n_s=0.9667$, $\sigma_8=0.8159$ and $H_0=67.74$ km/s/Mpc.

\subsubsection{Physical models in \textsc{Thesan}}
\label{sec:ThesanModelling}

The \textsc{Thesan} simulations build upon the galaxy formation model of \textsc{IllustrisTNG} to model star formation. This model assumes star formation occurs in dense gas, following the Kennicutt-Schmidt relation, and employs a two-phase effective equation of state for the interstellar medium. While \textsc{Thesan} inherits the fundamental star formation prescription from \textsc{IllustrisTNG}, the implementation of radiative transfer is the distinctive feature setting \textsc{Thesan} apart from \textsc{IllustrisTNG}.

Instead of the spatially uniform UV background used in \textsc{IllustrisTNG} \citep{Faucher-Giguere}, \textsc{Thesan} incorporates a self-consistent radiation transport scheme using the model extension introduced in \textsc{Arepo-RT} \citep{ArepoRT,Thesan}. The \textsc{Arepo-RT} code links radiation pressure to the kinetic energy of ionisation fronts via the Eddington tensor, thereby removing the dependence of calculations on the local density of ionising sources \citep{ArepoRT}. It also employs a subcycling algorithm, in which radiative transfer is resampled multiple times within each hydrodynamical timestep. This structure enables the simulation to resolve photon energy bins and couple them to gas dynamics in real time. As a result, ionisation fronts form dynamically and interact directly with their surroundings \citep{Garaldi,Garaldi2,XShen2}. Together with non-equilibrium thermochemistry, this approach allows \textsc{Thesan} to model reionisation and galaxy formation in a fully self-consistent manner. It affects gas heating, cooling, and ionisation, providing a more self-consistent treatment of reionisation and its effects on gas in low-mass haloes at high redshifts \citep{Garaldi,Zier}.

\textsc{Thesan} also models \ac{smbh} growth and \ac{agn} feedback, again adopting the framework utilised in \textsc{IllustrisTNG} \citep{Garaldi}. This model includes black hole seeding, modified \citet{Bondi} accretion, and kinetic and thermal \ac{agn} feedback modes \citep[for details see][]{Weinberger}. However, the implementation of explicit radiation transfer in \textsc{Thesan} once again distinguishes the result of this model from that of \textsc{IllustrisTNG}.

In \textsc{Thesan}, gas near a black hole is exposed to a variable radiation field. The intensity and spectrum of this field depend on the distance and properties of local sources, including the black hole itself, nearby stars, and other \acp{agn} \citep{Garaldi,Garaldi2}. The simulation explicitly models the spectral energy distribution of each \ac{agn}, where they are treated as significant sources of ionising radiation \citep{Thesan,Garaldi2}. This more realistic radiation environment alters the gas conditions surrounding black holes. It introduces anisotropic feedback channels by modifying the supply of accretion fuel, influencing host galaxy evolution, and accounting for both \ac{agn} radiation and self-shielding effects in a self-consistent framework \citep{Garaldi,Garaldi2,Zier}.

For instance, more accurate self-shielding in \textsc{Thesan}, facilitated by radiative transfer, can allow the survival of cold, low-density gas phases within galaxies \citep{ThesanZoom,Zier}. If black holes reside in regions with more of this cold, dense gas, it could potentially increase the availability of accretion fuel compared to a scenario where a uniform UV background more readily ionises such gas \citep{ThesanZoom}. Conversely, intense radiation from nearby sources, such as starburst regions or other \acp{agn}, could photo-heat and ionize the gas around a black hole in \textsc{Thesan}, potentially reducing the accretion rate \citep{Garaldi}.

Although the radiative transfer physics introduces this additional complexity, \textsc{Thesan}’s \ac{agn} model includes the same transition to kinetic feedback at high black hole masses as in \textsc{IllustrisTNG}. Even with the suppressed feedback efficiency affecting both thermal and kinetic modes, identified as a numerical issue in the current release \citep{ThesanAddendum}, quenching occurs once black holes reach sufficient mass. Thus, while the subgrid models were originally intended to be equivalent between \textsc{Thesan} and \textsc{IllustrisTNG}, the numerical issues identified in the simulation \citep{ThesanAddendum} mean that the effective black hole physics differs substantially. Combined with the interplay of radiative transfer and feedback, this produces qualitatively different evolutionary pathways for galaxies and their central black holes, particularly during cosmic reionisation.

In this sense, the suppressed \ac{agn} feedback in the public \textsc{Thesan} release overlaps with an area of genuine observational uncertainty. Accretion rates in the early universe are highly uncertain, with models spanning sub-Eddington to strongly super-Eddington regimes \citep{Steinborn, Bennett, Prole}. Observations of high-redshift quasars and \textit{JWST} targets suggest phases of rapid, potentially weakly-coupled black hole growth in at least some systems \citep{Maiolino, WWang}. Combined with the diversity in subgrid prescriptions across simulations, there is no consensus on the correct implementation of \ac{agn} feedback in the early universe. Within this broader context, \textsc{Thesan} effectively provides a controlled numerical experiment exploring the limiting case of globally suppressed \ac{agn} feedback and accelerated black hole growth, offering insight into an observationally relevant regime during cosmic reionisation.

\subsection{Halo and galaxy characterisation}
\label{sec:properties}

\subsubsection{Structural and dynamical halo properties}

In order to study the baryonic components of \acp{mqg}, we extract stellar, gas and black hole masses, stellar and gas half-mass radii, metallicities and star formation rates from the \acp{mpb} of the merger trees. In order to analyse the halo properties of \ac{mqg} hosts, we do the same for \textsc{SubFind} halo mass $(M_h)$, half-mass radius $(R_h)$ and specific angular momentum $(L_h)$. Unlike the complete \textsc{SubFind} mass for haloes, galaxy properties are defined within twice the stellar half-mass radius, to avoid contamination from satellite sources, which can obscure the true quiescent state of some galaxies. 

Similarly to \citet{Chittenden}, we use two of these variables to compute a proxy for the halo's virial velocity: a measure of the halo's central potential and internal particle dynamics. Unlike the rotation curve maximum velocity provided in the merger trees, this proxy is not significantly biased by the prescence or abscence of baryons in the simulation, which is crucial for application of the model to N-body simulations. The proxy $V_h$ is defined in terms of halo mass $M_h$ and halo half-mass radius $R_h$ as follows:

\begin{equation}
V_h = \left( \frac{M_h}{M_\odot} \right)^\frac{1}{2} \left( \frac{R_h}{\text{ckpc}} \right)^{-\frac{1}{2}},
\label{eq:vh}
\end{equation}where constant factors such as the Newtonian gravitational constant are omitted.

\subsubsection{Local and large scale overdensities}

We include overdensity as an additional measure of each halo’s local environment. These values are computed explicitly from the halo catalogues at each snapshot. To ensure future compatibility with pure dark matter simulations, we calculate overdensities using only the dark matter component. This approach differs from the total gas overdensities provided in \textsc{Thesan}'s supplementary data catalog. Although the dark matter and gas overdensities may be correlated, the gas-based values can reflect local biases introduced by variations in gas fraction or star formation rate.

\citet{Chittenden} computed overdensity using a nearest neighbour search with periodic boundary conditions. However, \citet{Chittenden2} found this method to be sensitive to mass resolution when applied to a different simulation. The main issue arose from edge effects near the boundaries of the calculation volume, which particularly affected low-mass haloes and underdense regions. To avoid these limitations, we estimate density at each snapshot using a Gaussian kernel density estimator. We apply a comoving bandwidth and evaluate the local density at the position of each target subhalo.

We investigate the role of environment on the evolution of massive galaxies on multiple scales, aiming to characterise both the effects of localised interactions and the influence of large scale structures such as cosmic filaments. Specifically, we evaluate overdensities with three apertures: 200 \acp{ckpc}, 1 \ac{cMpc} and 3 \acp{cMpc}; which we denote as $\delta_\text{200ckpc}$, $\delta_\text{1cMpc}$ and $\delta_\text{3cMpc}$. \citet{Kimmig} show that \acp{mqg} in the \textsc{Magneticum Pathfinder} simulation from $z=2.79$ to $z=4.23$ reside in underdense regions up to a 2 \ac{cMpc} aperture when compared with massive star-forming galaxies; a difference which becomes negligible on larger scales and at higher redshifts. In contrast, results from the \textsc{Colibre} simulations indicate that \acp{mqg} preferentially reside in overdense environments prior to selection, particularly on local ($\sim0.3$ \ac{cMpc}) and intermediate ($\sim1$ \ac{cMpc}) scales. These overdensities reflect early positions within dense nodes of the cosmic web and correspond to more massive host haloes with deeper potential wells, more closely resembling the environmental trends seen in \textsc{Thesan}.

In \textsc{Colibre}, the environment acts not as a direct quenching agent but as a catalyst for internal processes: overdense regions enhance gas inflows and merger rates, accelerating stellar mass assembly and \ac{smbh} growth \citep{ChandroGomez}. The resulting powerful \ac{agn} feedback, identified as the primary quenching mechanism at these redshifts, drives outflows and shock-heats the circumgalactic medium, suppressing gas cooling and depleting the fuel supply. These outcomes are shaped by the simulation’s subgrid modelling, which explicitly resolves a cold, chemically enriched ISM phase, and permits super-Eddington accretion, both of which promote rapid black hole growth. This contrasts with \textsc{Magneticum}, where strong \ac{agn} coupling can disrupt filaments in comparatively underdense regions. \textsc{Thesan}, like \textsc{Colibre}, associates quenching with overdense environments, though environmental distinctions may evolve differently by the selection epoch. Together, these results suggest that small scale environment influences \ac{mqg} formation in a manner which remains tightly coupled to the adopted feedback implementation.

\subsubsection{Assembly timescales and growth metrics}
\label{sec:beta}

For each of the four halo properties and three calculated overdensities, we implement measures of the timescales on which these quantities grow to their final values, and the time at which the masses, structures and enviornments are achieved. Traditionally, these quantities are used to measure mass assembly, yet we are interested as well in the formation of gravitational potentials and filamentary networks which lead to the assembly and quenching of massive galaxies. As many quantities in the halo formation history are inherently correlated, however, their temporal geometries can be exploited in the same way to parameterise the growth of these halo properties.

To characterise the proportionate speed and time scale of a halo's growth, we use the specific accretion gradient introduced by \citet{Montero-Dorta}, denoted $\beta$. These authors define the mass accretion gradient as the slope of a straight line fit to the logarithmic specific mass accretion rate of a halo as a function of logarithmic time. They show its superiority to the more commonly used half-mass formation time in predicting stellar masses, gas fractions, photometry and assembly bias in \textsc{IllustrisTNG} data. \citet{Chittenden} also use the $\beta$ gradient of halo mass to predict star formation and metallicity histories in \textsc{IllustrisTNG}. This parameter helps constrain both the stellar-halo mass and mass-magnitude relations, and serves as a diagnostic for identifying poorly resolved haloes, due to the Gaussian distribution of its values.

\begin{figure}
\includegraphics[width=\linewidth]{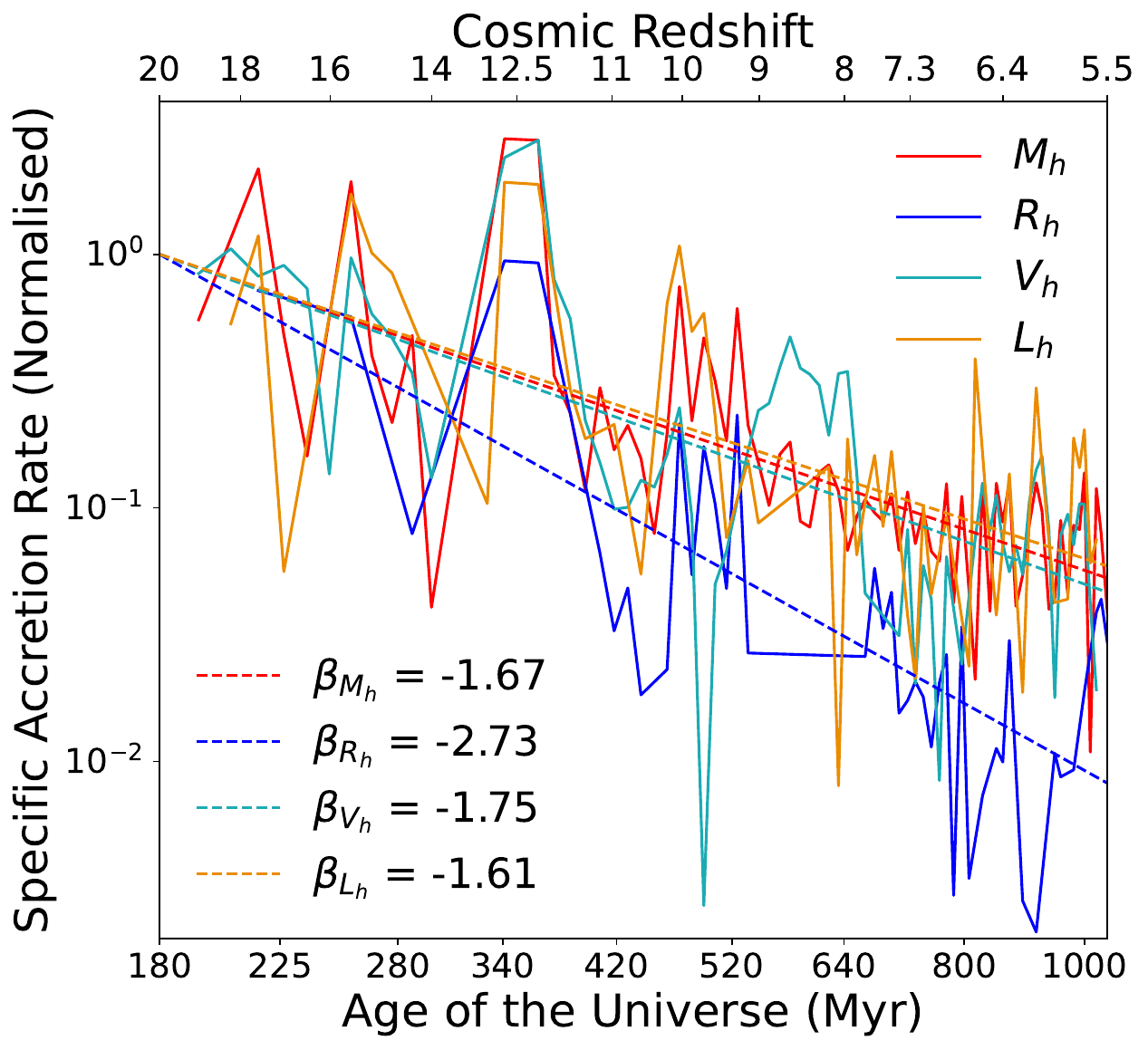}
\vspace*{-20pt}
\caption{Specific accretion gradients for four properties of an example halo in \textsc{Thesan-1}, where $\log_{10} ( M_h / M_\odot) \simeq 11.47$ at $z=5.5$. All specific accretion units have been scaled according to the best fit value at $z=20$. For this example, each of these gradients are of similar value, with the exception of the steeper radial growth history gradient; which may indicate a tidal interaction with large scale structure, considering that the 3 \acp{cMpc} scale overdensity for this sample is exceptionally high for haloes of this mass.}
\label{fig:fourbetas}
\end{figure}

As we are interested in the development of a halo's size, potential, environment and spin as well as its mass, we compute $\beta$ gradients for all aforementioned halo properties.  In \cref{fig:fourbetas}, we define the gradients using an example halo from \textsc{Thesan-1}. The similar structure of the formation histories for these baseline properties enables consistent fitting across all cases. The mass accretion gradient $\beta_{M_h}$ directly reflects the rate of halo assembly, however, other gradients carry additional physical meaning beyond simple growth rates.

Firstly, the angular momentum gradient $\beta_{L_h}$, like $\beta_{M_h}$, follows a Gaussian distribution, but this Gaussian has a larger width, and its mean corresponds to a shallower $\beta$ gradient. This is because a halo can lose a significant fraction of its angular momentum, such as through retrograde interactions with its environment. Extreme values of $\beta_{L_h}$ are therefore not necessarily indicative of poor mass or spatial resolution, but simply a halo which does not consistently gain this angular momentum.

Second, environmental gradients show a positively skewed distribution, which appears more Poissonian than Gaussian. Some haloes occupy environments that grow steadily over time, whereas others remain in regions with little or no environmental evolution, especially on large spatial scales. This is because small-scale environments tend to change rapidly, influenced by interactions between neighbouring haloes. In contrast, megaparsec-scale structures evolve more slowly, as they form through the gradual clustering of galaxies and groups.

For each of the above cases, the evaluated gradients are close to zero where there is little to no net growth in the spin or environment of this halo.

As in \citet{Chittenden}, we use the $\beta$ gradients to identify and exclude haloes with unphysical or poorly resolved accretion histories. Specifically, we discard any halo for which a $\beta$ value deviates by more than 5$\sigma$ from the best-fit Gaussian distribution, as such outliers often indicate resolution artefacts or irregular behaviour which is not representative of the underlying population.

In addition to the specific accretion gradients of these properties, we include two other scalar variables per property to our analysis: scaled accretion timescale, and half-peak formation redshift. We find in the \textsc{Thesan} data that haloes and galaxies begin to grow and acquire significant fractions of their mass at a wide range of times in the simulation's time domain, which is less true for simulations which terminate at much lower redshifts. The $\beta$ values, while practical as measures for sustained accretion, do not strictly capture this.

The scaled accretion times, denoted $a$, were introduced by \citet{Shi}, and are defined as the ratio of Hubble scale factors between two key snapshots: the time of maximised halo mass to the time that half of this value is reached. In combination with the half-peak formation redshift, these indicate when the main body of the halo is established, and how it continues to grow after this. In contrast with $\beta$, these quantities measure the critical times in the halo's history rather than the global rate of assembly, and so distinguish early-forming and late-forming haloes from rapidly and slowly growing haloes.

\subsubsection{Merger histories}

The interactions of galaxy-hosting haloes with their surroundings, such as the acquisition of secondary structures from the infall of adjacent haloes, has long been established as a significant phenomenon influencing the evolution of galaxies. In our analysis, we consider key properties of the merger histories of massive haloes as a means to differentiate star-forming and quiescent galaxies.

Following the methods of \citet{McGibbonKhochfar2}, we select merging progenitors by identifying subhaloes whose main descendant is on the \ac{mpb} of the target subhalo, while not being a member of the \ac{mpb} itself. We refer to this dataset as the \ac{spl} of the merger tree: the list of all subhaloes which merge directly with subhaloes along the \ac{mpb}. For each subhalo, we compute the following merger quantities from its \ac{spl}:

\begin{itemize}
\item $N_\mu$: Total number of mergers.
\item $N_{\mu>\sfrac{1}{4}}$: Total number of major mergers, defined hereafter as mergers whose mass ratio exceeds $\sfrac{1}{4}$.
\item $\mu_\text{max}$: Largest merger ratio in the subhalo's history.
\item $\mu_\text{med}$: Median merger ratio throughout the subhalo's history.
\item $\bar{z}$: Characteristic merger time, defined as the average redshift of all mergers weighted by their mass ratio.
\item $z_{\mu_\text{max}}$: Redshift at which the merger with the largest mass ratio takes place.
\item $z_{\mu>\sfrac{1}{4}}$: Redshift at which the most recent major merger takes place; if none exist, equal to $z_{\mu_\text{max}}$.
\end{itemize}

\section{Selection and Classification of Massive Quiescent Galaxies}
\label{sec:selection}

\subsection{Dimensionality Reduction and Parameter Clustering}

\subsubsection{Principal component analysis}

\Ac{pca} \citep{MurtaghHeck, Kendall, Jolliffe} is a linear data compression technique used to reduce the dimensionality of a dataset while preserving as much variability as possible. It works by identifying the \acp{pc} along which the data varies the most, i.e. the eigenvectors of the covariance matrix of the data, and the corresponding eigenvalues which indicate the amount of variance along each direction. \ac{pca} projects the original data onto these principal components, resulting in a new set of uncorrelated variables ordered by the amount of variance they capture. The first few principal components typically capture the most significant patterns in the data, allowing for a more compact representation while minimising information loss.

In this study, we use \ac{pca} not simply as a dimensionality reduction tool, but as a framework for identifying the combinations of halo, galaxy, and environmental properties that most strongly differentiate \aclp{mqg} from the broader galaxy population. We apply \ac{pca} to all \textsc{Thesan-1} galaxies with a final stellar mass above $10^7 M_\odot$. This provides a sample of 35,963 haloes following the gradient quality cuts described in \cref{sec:beta}, which eliminate a total of 467 samples. While this includes a large number of low-mass galaxies irrelevant to our final analysis, it allows the probabilistic clustering algorithm described in \cref{sec:GMM} to identify a physically meaningful group containing all \acp{mqg} in the simulation. This allows us to extract a predictive subset without directly imposing any threshold criteria during clustering.

By examining the dominant contributing variables in each \acl{pc}, we quantify which combinations of physical properties, such as early halo assembly, compactness, \ac{smbh} mass, and environmental density, systematically correlate with quenching. In this way, \ac{pca} serves a dual purpose: it enables efficient clustering in high-dimensional space while also offering a physically interpretable decomposition of the parameter space. This statistical foundation ensures that the identification of rare populations like \acp{mqg} is robust, reproducible, and free from arbitrary thresholds.

\subsubsection{Variable normalisation}

The \ac{pca} approach requires normalisation of the data, which cannot be done equivalently for all data due to extreme differences in their data distributions. In \citet{Chittenden}, most variables are normalised by \ac{gqt}, where the distribution of each variable is mapped onto a normal distribution, regardless of its initial distribution. They show that the normal distribution was more suitable than a uniform distribution due to a steep gradient between the original and normalised data, which made it unsuitable for optimising a data model. In \textsc{Thesan}, we see this for normal distributions as well, and so we apply an alternative method to quantile transformation.

Aside from $\beta$ gradients and formation times, halo and merger properties are strictly positive-valued, and their distributions often appear approximately Gaussian or Poissonian along a logarithmic axis. By fitting a straight line between the logarithmic value of a variable and its normalised percentile score in the distribution, each variable is translated to a distribution between 0 and 1 which resembles its true distribution in logarithmic space. We refer to this mode of normalisation as \iac{lt}. $\beta$ gradients and formation times, on the other hand, follow basic distributions on linear axes, and so are only transformed by a minimum-maximum scaler.

\subsubsection{Gaussian Mixture Models and subsample isolation}
\label{sec:GMM}

To further analyse the \ac{pca} distribution, we make use of a \ac{gmm} to isolate the \ac{mqg} haloes as much as possible. \Acp{gmm} are probabilistic models used to represent a dataset as a mixture of several Gaussian distributions, each with its own mean, covariance, and weight. \Acp{gmm} assume that the data is generated from a combination of these Gaussian distributions, and the goal is to identify the parameters of each Gaussian which best explain the data.

The \ac{em} algorithm is commonly used to fit a \ac{gmm} to the data. In the expectation step, the algorithm estimates the probability that each data point belongs to each Gaussian component, and in the maximisation step, it updates the parameters of the Gaussians to maximise the likelihood of the data given these assignments. This iterative process continues until the model converges, resulting in a set of Gaussian distributions which collectively describe the underlying structure of the data.

\Acp{gmm} are particularly useful for data clustering, as they can model complex, multimodal data distributions. In order to identify the halo properties specific to \acp{mqg}, we run a \ac{gmm} fitting algorithm using the Gaussian mixture module provided by the SciKit-Learn Python library \citep{SKLearn}.

For the sake of convergence, the \ac{gmm} is trained on the seven largest \acp{pc}, constituting $\sim 62 \%$ of the variance in the 35-parameter dataset. As we are specifically interested in the \ac{gmm} cluster\footnote{In the context of this paper,a  'cluster' refers to a parametric cluster in principal component space; not an astrophysical galaxy cluster, unless explicitly stated otherwise.} which contains \acp{mqg}, instead of running a single \ac{gmm} fitting algorithm while tuning the number of clusters or iterations, we recursively run the algorithm on the \ac{mqg}-containing cluster, each time using a random cluster number between 3 and 6, inclusive. This iteration onto smaller and smaller clusters continues until the algorithm can no longer create a smaller cluster than that which contains all \acp{mqg}. Any \ac{gmm} clustering result which splits this population is discarded.

\subsection{Evolutionary bifurcation in massive galaxies}

\begin{figure}
\includegraphics[width=\linewidth]{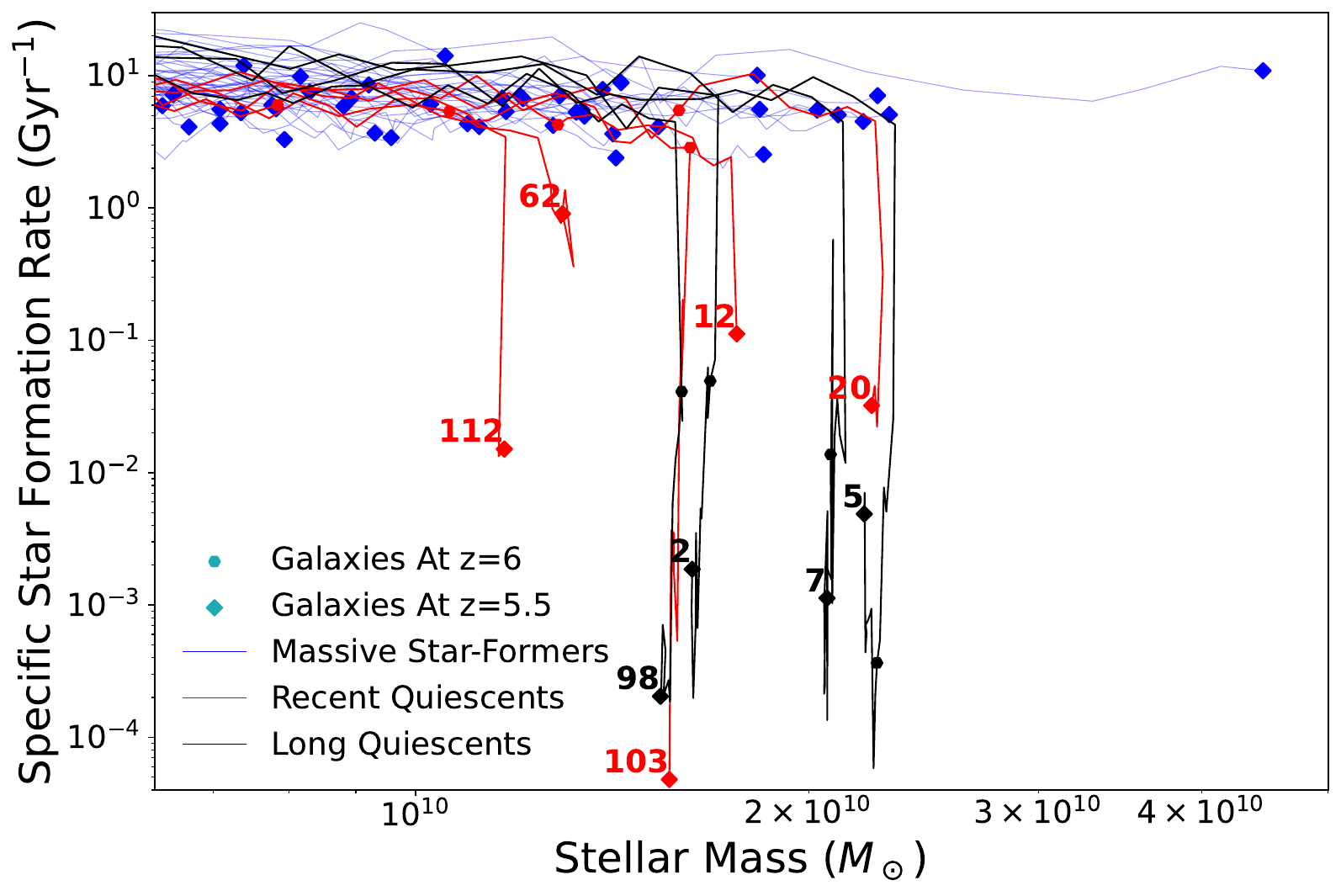}
\vspace*{-20pt}
\caption{The evolutionary trajectories of the 0.2\% most massive galaxies in terms of stellar mass and specific star formation rate over time, categorised into star-forming galaxies (blue), recently quenched galaxies (red), and long quiescent galaxies (black). The state of each quiescent galaxy at $z=6$ and $z=5.5$ is marked with a hexagon and diamond point respectively, to highlight the speed with which these galaxies grow and quench. The points at $z=6$ are omitted for star-forming galaxies for clarity, and given that they are not significant to their evolution or classification, as they are for quiescent galaxies. The rapid cessation of star formation observed in massive quiescent galaxies is distinctive to this group; other high-mass galaxies persist on the star-forming main sequence throughout the simulation. Additionally, the massive quiescent galaxies show minimal to no resurgence in star formation, implying that transient quiescence is not a visible phenomenon among high mass galaxies within the time domain of the \textsc{Thesan} simulations. Each quiescent galaxy is labelled with its halo's FoF table index at $z=5.5$.}
\label{fig:MssSFR}
\end{figure}

A defining feature of \aclp{mqg} in \textsc{Thesan-1} is the abrupt drop in specific star formation rate. This decline occurs on timescales shorter than the 10 Myr interval between snapshots, and typically coincides with a reduction in \ac{smbh} accretion. The cause is the depletion of cold gas reservoirs, driven by a combination of stellar and \ac{agn} feedback \citep{XShen2}. As a result, these quiescent galaxies fall well below the star-forming main sequence and have no analogue in other high redshift simulations. \citet{Carnallmnras2}, using \textit{JWST} data, fit model star formation histories to observed early-forming \acp{mqg} and find trajectories that closely resemble those in \textsc{Thesan}. This suggests that the adventitious feedback presciption in \textsc{Thesan} could explain their evolution.

However, there are a limited sample of \acp{mqg} in \textsc{Thesan-1}, previously shown in \cref{tab:MQs}, where it can be seen that most of our \acp{mqg} quench very close to the end of the simulation run at $z=5.5$. We identify nine galaxies which concomitantly exhibit a rapid drop in star formation, grow to $\gtrsim 10^{10} M_\odot$ in stellar mass, and fall to $\lesssim 1 \text{Gyr}^{-1}$ in specific star formation rate by $z=5.5$. However, there are only four in the $z=5.5$ sample of nine which meet these criteria by $z=6$. We refer to this subset as \acp{lqg}, to distinguish from the remaining \acp{mqg} population, which we term \acp{rqg}. This distinction allows us to compare the properties of \acp{mqg} and their haloes at the time of quenching and following a sustained period of quiescence, to provide insight on their immediate post-quenching evolution.

Some galaxies are known to undergo temporary quiescence, where the star formation rate drops to a low value and later resumes \citep[e.g.][]{Looser}. However, this behaviour does not appear in massive galaxies within \textsc{Thesan}. In \cref{fig:MssSFR}, we present the evolution of stellar mass and specific star formation rate for the 0.2\% most massive galaxies at $z = 5.5$. The results show that the rapid drop in star formation is a feature unique to the \acp{mqg}. Other high-mass galaxies remain on the star-forming main sequence throughout the simulation, and no \acp{mqg} exhibit a significant recovery in star formation after quenching. Although the \acp{agn} in these systems slow their accretion, they continue to grow and heat the surrounding gas. This sustained heating likely suppresses any chance of star formation resuming. These trends suggest that temporary quiescence is not prevalent for high-mass galaxies at this epoch.

However, there exists one quenching galaxy (FoF ID 62) whose specific star formation rate does not decline as dramatically as the other eight galaxies. Despite its distinction from the main sequence, its star formation activity fluctuates just below $1 \text{Gyr}^{-1}$. This galaxy and its halo do grow similarly to the remaining \acp{mqg}, yet it exhibits a smaller \ac{agn} and less turbulent CGM, suggesting an as-yet weaker but developing phase of quenching.

A separate analysis of properties of \acp{rqg} and \acp{lqg} provides an advantage, in that it distinguishes halo and environmental properties which depend on the time for which the galaxy has been quiescent. This provides some insight into their evolution post-quenching, and possibly into observing descendants of these galaxies at lower redshifts.

\section{Characterising Massive Galaxies And Haloes}
\label{sec:comp}

\begin{figure}
\includegraphics[width=\linewidth]{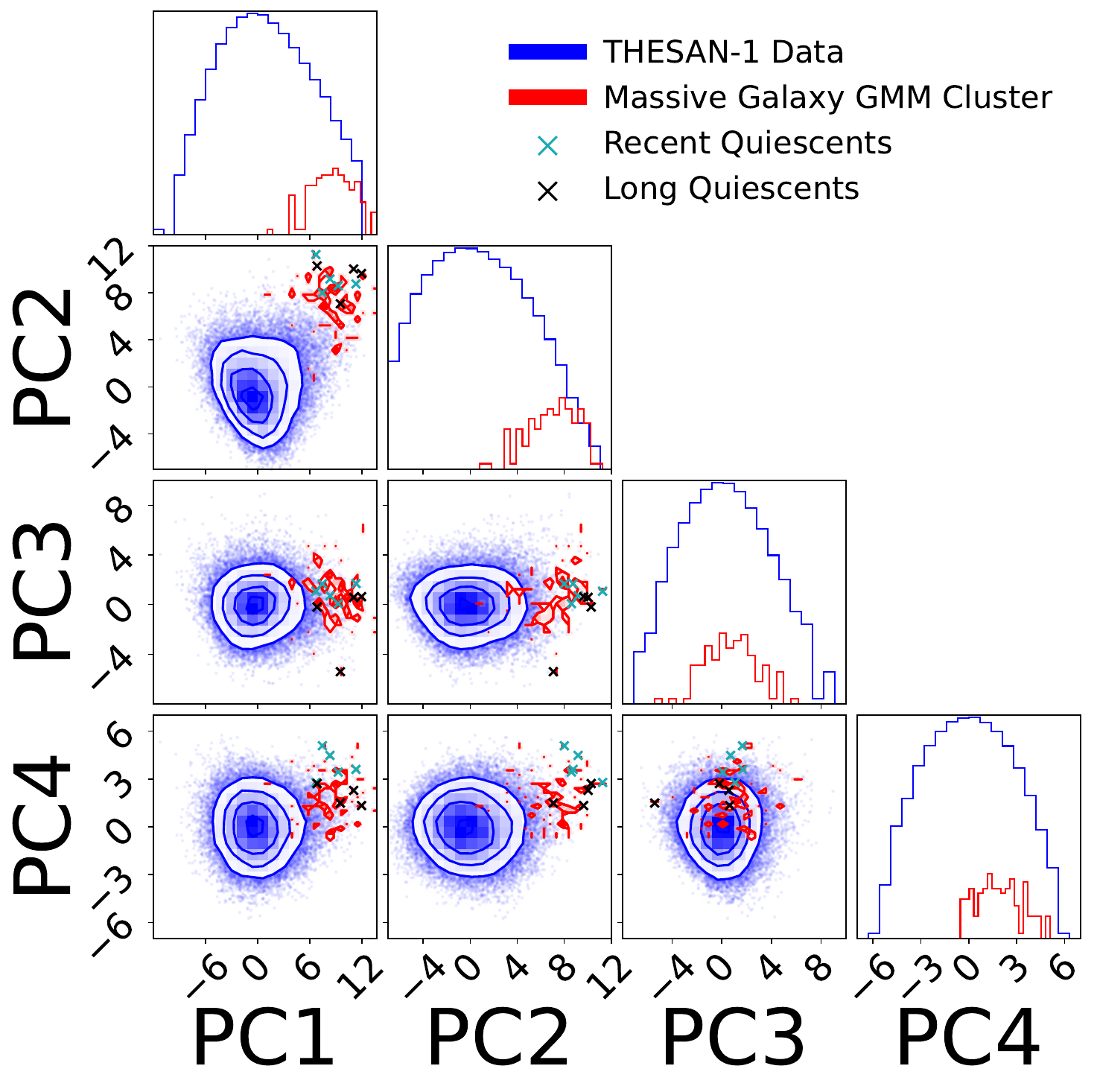}
\vspace*{-20pt}
\caption{A visualisation of the principal component space showing the four most important components, showing the distribution of \textsc{Thesan} galaxies in blue, and the \ac{gmm} cluster containing massive quiescent galaxies in red. The quiescent galaxies themselves are shown as teal cross-points for recently quenched samples, and black cross-points for long quiescent galaxies. This parameter space shows that the \ac{gmm} cluster is clearly separated from the rest of the galaxy population in the space of the first two principal components, while the quiescent galaxies become separated in subsequent dimensions. Thus, these main components can signify which halo and environment properties are important to the evolution of different massive galaxies.}
\label{fig:clustercorner}
\end{figure}

\subsection{Features of the massive galaxy GMM cluster}
\label{sec:fullcluster}

We show in \cref{fig:clustercorner} how the \ac{gmm} cluster containing \acp{mqg} compares with the general \textsc{Thesan} data for the four largest \aclp{pc}. The \ac{gmm} cluster containing all \acp{mqg} is clearly separated from the overall galaxy population in the \ac{pc} space, but this is particularly so for the \acp{mqg} themselves. The median \ac{pc} value for \acp{mqg} is in the top 0.283\% of PC1 values, and the top 0.057\% of PC2 values; compared with 0.455\% and 0.307\% for the \ac{gmm} cluster as a whole. The GMM cluster has a median halo mass of $2.83\times 10^{10} M_\odot$, compared with $4.81\times 10^{10} M_\odot$ for haloes hosting \acp{mqg}.

Clearly, these components are practical for selecting haloes which could host \aclp{mqg}, with the \ac{gmm} cluster being visibly distinct from the main distribution of haloes. Furthermore, it can be seen in \cref{fig:clustercorner} that there is a separation of recent and long quiescent galaxies in \acp{pc} 1 and 4, which may indicate the existence of halo properties which align with the period of a galaxy's quiescence.

\begin{figure}
\includegraphics[width=\linewidth]{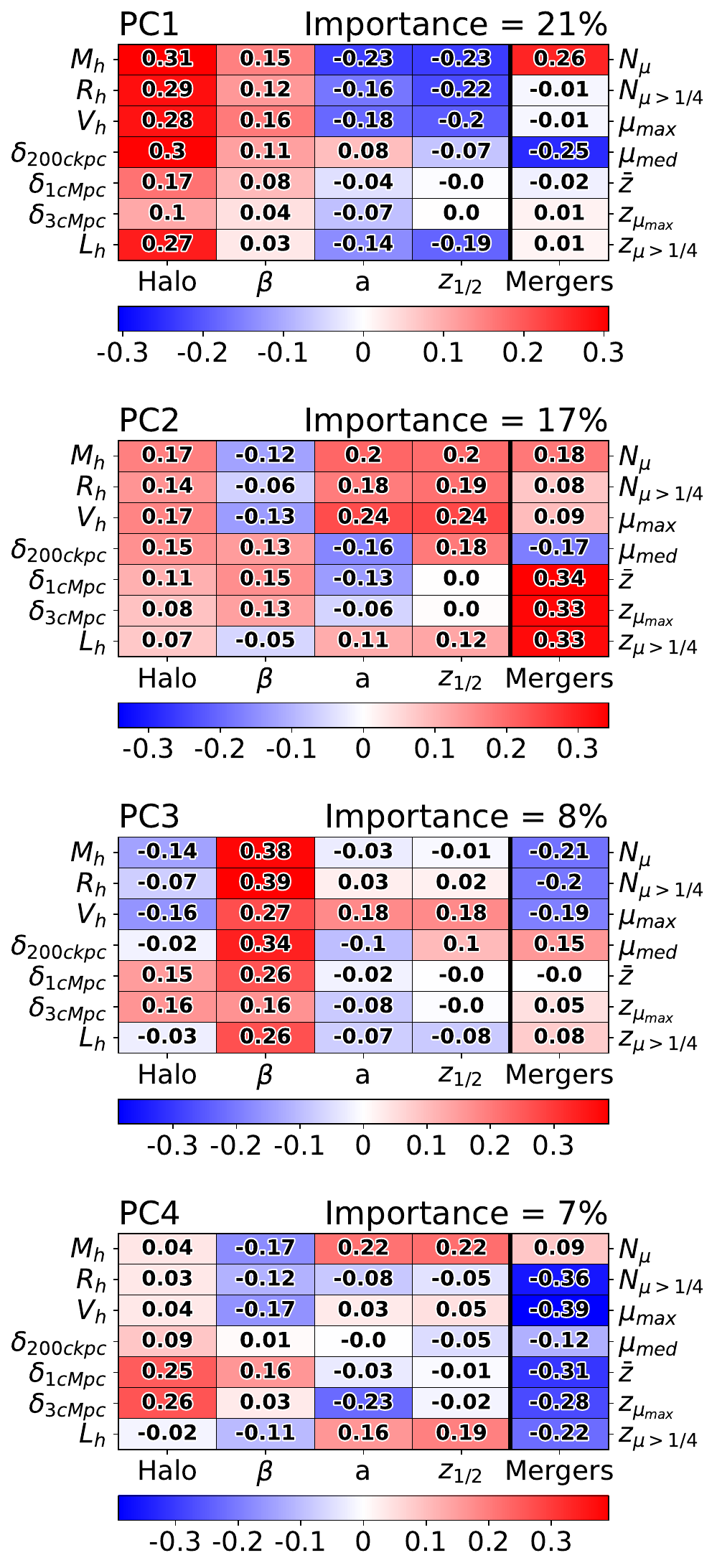}
\vspace*{-20pt}
\caption{A visualisation of the first four eigenvectors of the principal component space, in a tabular form which shows the halo and environment quantities on the vertical axis, and baseline and historical properties on the horizontal axis. The contributions of each physical quantity to its eigenvector are indicated by the colour scale, while the numbers in each cell give a precise value. The PCA eigenvalues are quoted as percentage importance values - quantifying the relative amount of variance in each component. The thick, black vertical line separates this tabular form from merger properties, which are labelled differently and are unconnected to the remaining variables. All notation is equivalent to that given previously in the main text of this paper. This figure shows how the baseline halo properties and early formation histories dominate the first two components, which distinguish the \ac{gmm} cluster in \cref{fig:clustercorner}, but the remaining variables relating to accretion gradients, large scale environments and the lack of mergers affect the smaller components, which affect quiescent galaxies within the \ac{gmm} cluster.}
\label{fig:pcaeig}
\end{figure}

In the SciKit-Learn \ac{pca} module, the eigenvectors are obtained by performing a singular value decomposition on the standardised data matrix, and represent the directions (principal components) along which the variance in the data is maximised. Mathenatically, the \ac{pca} eigenvectors are obtained from the solutions of the eigenvalue equation $\mathbf{\Sigma} \bm{\nu}_j = \lambda_j \bm{\nu}_j$, where the covariance matrix $\mathbf{\Sigma}$ is defined by the dataset $\mathbf{X}$, of sample size $N$, and its feature-wise mean $\mathbf{\bar{X}}$ as follows:

\begin{equation}
\mathlarger{\mathbf{\Sigma}} = \frac{\left( \mathbf{X} - \mathbf{\bar{X}} \right)^\text{T} \left( \mathbf{X} - \mathbf{\bar{X}} \right) }{N-1}
\label{eq:covarmat}
\end{equation}

The four \ac{pca} eigenvectors with the highest variance, shown in \cref{fig:pcaeig}, are presented in tabular form to group physical variables by halo and environment properties, and by baseline, historical, and merger-related features. This decomposition reveals which parameter combinations best separate \acp{mqg} from the broader galaxy population. It also provides a selection framework that is both physically interpretable and statistically robust.

The first \ac{pc} reflects the halo’s mass, radius, and small-scale environment, and defines the overall scale of the host structure. The second component shows strong correlations with the growth times of mass and potential, as well as with the redshifts of major mergers. In the case of \acp{mqg}, this component tracks the formation epoch of their earliest progenitors, since these systems lack significant merger activity. Together, these components trace the compact and early-forming nature of the most extreme haloes.

The simulation data outside the \ac{gmm} cluster include some haloes with comparable mass. However, galaxies within the \ac{gmm} cluster tend to have higher stellar masses, earlier mass and potential formation times, and larger median merger redshifts ($\bar{z}$). In contrast, haloes outside the cluster show higher merger ratios. The \ac{gmm} cluster haloes also host massive galaxies with larger stellar/halo mass ratios\footnote{The GMM cluster's median $M_s/M_h$ value is approximately twice as large compared with the 1\% most massive haloes in our \textsc{Thesan} dataset.}, larger stellar and gas phase metallicities, and smaller stellar half-mass radii. These distinctions suggest that the \ac{gmm} clustering algorithm effectively separates different modes of the galaxy-halo connection, without the explicit inclusion of baryonic data.

The other components in \cref{fig:clustercorner} are driven by accretion gradients, maximum and median merger ratios and the number of major mergers. This suggests that \ac{gmm} cluster galaxies and \acp{mqg} on the whole reside in high mass, early-forming haloes with rapidly growing potentials, whereas the growth timescales and merger incidences influence the quiescence time to a lesser extent.

Halo mass and potential, the most important variables for hosting massive galaxies, exhibit small contributions to the second and third \acp{pc}. In contrast, \ac{pc} 1 shows that these variables lie nearly orthogonal to major merger count, maximum merger ratio, large-scale environment, and the growth time of angular momentum. This suggests that the lesser principal components capture variation in these secondary properties independently of halo mass. \Ac{pc} 4 reveals a strong correlation between large-scale environment, major merger frequency, and maximum merger ratio. This implies that external structure plays a significant role in driving the stochastic nature of halo growth.

This suggests that massive galaxies, whether star-forming or quiescent, reside in large haloes and dense small-scale environments, growing rapidly at early times from the coalescence of subhaloes of similar masses. Despite the lack of a clear offset of the massive galaxy \ac{gmm} cluster in \acp{pc} 3 and 4, there is some offset of the \ac{rqg} and \ac{lqg} points from the centre of the \ac{gmm} cluster on these axes. Therefore, the \acp{mqg} are distinguished from the cluster by extreme values of accretion gradients, large scale environments and major merger numbers. These variables could, in future work, be used to identify objects capable of hosting \acp{mqg} in the populations of simulations and deep surveys.

\subsection{Distinguishing quiescent galaxies within the GMM cluster}
\label{sec:incluster}
\subsubsection{Halo properties and cosmic environments}

The \ac{gmm} cluster has isolated ninety haloes with the highest masses, densest local environments and earliest formation times, which are critical factors for hosting \aclp{mqg}. The \ac{gmm} clustering algorithm alone, however, cannot distinguish \acp{mqg} completely, and still encompasses star-forming galaxies of similar mass.

Given the small sample size of haloes in the \ac{gmm} cluster, we aim to identify significant differences in \ac{mqg} properties by comparing the mean and median percentile scores of halo and galaxy quantities in the \ac{rqg} and \ac{lqg} subsets against those in the full \ac{gmm} cluster. We draw 10,000 random samples for each population: star-forming galaxies, \acp{rqg}, and \acp{lqg}, and evaluate the three quartiles of the resulting distributions for each statistical moment. We compute percentile scores relative to the full simulation dataset to place each population in context with the general galaxy population. This method removes the influence of skewed or arbitrary distributions on the inferred statistical significance.

\begin{figure*}
\includegraphics[width=\linewidth]{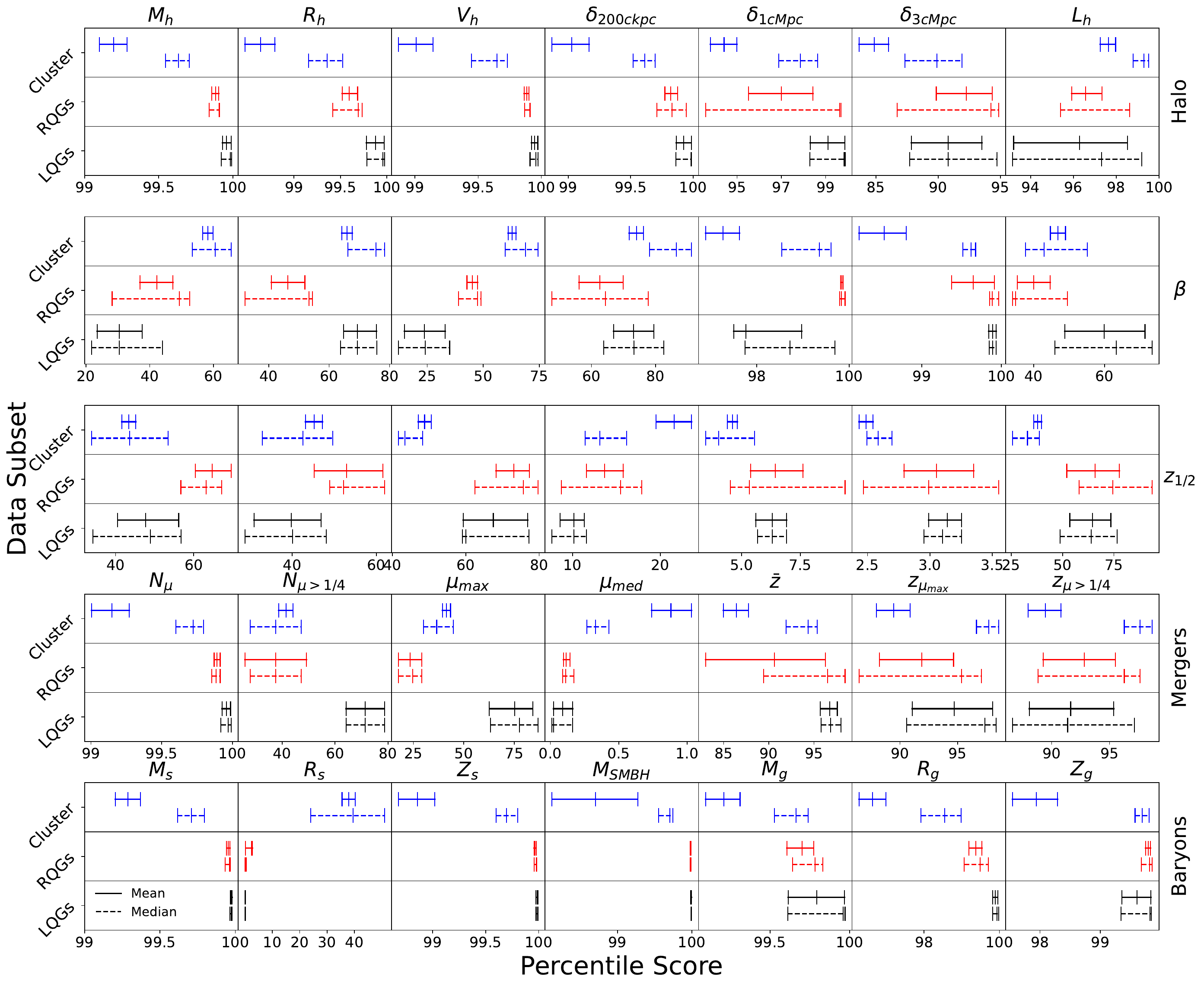}
\vspace*{-20pt}
\caption{Bootstrapped mean (solid lines) and median (dashed lines) percentile scores for key variables in the \textsc{Thesan} data, showing mean values with error bars from ten thousand bootstrap samples. While the bottom two rows show merger-related variables and baryonic properties, as indicated by the labels on the right, the first three rows constitute a tabular arrangement of figures. The labels at the top indicate the halo or environmental property, and the labels on the right indicate the baseline or historical variable category; i.e. the second row and third column shows the $\beta$ gradient of the proxy for maximum circular velocity and halo potential. These figures compare the percentile scores for star-forming \ac{gmm} cluster galaxies (blue) against recently quenched galaxies (red) and long quiescent galaxies (black). In many cases, the \ac{gmm} cluster scores are close to the extrema of the \textsc{Thesan} distribution, yet the recently quenched galaxies exceed the scores of the \ac{gmm} cluster, while the long quiescent galaxies exceed even them. This shows that galaxies which quench earlier have also acquired the highest stellar mass, through early and fast growth of their haloes and environments.}
\label{fig:pcastats}
\end{figure*}

In \cref{fig:pcastats} we show bootstrapped mean and median percentile scores for key components of the \textsc{Thesan} data, for all galaxies whose stellar mass exceeds $10^7 M_\odot$ at $z=5.5$. This compares the complete massive galaxy \ac{gmm} cluster (blue) with \acp{rqg} (red) and \acp{lqg} (black). We see consistently that \acp{mqg} reside in haloes of higher mass, radius and circular velocity at $z=5.5$, with the highest values taken by \acp{lqg}. As indicated by the panels containing stellar mass, radius and metallicity, the galaxies themselves are particularly massive, compact and metal-rich when compared with the \ac{gmm} cluster, and are even comparatively small to low-mass galaxies, as indicated by their low percentile score. On the contrary, the panels which contain the equivalent gas quantities show that the gas component of the halo-galaxy ensemble is massive, metal-rich and extended.

One aspect of the \acp{rqg} and \acp{lqg} is their extreme \ac{smbh} mass with respect to the \ac{gmm} cluster. \citet{XShen2} show that the star formation activity of \textsc{Thesan} galaxies is tightly correlated with this \ac{smbh} accretion, which indicates that this is a key part of the internal quenching process. It could be argued that the \acp{agn} reaching a threshold mass leads to quenching by introducing kinetic feedback; yet the \acp{mqg} are not unique in this regard, with some of their star-forming counterparts having similar \ac{smbh} mass. However, we notice that \acp{mqg} have a preferentially higher \ac{smbh} mass fraction (shown in \cref{fig:ghe}). Due to the strong correlation between \ac{agn} activity and \ac{smbh} mass observed in \textsc{Thesan}, it appears that the \ac{agn} feedback effect is larger in proportion to the gas reservoir. It is plausible that this rapid quenching quenches compact galaxies more efficiently, operating over longer timescales in more extended galaxies.

Another feature of \acp{mqg}, one which distinguishes \aclp{rqg} from \aclp{lqg}, is the enhancement of dark matter overdensities on different scales. We show in \cref{fig:pcastats} that on small (200 \acp{ckpc}) scales, the same hierarchy between subsets as seen with masses exists. Yet on intermediate (1 \ac{cMpc}) scales, the \ac{rqg} densities show no difference in relation to the \ac{gmm} cluster. On large (3 \acp{cMpc}) scales, this information is lost for \acp{lqg} as well. Hence, there appears to be a dependence on the time in which a massive galaxy has been quenched on the environmental densities measured on larger scales. This agrees with the distinction made by the \aclp{pc}, where the component which characterises dense large scale environments in absence of mergers separates the quiescent galaxies independently of their mass.

\citet{RemusKimmig} explore the gas densities surrounding \acp{mqg} in the \textsc{Magneticum} simulation at $z=3.4$; approximately 840 Myr post-reionisation, according to their cosmology. They also find that the galaxies' immediate surroundings contain dense gas, but they do not typically reside at nodes of the cosmic web. They too suggest that a dense large scale environment is not paramount for massive galaxy quenching, while the prescence of massive \acp{smbh} in such environments rules out the simplified picture of quenching owing solely to \ac{smbh} mass.

\citet{Kimmig} further show that small-scale densities around \textsc{Magneticum} \acp{mqg} are lower than those of star-forming galaxies of the same stellar mass, likely due to filament disruption from strong \ac{agn} feedback. This contrasts with \textsc{Thesan}, where \acp{mqg}, particularly \acp{lqg}, inhabit the densest environments in the simulation, and where the link between quiescence and overdensity persists to higher redshift than previously reported for \textsc{Magneticum} or \textsc{IllustrisTNG}.

The \textsc{Colibre} simulations instead align more closely with \textsc{Thesan}. In \textsc{Colibre}, \acp{mqg} preferentially reside in overdense regions on small and intermediate scales, especially prior to selection ($z>4$), though these contrasts diminish by $z\sim3$ as structures merge \citep{ChandroGomez}. Unlike \textsc{Magneticum}, \textsc{Colibre} links dense environments to accelerated \ac{smbh} growth: overdensities enhance gas inflows, drive rapid black hole growth (aided by cold ISM resolution and super-Eddington accretion), and trigger powerful \ac{agn} feedback which depletes molecular gas and dust. Thus, while \textsc{Magneticum} suggests quenching can occur in local underdensities due to efficient coupling, \textsc{Colibre} and \textsc{Thesan} associate quenching with overdense fuel reservoirs. The environmental dependence of high redshift \ac{mqg} formation therefore remains strongly model-dependent and closely tied to subgrid feedback prescriptions.

\subsubsection{Assembly and quenching of massive galaxies}
\label{sec:growthhistory}

We find from bootstrapped accretion gradient values in \cref{fig:pcastats} that the hierarchy between \ac{gmm} cluster galaxies, \acp{rqg} and \acp{lqg} exists for accretion gradients for halo mass and circular velocity. This highlights the faster growth of haloes and their potential wells, which allows them to host massive \acp{agn}. These trends align with the principal components identified in \cref{fig:clustercorner,fig:pcaeig}, where PC1 and PC2 capture variations in baseline halo properties and early growth timescales.

The characteristic redshifts of mergers are a key part of the \acl{pc} space which defines the \ac{gmm} cluster, yet the difference between these values between subsets of the \ac{gmm} cluster is less significant. \Ac{gmm} cluster haloes have similar merger histories irrespective of the galaxies they host, which suggests that it is growth by smooth accretion, reflected in accretion gradients and early formation times, which distinguishes these haloes.

\begin{figure*}
\includegraphics[width=\linewidth]{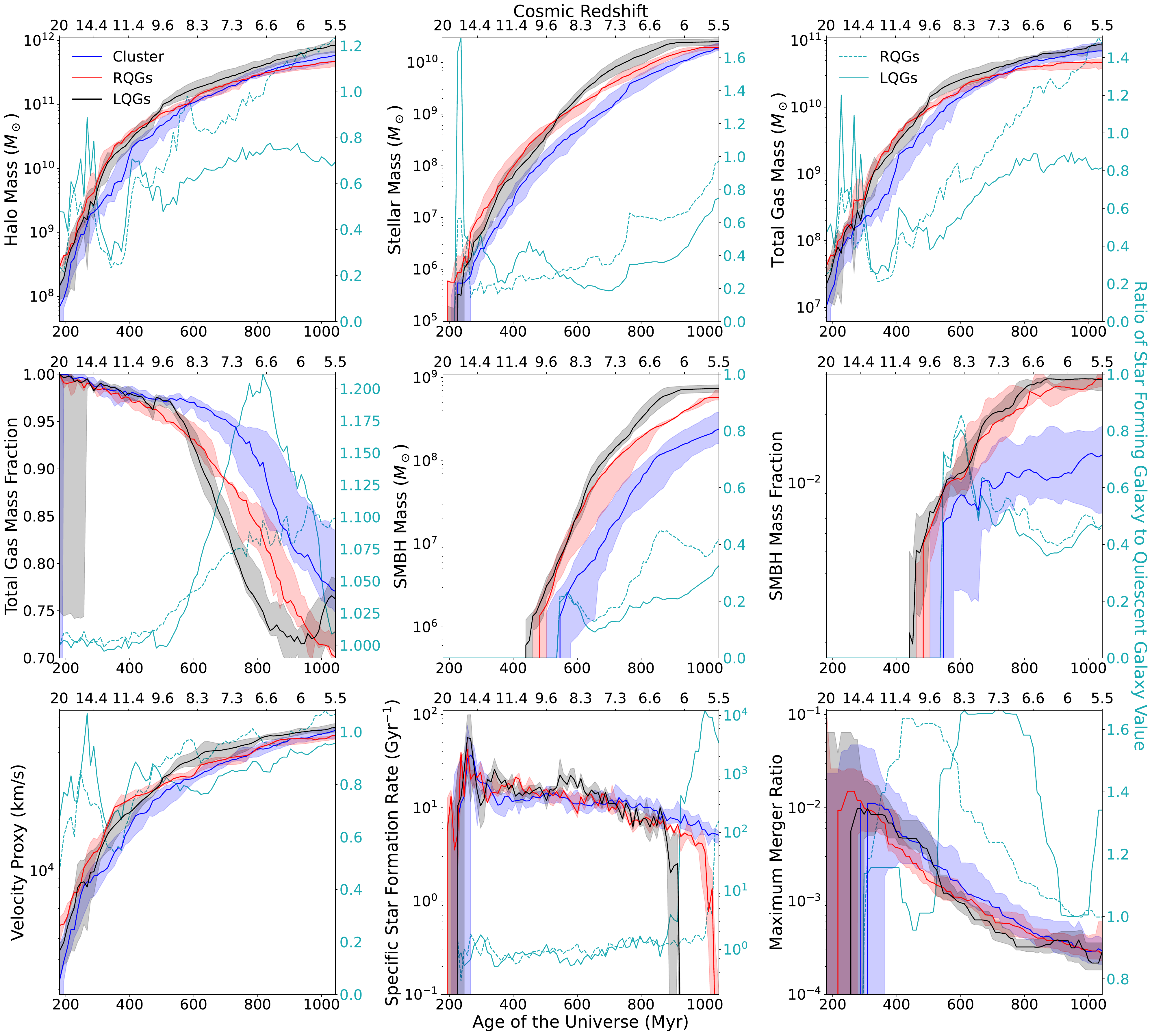}
\vspace*{-20pt}
\caption{The median (solid lines) and interquartile ranges (shaded regions) of different evolutionary properties of star-forming \ac{gmm} cluster galaxies (blue), recently quenched galaxies (red) and long quiescent galaxies (black). The highest merger ratio per snapshot, shown in the lower right panel, has been smoothed with a median filter for clarity, showing that all objects within the \ac{gmm} cluster have insignificant merger activity throughout the simulation. Each variable is complemented by a teal line showing the temporal ratio of recently quenched galaxies (dashed line) and long quiescent galaxies (solid line) to the median line of the remainder of the massive galaxy \ac{gmm} cluster. This shows visually that the earliest-quenching galaxies grow faster than their host haloes, and while \acp{agn} begin to grow later, their masses grow fastest for long quiescents as well. We see in addition that that gas and \acp{agn} mass fractions are distinct from the \ac{gmm} cluster, implying the presence of gas depletion by \acp{agn} feedback in the quiescent galaxies.}
\label{fig:ghe}
\end{figure*}

In \cref{fig:ghe}, this fast and early accretion can be seen visually, with the median halo mass for both quiescent populations growing more rapidly at early times than the \ac{gmm} cluster as a whole. Gas mass, stellar mass and \ac{smbh} mass grow similarly swiftly. These trends underscore a key finding of our PCA-GMM approach: that early mass assembly and potential deepening are predictive of quenching: not merely because they enable high stellar mass, but because they also facilitate early and efficient black hole growth.

It can be seen here that both gas and \ac{smbh} mass fractions evolve distinctly for quiescent galaxies. The proportion of gas in the subhalo drops sharply, and the \acp{agn} continues to grow in proportion to stellar mass. On the other hand, the \ac{gmm} cluster overall exhibits a steady fraction from around 600 Myr onward. This clearly illustrates the exhaustion of the star-forming gas supply, either by rapid star formation, \acp{agn} feedback, and their combined feedback effects demonstrated by \citet{XShen2}. \Acp{lqg}, however, experience a small uptick in gas fraction at around the time of their quiescence, which can be attributed to the acquisition of gas from the large scale environment. This trend remains when we compare total gas masses within the full galactic aperture, but it disappears when we consider only centrally located gas particles. This confirms that the excess gas originates from external accretion rather than internal retention.

In this analysis, we do not differentiate between gas phases, as our focus is on the total gas reservoir. However, observationally, the detectability of this accreted gas depends on its phase. Cold molecular gas, typically traced by \ion{C}{I} and \ion{C}{II} emission, is a key component of star-forming reservoirs, but may be depleted in quiescent galaxies. \citet{Umehata} have demonstrated ALMA's capability to observe molecular gas in quiescent galaxies at $z\sim 3$, and while \ac{mqg} populations are nonetheless prevalent in ALMA data \citep{Chworowsky}, \citet{Umehata} remains the first direct detection of molecular gas at this redshift.

\begin{figure}
\includegraphics[width=\linewidth]{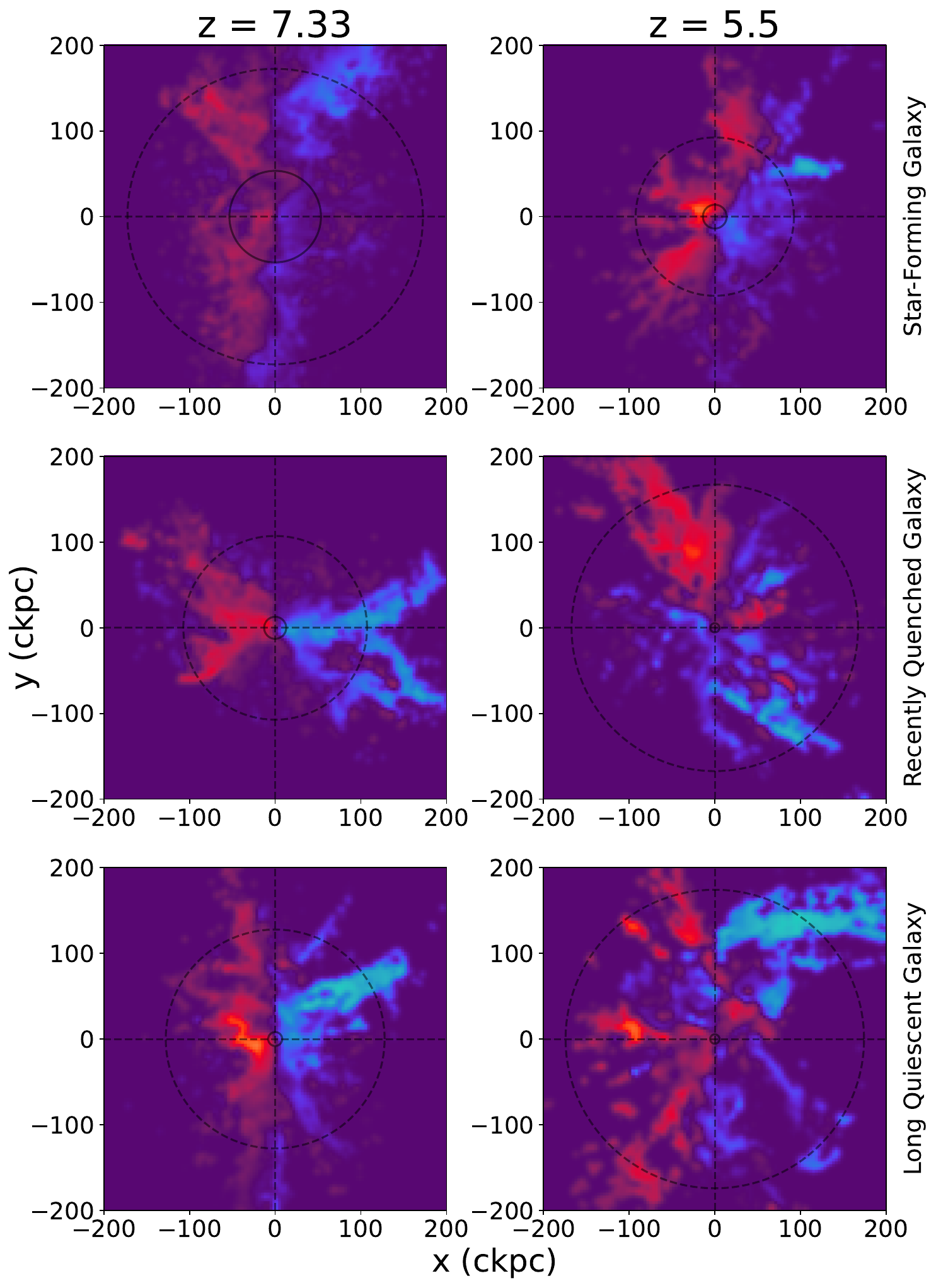}
\includegraphics[width=\linewidth]{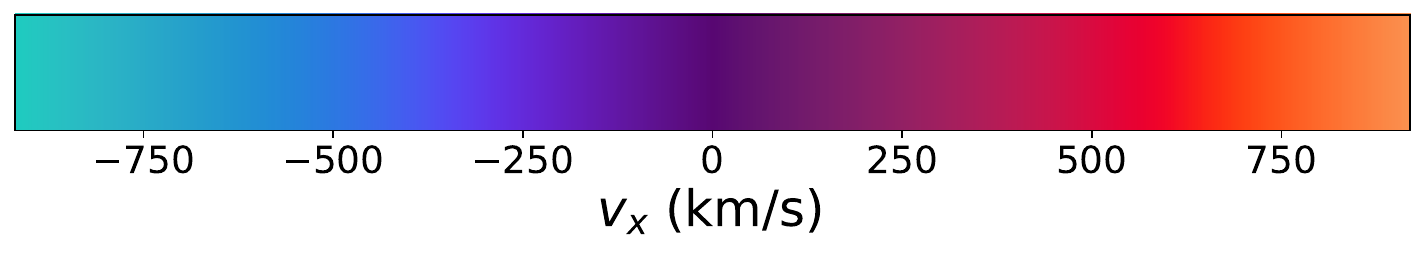}
\vspace*{-20pt}
\caption{A map of the gas distribution around three massive galaxies at two snapshots, coloured according to the velocity of the gas along the horizontal axis. In each column, we see a massive star-forming galaxy (top row), a recently quenched galaxy (middle row) and a long quiescent galaxy (bottom row); while the columns themselves represent two different redshifts, before and after quenching, if applicable. Each object is typical of the visually inspected sample of massive galaxies in the massive galaxy \ac{gmm} cluster. In each figure, there is a small solid circular line, whose radius is twice the stellar half-mass radius of the galaxy; and a large dashed circular line, whose radius is double the half-mass radius of the gas component of the halo. We see in this figure that the gas is being accreted onto each galaxy prior to quenching, while the gas becomes dispersed as the second galaxy quenches, and is completely disordered in the third. The rapid quenching of these galaxies is a violent and near instantaneous event which extends into the circumgalactic medium, dramatically altering the conditions of the gas which may be infalling to the galaxy. We also see the physical extent of the quenching galaxies contract as a result of this contrived instability, as the gaseous halo expands due to rapid blowout.}
\label{fig:vmap}
\end{figure}

Atomic hydrogen (\ion{H}{I}) remains a key tracer of the cosmic environment, and during the epoch of reionisation, its large-scale distribution can be probed via the redshifted 21 cm line using intensity mapping techniques \citep{Bera}. Ionised gas, observable through rest-frame optical and UV lines, is accessible to \textit{JWST}, but may not fully capture the total gas reservoir \citep{Williams}. Future radio observations, such as \ion{H}{I} mapping with next-generation facilities, will be valuable tracers of the large-scale environmental gas that may contribute to this effect.

The \ac{smbh} masses of \acp{mqg} show a clear hierarchy at all times, with the more established quiescent galaxies forming \acp{smbh} at earlier times and growing at greater rates. This is in line with the steep halo mass and velocity $\beta$ values associated with these haloes, considering that a deep potential is crucial for the stability of a rapidly growing \acp{smbh} \citep{Hopkins}. The faster growth of \acp{smbh} for \acp{lqg} can be argued to advocate their earlier quenching.

\begin{figure}
\includegraphics[width=\linewidth]{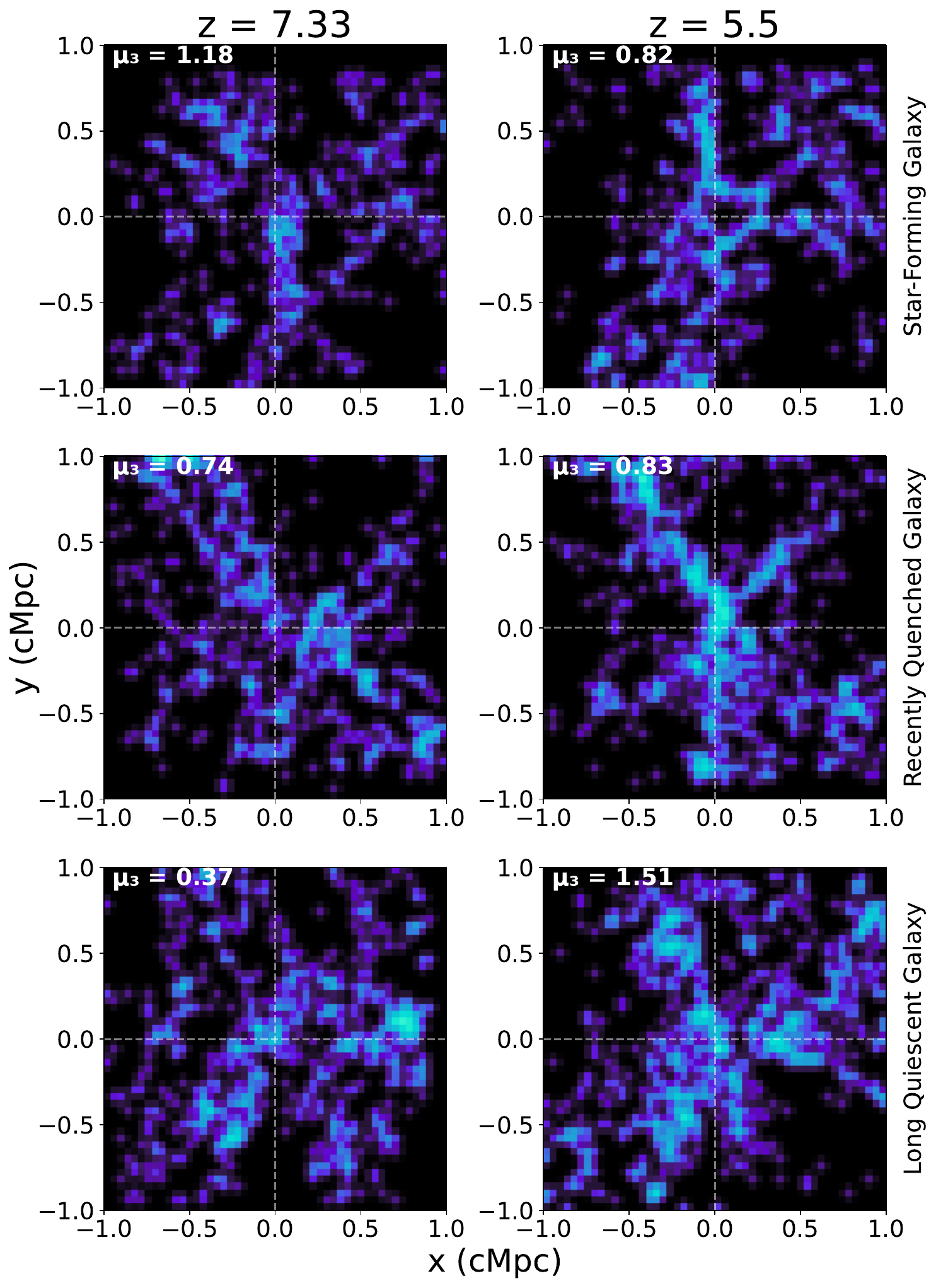}
\includegraphics[width=\linewidth]{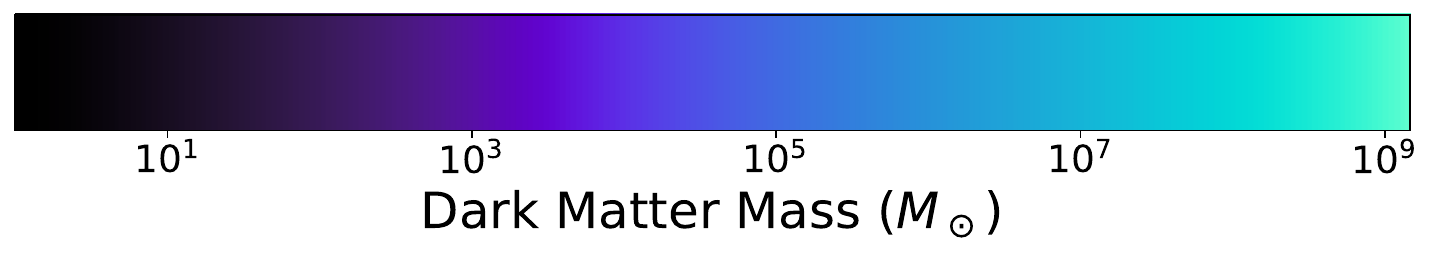}
\vspace*{-20pt}
\caption{Dark matter mass distributions around the massive galaxies shown in \cref{fig:vmap}, using the same samples and same redshifts as before, yet now shown over a 2 \ac{cMpc}-wide aperture to visualise the external environment surrounding their haloes, rather than the circumgalactic medium surrounding the galaxies themselves. Radial skew values, defined in \cref{sec:growthhistory}, are shown for each image. In this example, the skew increases with time for the quenched galaxies, but decreases for the star-forming galaxy. We also see in this figure that a number of large haloes in the vicinity of the quiescent galaxies drift closer to the target halo at the centre, as measured by the skew. This concentration of mass around the quiescent galaxy's host is likely to provide sufficient accreting material to drive the halo's fast accretion.}
\label{fig:dmap}
\end{figure}

\begin{figure*}
\includegraphics[width=\linewidth]{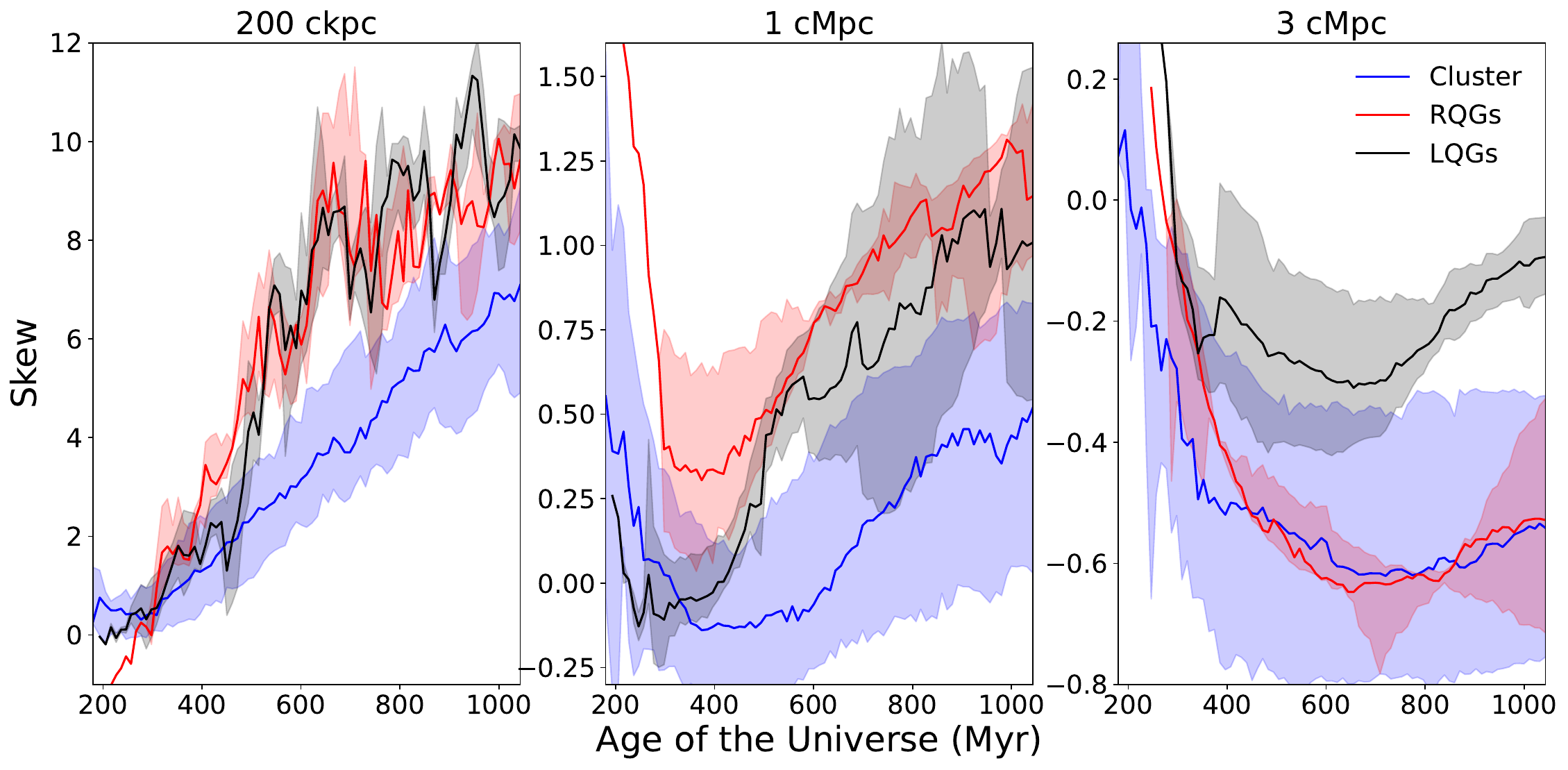}
\vspace*{-20pt}
\caption{The median (solid lines) and interquartile ranges (shaded regions) of the radial skews surrounding star-forming \ac{gmm} cluster galaxies (blue), recently quenched galaxies (red) and long quiescent galaxies (black), evaluated over apertures of 200 \acp{ckpc} (left), 1 \ac{cMpc} (centre) and 3 \acp{cMpc} (right). Similarly to overdensities evaluated over these three apertures, there is a discernible enhancement of skews for quiescent galaxies with respect to the massive galaxy \ac{gmm} cluster, particularly when evaluated over small apertures. The growth of these skews over time traces the accumulation of large structures around the galaxies' hosts, as seen in \cref{fig:dmap}, provoking the increase in smooth accretion which fuels the halo's expansion and the galaxy's quenching.}
\label{fig:skew}
\end{figure*}

We also see from our investigation of the haloes' merger histories that almost no significant merger events take place in the history of \ac{gmm} cluster galaxies. However, we can see from \cref{fig:vmap} that the gas within the circumgalactic medium is disrupted significantly during the quenching process. As we compare a typical example of a massive star-forming galaxy against a recently quenched and long quenched example, we see the converging motion of gas towards the centre of the galaxy transform into a turbulent flow around it, provided that the galaxy quenches. This suggests that while there is little influence from infalling progenitors in affecting the state of the massive halo, the quenching process is tied to a significant disturbance to the dynamics of the halo's environment. Along with the excess halo mass of \acp{lqg}, this disturbance may allow these environments to grow in density.

Looking to larger scales, we show the dark matter density surrounding the same haloes in \cref{fig:dmap}, again seeing systematic differences in the development of the surrounding structure. The complete density of these regions does not appear to evolve significantly in the time interval shown, but what is curious is how the largest haloes in these images have converged towards the massive galaxy at the centre. For each quiescent galaxy, some large concentrations of matter are seen to move closer to the target. While this may not constitute a merger, it could advocate the rapid accretion of matter onto the target halo; a necessary feat for the established rapid \ac{smbh} growth. This can be measured over time using the radial skew parameter introduced in \citet{Chittenden}, which quantifies the asymmetry in the radial distribution of dark matter around a target halo, focusing on deviations from symmetry between the inner and outer regions. With weighted statistical moments of any dataset $x_j$, combined with weights $w_j$, the weighted statistical moments of this dataset are defined as follows:

\begin{align}
&\mu_1 = \frac{\sum_{j=1}^N w_j x_j}{\sum_{j=1}^N w_j} , \\ &\mu_2 = \frac{\sum_{j=1}^N w_j (x_j - \mu_1)^2}{\sum_{j=1}^N w_j} , \\ &\mu_n = \frac{\sum_{j=1}^N w_j \left( \frac{x_j - \mu_1}{\sqrt{\mu_2}} \right)^n}{\sum_{j=1}^N w_j}, \forall n \geq 3 ;
\label{eq:moments}
\end{align}therefore, the radial skew parameter of the dark matter distribution, denoted $\mu_3$, is calculated from the distances of subhaloes from the target object. The masses of each subhalo are treated as weights in the expression above, where $n=3$.

A more positive skew reflects a concentration of dark matter towards the target halo. Conversely, a more negative skew indicates a distribution of mass away from it. As a mass-weighted quantity, the radial skew parameter traces the directionality and timing of dominant mass inflows, which in this work offers insight into how material assembles around a halo in the absence of mergers. Originally introduced in \citet{Chittenden}, it proved particularly effective in machine learning models for predicting galaxy chemical enrichment, as it captures asymmetries in halo growth and halo-halo interactions that are not encoded in simpler metrics like local density or concentration. In this study, its continued effectiveness in distinguishing \ac{mqg} environments reinforces its value as a physically interpretable tracer of environmentally driven evolution; especially in rare, merger-poor systems where traditional measures may fall short.

In \cref{fig:dmap}, we show that skew values, evaluated over a 1 \ac{cMpc} aperture, do in fact increase for \aclp{mqg} over the course of their quenching. \Cref{fig:skew} compares skew histories for the full population of \ac{gmm} cluster galaxies and \aclp{mqg}, showing that the environments around the latter grow to become more concentrated over time, particularly on small to intermediate scales. This persistent enhancement in radial skew relative to the \ac{gmm} cluster indicates a long-term concentration of surrounding dark matter towards the target halo. Since this skew traces the net asymmetry and mass-weighted infall of surrounding substructure, its continued elevation supports the interpretation that sustained, directed matter inflow contributes to prolonged \ac{smbh} growth, which ultimately leads to quenching.

As with environmental densities, the distinction of quiescent galaxies from their star-forming counterparts becomes weaker when evaluating skew histories over increasing apertures. However, it is curious that \acp{lqg} exhibit a strong skew when those quenched recently exhibit similar skew histories to the remainder of the \ac{gmm} cluster. At the larger scales to the right of \cref{fig:skew}, the overall skew signal is diminished and exhibits lower variability, as expected due to averaging over more isotropic environments. Nonetheless, \acp{lqg} retain a consistent excess in skew relative to the \ac{gmm} cluster average, which persists after their quenching. This residual asymmetry may reflect sustained anisotropic matter inflows or proximity to large-scale filamentary structures, which could facilitate continued growth in halo mass and local density. The consistent offset of \ac{lqg} skews even after quenching may suggest that this continues to factor into the haloes' growth, which could explain the apparent growth in local densities after quenching.

\begin{figure}
\includegraphics[width=\linewidth]{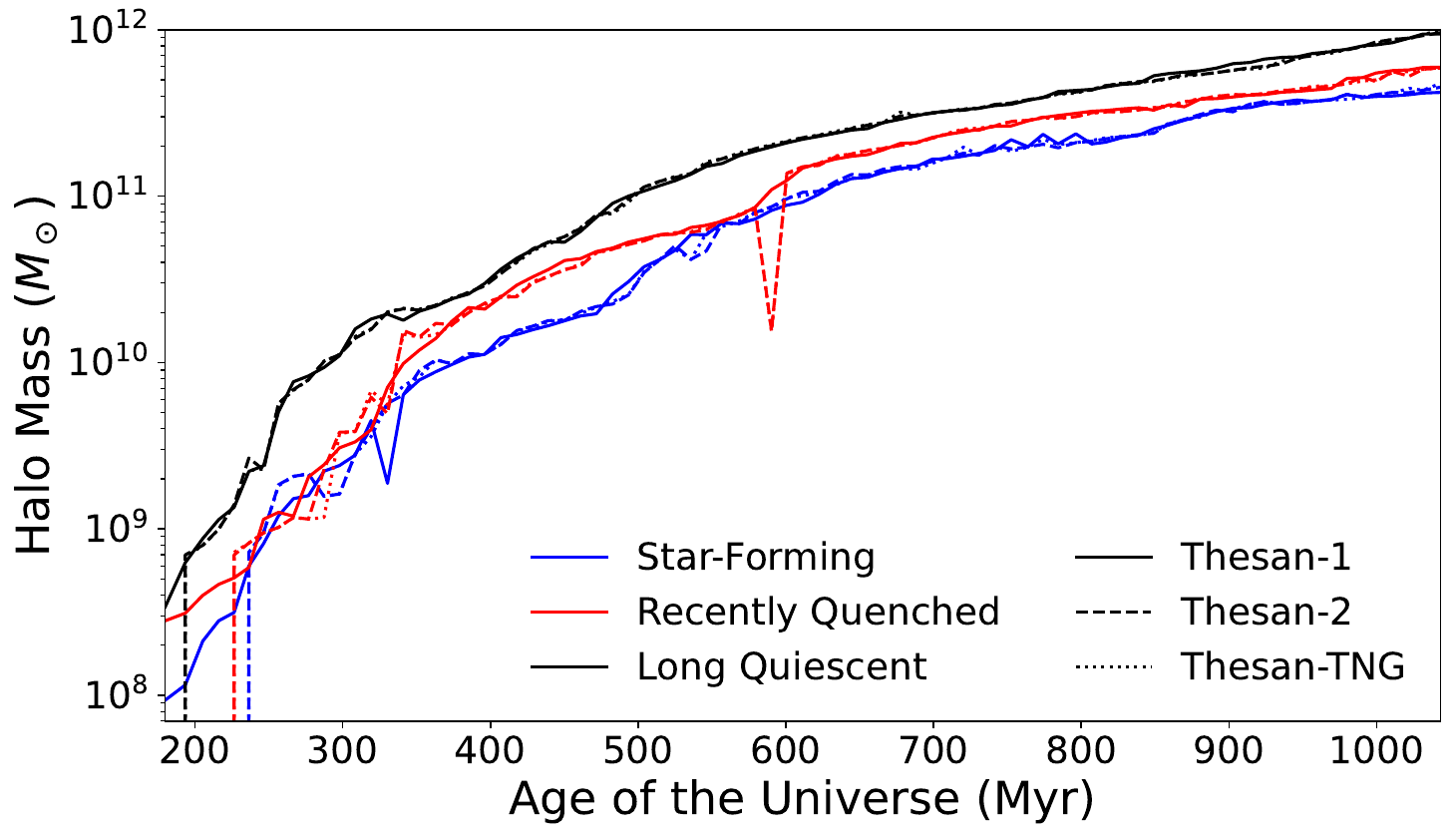}
\includegraphics[width=\linewidth]{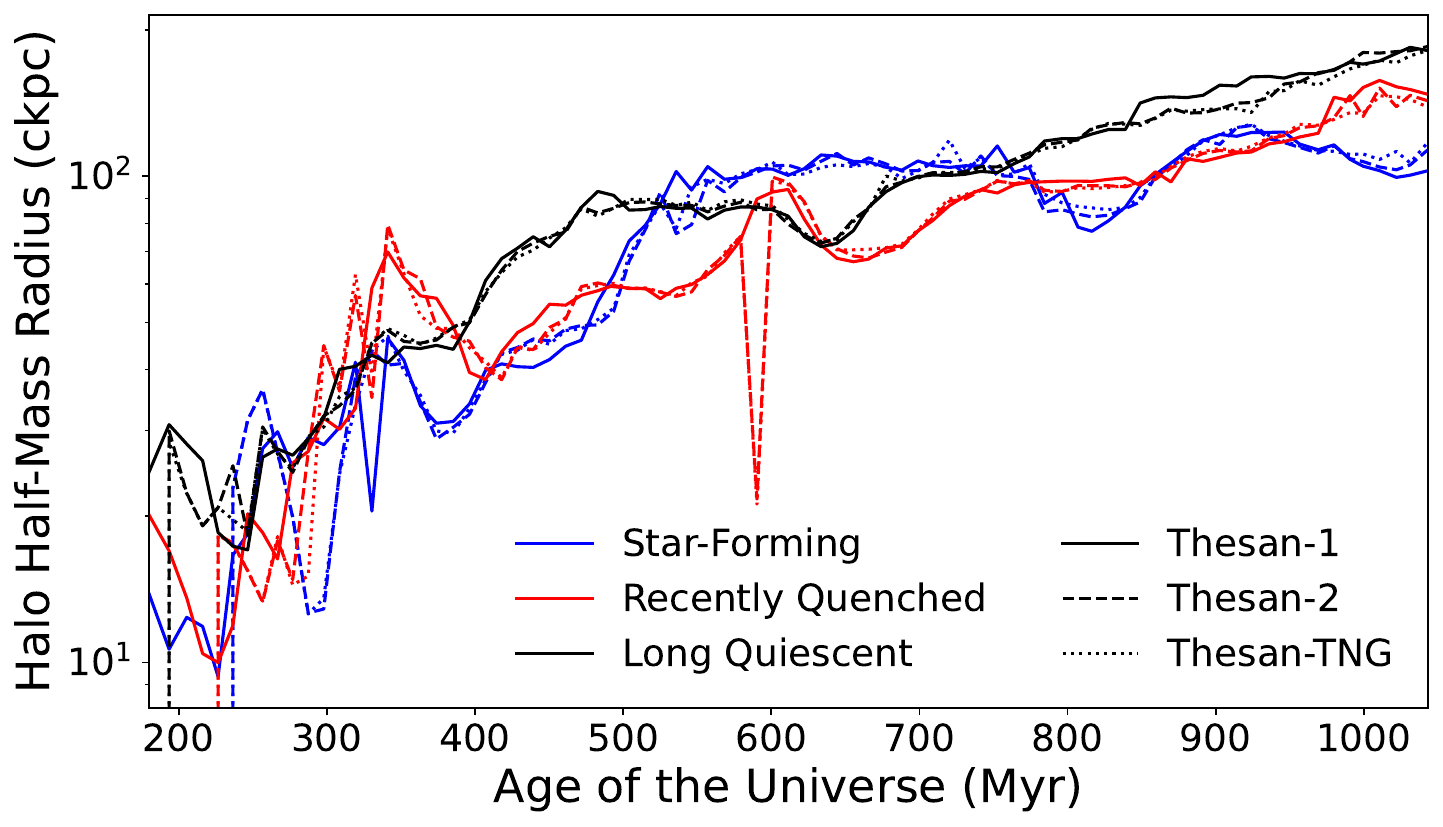}
\includegraphics[width=\linewidth]{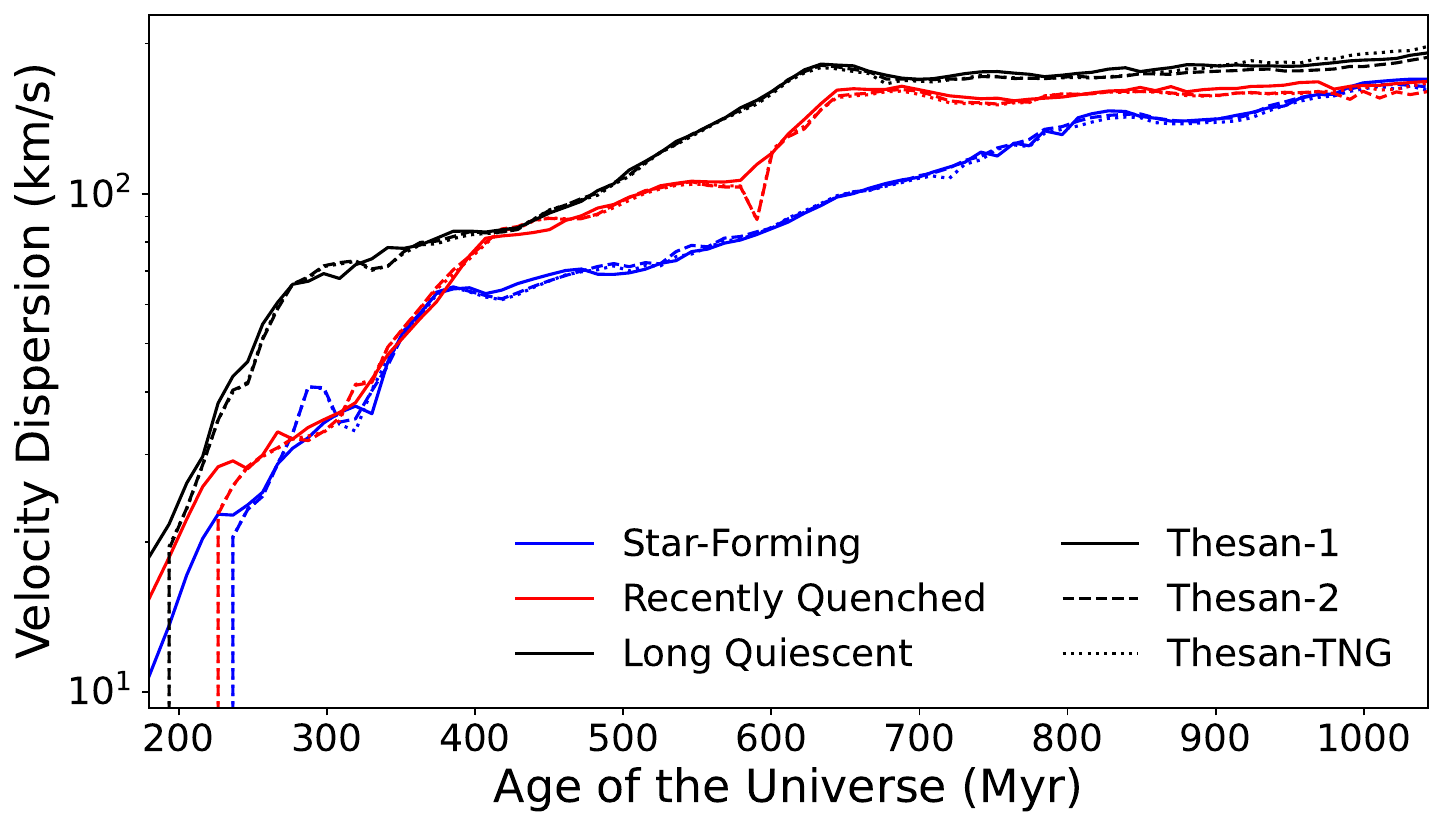}
\vspace*{-20pt}
\caption{The full history of the halo mass (upper panel), dark matter half-mass radius (central panel) and velocity dispersion (lower panel) of the three massive galaxies whose environments were shown in \cref{fig:vmap,fig:dmap}. The variables seen in the primary \textsc{Thesan-1} simulation (solid lines) are compared with their cross-matched results in the low resolution run \textsc{Thesan-2} (dashed lines), and the \textsc{Thesan-TNG} run, which employs the original \textsc{Arepo} code, without the radiative transfer model used in the fiducial \textsc{Thesan} model to understand cosmic reionisation. There is no significant difference between the halo properties of these three simulations, proving that resolution, hydrodynamics and \ac{agn} feedback are largely irrelevant to the structural evolution of massive haloes.}
\label{fig:simcomp}
\end{figure}

\begin{figure}
\includegraphics[width=\linewidth]{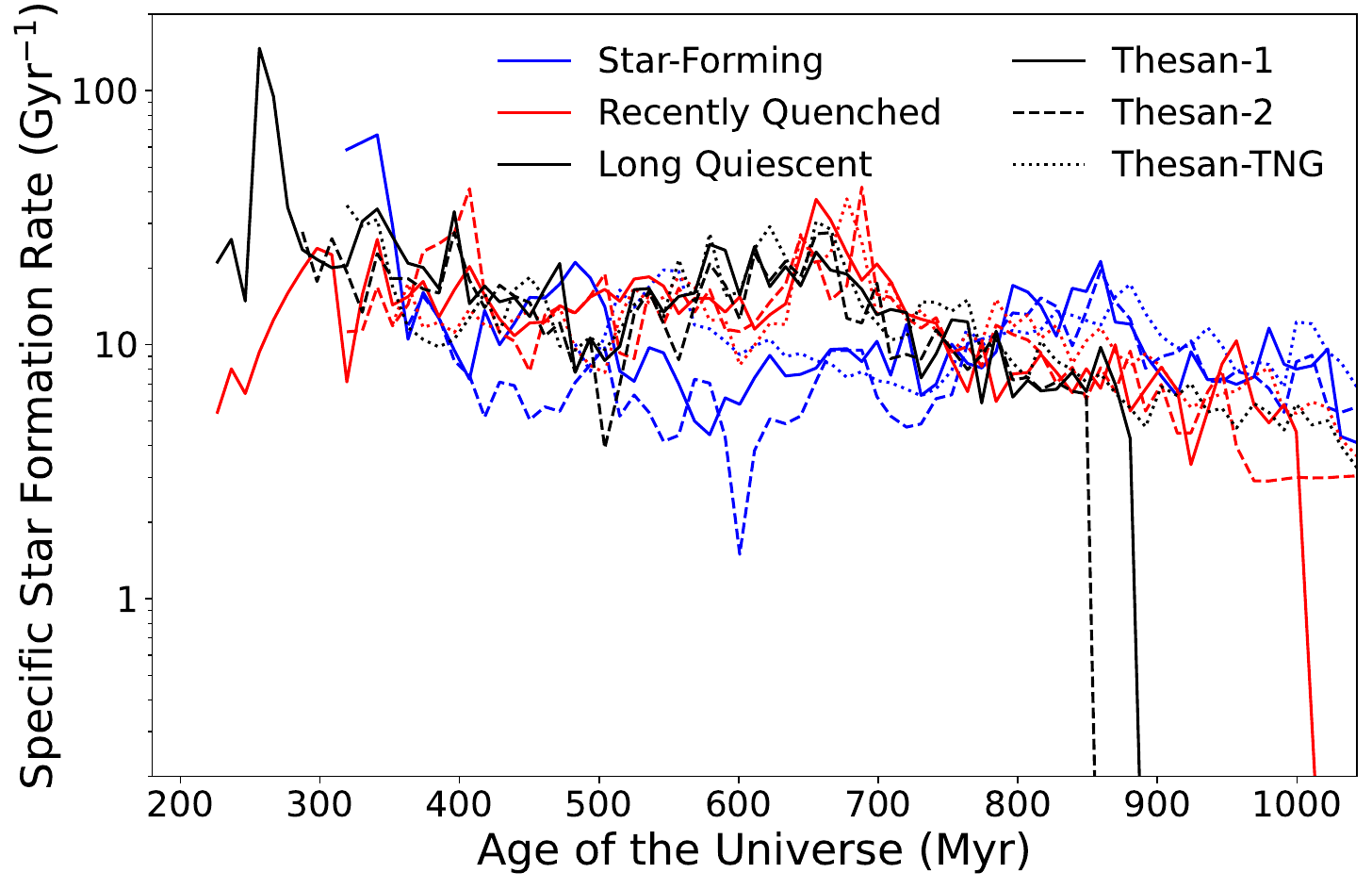}
\includegraphics[width=\linewidth]{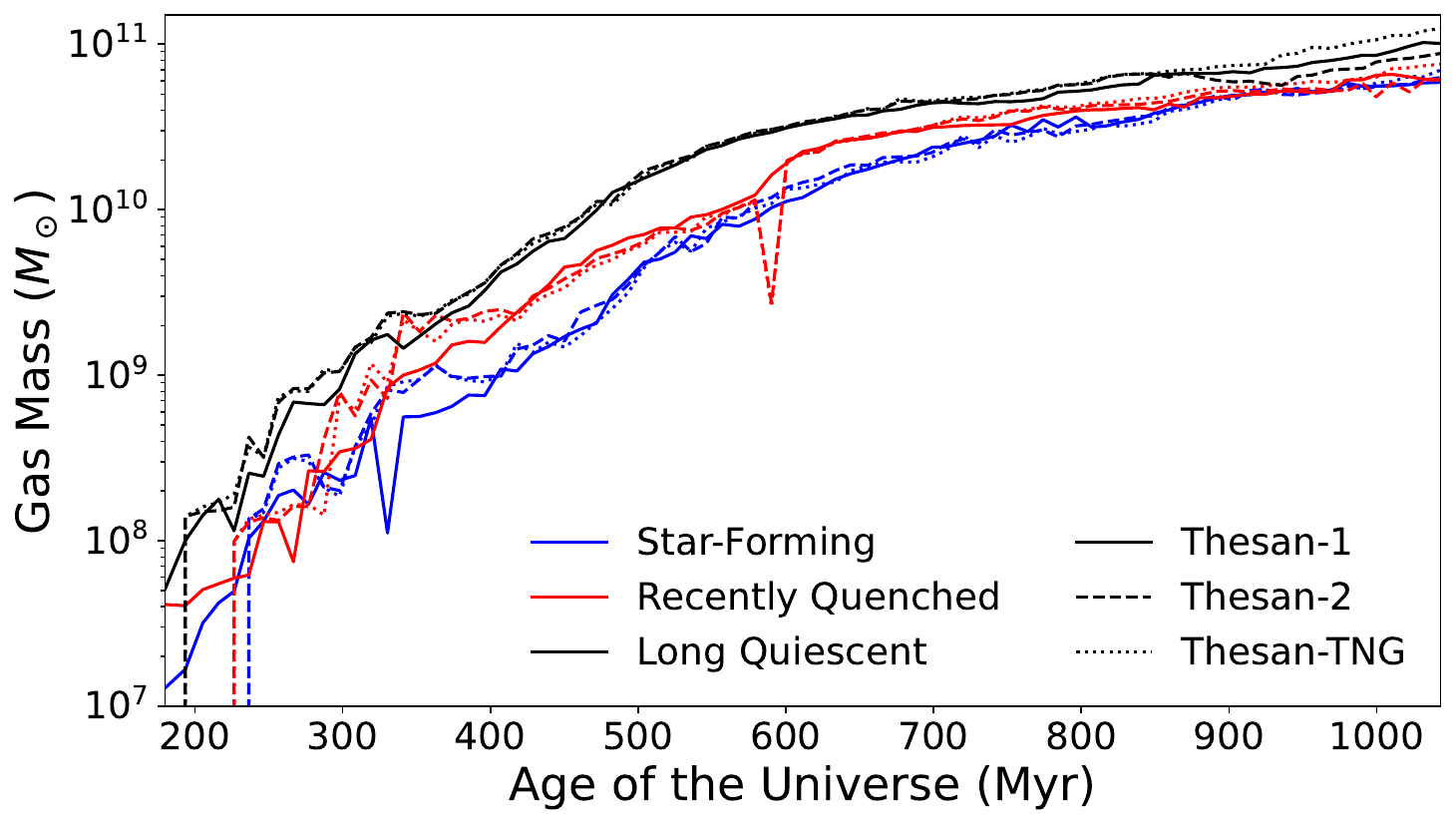}
\includegraphics[width=\linewidth]{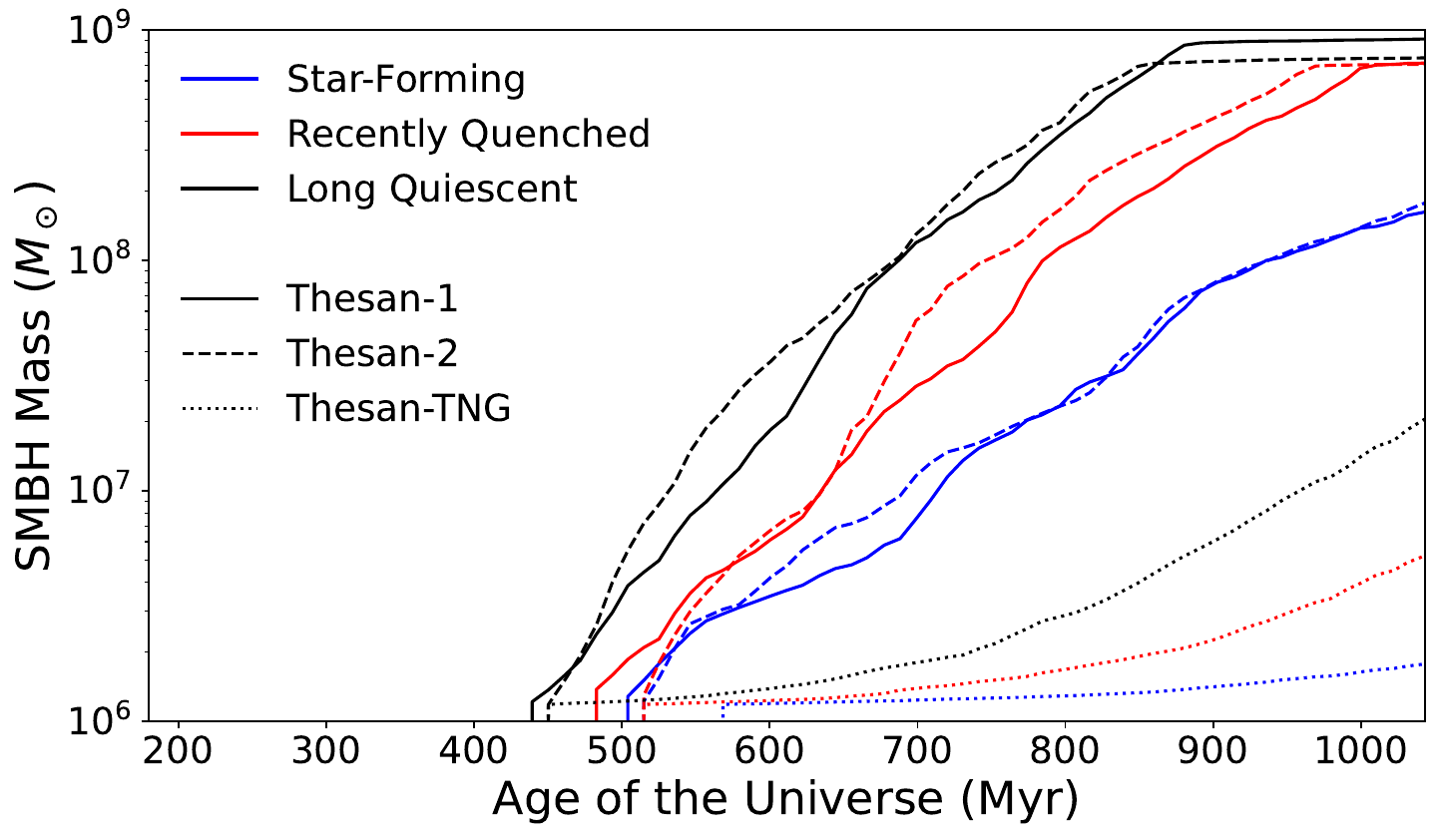}
\vspace*{-20pt}
\caption{The evolution of specific star formation rate (top panel), gas mass (middle panel), and \ac{smbh} mass (bottom panel) for three massive galaxies from earlier schematics is illustrated, comparing results from the primary \textsc{Thesan-1} simulation (solid lines) with its low-resolution counterpart, \textsc{Thesan-2} (dashed lines), and the \textsc{Thesan-TNG} run (dotted lines), which utilises the \textsc{IllustrisTNG} subgrid physics, excluding the radiative transfer model integral to the \textsc{Thesan} suite for modeling cosmic reionisation. While baryonic mass components such as star formation and gas mass are relatively consistent across all simulations, prior to the onset of quenching, \acp{smbh} masses exhibit significant discrepancies. Specifically, in the \textsc{Thesan-TNG} run, \acp{smbh} growth proceeds more slowly and culminates in substantially lower masses by the end of the simulation. Additionally, galaxies in \textsc{Thesan-TNG} corresponding to quenched galaxies in the fiducial runs display continued stellar mass growth, reflecting the inability of \ac{agn} feedback in these models to effectively suppress star formation. This stark contrast highlights the result of different \ac{agn} feedback coupling strengths in regulating \ac{agn} growth and the effectiveness of quenching in massive galaxies.}
\label{fig:simcomp2}
\end{figure}

\subsection{Comparing \textsc{Thesan} simulation models}
\label{sec:compthesantng}

An important aspect of the population of \aclp{mqg} shown in this study is that no such objects exist in the \textsc{IllustrisTNG} simulations; or the \textsc{Thesan-TNG} simulation: a simulation where the \textsc{IllustrisTNG} model was run with the initial conditions of \textsc{Thesan}. Looking to explain massive galaxy quenching by means of the galaxy formation model or the halo properties which influence massive galaxies, we look at haloes in the lower resolution \textsc{Thesan-2} simulation, alongside the \textsc{Thesan-TNG} run, which have been cross-matched to \ac{mqg} hosts and \ac{gmm} cluster haloes in \textsc{Thesan-1}.

In \cref{fig:simcomp} we show the evolution of halo properties for the same sample of three massive galaxies shown in \cref{fig:vmap,fig:dmap}, across the three \textsc{Thesan} runs. There is very little difference between any of these curves when compared across simulations; and by construction, the final values, $\beta$ gradients and formation times of these quantities are not significantly affected. The halo properties which drive galaxy growth and quenching are consistent across the different \textsc{Thesan} runs, despite variations in resolution and hydrodynamic method. This suggests that these numerical choices do not significantly influence \ac{mqg} formation. We conclude that the difference in \ac{agn} feedback facilitates the early growth and subsequent quenching of \acp{mqg}. The absence of significant differences in halo properties supports future efforts to link haloes to galaxies. Models applied to N-body simulations can adopt arbitrary baryonic prescriptions without introducing changes to halo structure, aside from known effects related to resolution, halo finder algorithms, and the computation of key variables, as discussed in \citet{Chittenden2}.

Looking at baryonic properties in \cref{fig:simcomp2}, however, there are visible differences. As well as the presence of quenching shown by specific star formation rates, there is a delayed growth in stellar mass for the \textsc{Thesan-TNG} simulation, while gas mass is unaffected, suggesting a reduction in feedback suppression. Yet, what is very notable is the severe delay in the \ac{smbh} mass growth in the \textsc{IllustrisTNG} run, compared with the simulations running the fiducial \textsc{Thesan} model. We find that the 20 largest \ac{smbh} masses in \textsc{Thesan-TNG} are over an order of magnitude lower than those in the fiducial simulations as a result of stronger \ac{agn} feedback, which can justify the abscence of massive galaxy quenching.

While some studies have reported an offset in the black hole-stellar mass relation at high redshift, suggesting that black holes are overmassive compared to their host galaxies \citep{Pacucci, Mezcua, Baker}, others argue for little to no evolution in the relation \citep{Suh, Li_2022, Zou}. Increasingly, however, observational results reveal black holes at $z\geq 4-11$ whose masses lie well above the local black hole to stellar mass relation \citep[e.g.][]{Maiolino, Ubler}. Such offsets are theoretically anticipated in dense, gas-rich early galaxies where \ac{agn} feedback couples only weakly to the surrounding ISM; low mass-loading factors and inefficient radiative or thermal coupling allow black holes to grow rapidly through near- or super-Eddington accretion \citep[e.g.][]{Pacucci2, Bassini, Bennett, YShi, Sanati}. Observational evidence for this weak coupling already exists; for example, the compact AGN 4C+19.71 at $z=3.5892$, where radiative feedback is inferred to couple at efficiencies below $10^{-4}$ \citep{WWang}. Against this backdrop, the public \textsc{Thesan} release, which unintentionally operates with reduced \ac{agn} feedback efficiency across both thermal and kinetic modes, provides a timely numerical experiment: it explores a physical regime in which black holes grow rapidly under suppressed coupling, eventually reaching masses where quenching becomes effective, mirroring conditions which may arise in the early universe.

\section{Discussion}
\label{sec:disc}

\subsection{Explaining the characteristics of MQGs}
\label{sec:haloenv}

\subsubsection{Key predictors of quenching in early massive galaxies}

By investigating the properties of massive haloes in \textsc{Thesan} which host \aclp{mqg}, we have identified key properties which can explain their unique modes of evolution. The \ac{gmm} cluster containing all such haloes showcases the most massive haloes with early growth and structure assembly, and merger-poor formation histories. Within the \ac{gmm} cluster, \acp{mqg} are separated from star-forming galaxies of similar mass by faster assembly of their mass and central potential, leading to the assembly of higher stellar and \acp{agn} masses, localised to within a smaller stellar radius. \Acp{rqg} are separated from \acp{lqg} by a heirarchical correlation between large scale environments and quenching redshifts, which can be seen both in bootstrapped statistics and their separation in \acl{pc} space.

Recent observations of solitary \acp{mqg} support several of the physical phenomena we identify in our analysis. For example, RUBIES-EGS-QG-1 has been found in a prominent 3 cMpc overdensity at $z=4.9$ \citep{deGraaff}, consistent with the dense environments of \acp{mqg} in \textsc{Thesan}. Resolved \textit{JWST} observations have also shown evidence of inside-out quenching at $z<2.5$ \citep{Lyu} and at $3<z<5$ \citep{Wright}, where star formation is suppressed first in the galactic centre, and then propagates outward. Additionally, studies on \acp{mqg} at $z=2.45$ \citep{Belli} and $z=4$ \citep{PFWu} report rapid quenching associated with strong gas outflows. These observational features align with the evolutionary pathways of \acp{mqg} in \textsc{Thesan}, which suggests that our results may help interpret the physical processes that govern early quenching. In particular, the mechanism of rapid black hole growth presented in \textsc{Thesan} may explain the early termination of star formation and the environmental conditions in which it occurs.

Another aspect of \acp{mqg} in \textsc{Thesan} is the intriguing dependence of different quiescent populations on environmental densities measured on different scales. Quiescent galaxies which have experienced faster halo and substructure assembly have quenched earlier, and simultaneously, their environmental densities exceed those of recently quenched galaxies, even on large scales. These overdensities are not transient but have persisted throughout the lifetimes of these galaxies, which suggests that early structure formation in dense regions established a long-term environment conducive to accelerated growth and early quenching.

As there is little difference between haloes and environments when comparing subgrid models, we argue that the distinctive behaviour of \textsc{Thesan} arises primarily from its effective suppression of feedback coupling. In this regime, gas can accrete more efficiently onto forming \acp{agn}, sustaining rapid \ac{smbh} growth through lack of self-regulation. Despite the weaker intrinsic feedback coupling strength, this effect maintains elevated \ac{agn} activity without fully expelling the circumgalactic gas, allowing early quenching through kinetic processes that remain locally confined rather than globally disruptive.

While \textsc{IllustrisTNG} and \textsc{Thesan} adopt identical black hole seeding criteria, the weaker feedback coupling in the latter alters the thermal and dynamical histories of galaxies. This facilitates earlier black hole growth and sustained accretion. The resulting \acp{mqg} therefore reflect an evolutionary pathway in which inefficient feedback and strong environmental inflows coexist, producing massive quiescent systems at high redshift through accretion flows that allow continuous growth at super-Eddington rates \citep{Pacucci2}, even without the full energy regulation seen in equilibrium feedback models. Interestingly, recent simulations have shown that explicitly permitting super-Eddington black hole accretion is sufficient to reproduce the observed abundances of massive quiescent galaxies at $z \geq 6$ \citep{Chaikin2026}, which may suggest that at least part of the \textsc{Thesan} result has a physical basis, even if it arises here from a numerical issue rather than a motivated physical model.

It remains true that the rapid growth of black holes, regardless of feedback coupling, evolve in environments which are not only gravitationally dense, but rich in radiation sources which can influence their growth and activity. \citet{Conaboy} report that the most massive haloes and densest environments in the \textsc{Sherwood-Relics} simulations exhibit the strongest Ly$\alpha$ transmission during reionisation. \citet{Neyer} show that ionised bubbles in \textsc{Thesan} tend to form in overdense regions, which, in \textsc{Flares}, also host the most luminous \acp{agn} \citep{Wilkins}. These regions may expose galaxies to elevated radiation pressure over extended timescales. Collectively, these findings support a scenario in which early-forming, overdense environments both accelerate the assembly of massive galaxies and shape the thermodynamic conditions which govern quenching.

\subsubsection{Observational challenges in tracing MQG environments}

Large-scale structure that formed within the first few hundred million years of cosmic time may have created the environments required for rapid accretion. The radial skews observed near the end of reionisation suggest that these haloes will continue to grow, as nearby massive haloes supply inflowing matter. Distant \acp{mqg} in the observable universe may occupy large filaments which act as such reservoirs; however, the evolution of these environments after reionisation remains uncertain. It is unclear whether the same relationship persists at lower redshifts, where \acp{mqg} are more accessible to observation.

\textit{JWST} observations have spectroscopically confirmed that \acp{mqg} from $3 \lesssim z \lesssim 5$ are compact and early-forming \citep{Carnallmnras, Carnallnature, Kawinwanichakij}; consistent with the galaxies identified in \textsc{Thesan}. However, \textit{JWST} is a less effective probe of the diffuse gas which surrounds these galaxies and reflects their larger environments. Although the telescope can detect internal kinematics at high redshift through rest-frame optical emission lines \citep{deGraaff2, DelPino, Ubler, Forrest2025}, its limited spectral resolution and the faintness of these lines reduce its ability to measure halo potential through stellar velocity dispersion \citep{Marshall}. In addition, the spatial resolution at these redshifts restricts the ability to resolve internal kinematics, especially in compact \acp{mqg} \citep{NanayakkaraPASA}. To verify the role of halo potential in \ac{mqg} formation, future radio observations may be required. Mapping \ion{H}{I} with next-generation instruments such as SKA-Low could reveal extended cold gas reservoirs and environmental interactions which constrain these halo potentials \citep{Bera}.

\subsection{Modelling massive galaxy quenching}
\label{sec:simvsim}

\subsubsection{MQG physics in \textsc{Thesan}}

With little difference between halo properties in each simulation, the key differences in subgrid physics can be attributed to the discrepancy between massive galaxy and \acp{agn} evolution. Curiously, while cosmic reionisation serves to suppress and delay the formation of small progenitors, there is little difference between the stellar mass functions of separate runs of the \textsc{Thesan} suite, including that of the dark acoustic oscillations model which enhances this effect \citep{Thesan}. While this radiative transfer implementation affects the thermal state of gas in low-mass haloes \citep{Garaldi}, the key distinction between \textsc{Thesan} and \textsc{IllustrisTNG} for massive galaxies lies in the suppressed \ac{agn} feedback efficiency. The lack of significant differences in halo properties across different \textsc{Thesan} runs (see \cref{sec:compthesantng}) suggests that the primary driver of \ac{mqg} formation is the altered \ac{agn} feedback prescription rather than the details of radiative transfer or reionisation physics.

Comparing two resolution runs of the fiducial \textsc{Thesan} model shows no change in \ac{smbh} growth. However, \citet{ArepoRT} argue that accurate \ac{agn} feedback modelling requires resolving small-scale processes. As described in \cref{sec:ThesanModelling}, the \textsc{Arepo-RT} code links radiation pressure to the kinetic energy of ionisation fronts using the Eddington tensor, which removes the dependence of the calculation on the local density of ionising sources. It also applies a subcycling algorithm that resamples radiative transfer several times within each hydrodynamical step. Together, these methods enable precise treatment of radiation field propagation and its interaction with gas and dust, which are crucial aspects of \acp{agn} feedback.

Despite the weakened \ac{agn} feedback, the inclusion of on-the-fly radiative transfer in \textsc{Thesan} adds essential realism to \ac{agn} feedback by evolving the local radiation field alongside gas dynamics. Unlike \textsc{IllustrisTNG}, which applies a uniform UV background, \textsc{Thesan} models highly anisotropic radiation from local sources, including the \ac{agn} itself. This leads to non-uniform ionisation and temperature structures, which alters how feedback couples to the surrounding gas. In the process of \ac{smbh} growth, it is possible that the non-equilibrium cooling solutions \citep{Garaldi,Garaldi2} effectively shield the gas surrounding the \ac{agn}; akin to what is observed in dense environments in the \textsc{Thesan-Zoom} simulation \citep{ThesanZoom}. In these regions, overlapping ionisation fronts and intense local sources require a fully self-consistent radiative transfer formalism to resolve the dynamics of infalling gas and interpret their effects on compact star-forming regions and \acp{agn}.

\subsubsection{Logistics of AGN feedback}

In contrast to the rapidly growing \acp{smbh} in \textsc{Thesan}, \citet{Mistral} demonstrate that the \textsc{Mistral} \ac{agn} feedback model, which incorporates radiatively efficient \ac{agn} winds through stochastic momentum injection, strongly alters galaxy properties at $z = 2$. This model also produces more \acp{mqg} than the \textsc{IllustrisTNG} feedback model at $z > 3$ (Marion Farcy, private communication), which underscores the importance of energetic feedback acting on large spatial scales. By comparison, \textsc{Thesan} achieves a similar population of massive quiescent galaxies despite exhibiting globally suppressed \ac{agn} energy coupling. The convergence of these contrasting feedback regimes suggests that both strong and weak coupling can yield quenching, either through direct expulsion of gas or through prolonged overgrowth of supermassive black holes that later regulate star formation once sufficiently massive. This dichotomy may reflect physically plausible pathways: while \textsc{Mistral} represents efficient radiative coupling, \textsc{Thesan}'s regime provides an analogue for scenarios in which black holes grow under suppressed feedback, as inferred observationally in some high-redshift systems \citep{WWang, Maiolino}.

While the \textsc{Mistral} model quenches galaxies through rapid, momentum-driven outflows, the \textsc{Thesan} galaxies follow a distinct evolutionary pathway. In \textsc{Thesan}, the globally suppressed \ac{agn} feedback allows black holes to accrete for extended periods, enabling them to reach high masses where quenching eventually becomes effective. This delayed self-regulation may be likened to chaotic cold accretion in dense environments \citep{Gaspari1}, or radiatively inefficient flows which permit super-Eddington growth \citep{Pacucci2, YShi}. Thus, the presence of massive quiescent galaxies in \textsc{Thesan} reflects a numerical regime in which suppressed \ac{agn} feedback across all modes fosters overmassive black holes through prolonged accretion, eventually reaching masses where quenching becomes effective; conditions which may overlap qualitatively with physically motivated scenarios supported by \textit{JWST} observations of overmassive black holes at $z \sim 4-11$ \citep{Maiolino, Prole}.

We find that the quenching of massive galaxies in \textsc{Thesan} is associated with a sharp decline in the recorded Eddington ratio by almost two orders of magnitude, over the same short timescales in which star formation rapidly drops. However, due to the reduced effective speed of light in the black hole routines, the simulation permits accretion rates of up to five times the physical Eddington limit \citep{ThesanAddendum}. As a result, the recorded Eddington ratios do not correspond to physical accretion efficiencies and should be interpreted only as internal diagnostics of accretion state within the simulation. We therefore refrain from making direct quantitative comparisons between these values and observationally inferred Eddington ratios. The transition to quenching nonetheless occurs at earlier times and higher black hole masses than in \textsc{IllustrisTNG} \citep{Kurinchi-Vendhan}. This transition is preceded by a prolonged phase of sustained accretion, during which quiescent galaxies maintain consistently high recorded Eddington ratios $(\lambda_\text{Edd} \sim 0.1)$ while their effective accretion rates may be substantially higher, enabling the growth of more massive black holes than in their star-forming counterparts. This supports the interpretation that suppressed \ac{agn} feedback in \textsc{Thesan} enables uninterrupted \ac{smbh} growth over cosmic time until black holes reach sufficient mass for quenching to occur. Recent work suggests that super-Eddington accretion may indeed be needed for black holes to grow sufficiently massive to effectively quench galaxies through \ac{agn} feedback \citep{Chaikin2026}. This behaviour is further facilitated by the dense environments and rapidly growing halo potentials of these systems, as feedback effectiveness is strongly modulated by black hole mass relative to halo potential depth \citep{Bennett}.

This may imply that \ac{agn} feedback in the low recorded Eddington regime plays a pivotal role in the quenching process; yet in the radiation-dense environments seen in \textsc{Thesan} \acp{mqg}, stellar and \ac{agn} feedback remain closely intertwined \citep{XShen2}. Supernova feedback may not make a noticeable impact on \acp{mqg} alone, yet it may serve a secondary role in regulating \ac{agn}-driven outflows, such as enhancing the removal of residual gas after quenching. Although these feedback modes may be difficult to distinguish observationally, the persistence of massive, quenched systems in \textsc{Thesan} demonstrates that even in a regime of artificially suppressed feedback coupling, self-regulation can ultimately culminate in effective quenching. The simulated galaxies therefore provide insight into the boundary between efficient and inefficient \ac{agn} self-regulation in the early universe, representing a limiting case rather than a predictive model.

\subsubsection{Future simulation studies}

The congruence of halo properties between simulations is encouraging for future work, where these rare haloes can be investigated in N-body simulations. The scarcity of \acp{mqg} in most cosmic simulations fails to adhere to the observed number density of such galaxies, which may be compensated by the massive \ac{agn} quenching illustrated in this work. Additionally, the small quantity of \acp{mqg} implies little diversity of evolutionary trajectories, which makes it difficult to draw conclusions about their evolutionary physics. However, the strong correspondence between halo and galaxy properties may enable the use of machine learning to predict the population statistics of \aclp{mqg} across the reionisation epoch. In addition, identifying \ac{mqg} hosts in N-body simulations may be selected for future zoom simulations, where alternative massive galaxy evolution models may be tested.

With this projected continuation of growth past the epoch of reionisation, it will be intriguing to consider their continued evolution, providing analogues to the quiescent galaxies observed from $z=3$ to $z=5$. In \textsc{Magneticum}, there exists a significant population of quenched galaxies at $z=3.4$, albeit with quenching redshifts subsequent to reionisation. Similarly to our findings, the haloes hosting \acp{mqg} are massive and fast-growing, and there is a strong correlation between \acp{agn} and star formation activity in massive galaxies. However, \textsc{Magneticum} differs from other simulations because its subgrid parameters were not calibrated to match observed stellar mass functions or quenching trends, but instead to reproduce the hot gas content of galaxy clusters \citep{Popesso}. Its stellar properties therefore emerge more self-consistently, albeit with a mild overprediction of quenching at lower redshifts \citep{Lustig,EncyclopediaMagneticum,Lagos}.

By contrast, the \textsc{Colibre} simulations, calibrated to reproduce $z=0$ stellar mass functions and black hole scaling relations, also produce robust analogues of $z=3-5$ quiescent galaxies, while following their evolution to $z=0$ \citep{ChandroGomez}. In \textsc{Colibre}, \acp{mqg} likewise inhabit massive, rapidly assembling haloes and host overmassive \acp{smbh}, with quenching tightly coupled to cumulative \ac{agn} energy injection. However, unlike \textsc{Magneticum}, these systems preferentially reside in overdense environments prior to quenching, where enhanced inflows accelerate black hole growth and trigger powerful feedback. Although a significant fraction experience later rejuvenation episodes, strong \ac{agn} feedback ultimately re-establishes quiescence in most descendants.

Conversely, \citet{Kimmig} find an excess of massive haloes in \textsc{Magneticum} which do not host \acp{mqg}, and \citet{RemusKimmig} show some quenched systems to reside in underdense or isolated regions\footnote{It is important to note that while \citet{Kimmig} claim small-scale environments to be underdense for \acp{mqg}, they quote relative density contrasts within a narrow stellar mass bin, which is less meaningful in \textsc{Thesan} due to fewer high mass galaxies. We do see a rough underdense contrast when applying these authors' calculations, but this is not statistically significant.}. These differences, also seen in comparisons with \textsc{IllustrisTNG}, are attributed to differing \ac{agn} feedback conditions. It is therefore plausible that environmental trends at reionisation redshifts depend sensitively on how feedback couples to gas supply, with \textsc{Magneticum} favouring efficient ejective regulation in comparatively underdense regions, and \textsc{Colibre} (like \textsc{Thesan}) linking quenching to overdense environments which support rapid \ac{smbh} growth.

\subsection{Explaining the observed MQG diversity}
\label{sec:simvobs}

\subsubsection{Multifaceted JWST data}

Confirming the validity of the current assortment of simulation models with observations may prove challenging. For instance, there is limited observational evidence favouring either of the environmental dependences of \acp{mqg} seen in \textsc{Magneticum} and in \textsc{Colibre} and \textsc{Thesan}. On the one hand, \citet{Jin} showcase a set of massive galaxies lying along a cosmic filament at $z\simeq 3.44$, where the \acp{mqg} are the most massive of all and reside in the densest environments, which \citet{Ito} claim are in the process of merging post-quenching. Similarly, \citet{Turner} demonstrate that the \citet{Glazebrook2024} galaxy resides in a tightly clustered environment at $z\simeq 3.2$; akin to the concentrated environments we have seen in \textsc{Thesan}. However, based on recent \textit{VANDELS} data, \citet{Ortiz} suggest that environmental influences on \acp{mqg} extend beyond density-driven gas and \ac{smbh} accretion, with large-scale structures also inducing mergers and ram pressure stripping that can suppress growth and alter observed environmental trends. The structural diversity seen in observational catalogues would thereby indicate that \acp{mqg} may follow multiple evolutionary pathways.

This work offers valuable insights into the hidden population of rapidly evolving primordial galaxies. However, it does not capture the full diversity of evolutionary pathways for \acp{mqg}. Despite the similar ages, masses, and formation times of \textsc{Thesan} galaxies with those in \citet{Carnallmnras} and \citet{Nanayakkara2}, most observed galaxies quenched over longer timescales, and after the end of reionisation. These observed galaxies also exhibit key differences in their formation histories, such as significant merger activity leading to extended disk profiles and massive bulges driven by merger-induced starbursts \citep{Kawinwanichakij}. \citet{Ito} suggest that the merger-driven growth of \acp{mqg} can take place after quenching, which could explain the diversity of morphologies and environments seen in \acp{mqg} at lower redshifts. While merger activity may play a role in morphological transformation and ex-situ stellar mass growth, \citet{Nipoti} demonstrates that mergers alone have little effect on central stellar velocity dispersion. This finding suggests mergers do not affect the potential for rapid \ac{smbh} growth and quenching. Instead, \citet{Nipoti} identifies envelope accretion from the disruption of diffuse satellites by low-mass progenitors $(\mu < \sfrac{1}{10})$ as a key contributor to increases in galaxy size and black hole mass. The finding that the central potential is primarily influenced by small progenitors agrees with the lack of a merger signal seen in our \ac{mqg} population, and may explain how \acp{mqg} can develop in lieu of merger-driven quenching.

Overall, this multeity of massive galaxy quenching pathways is difficult to reconcile with simulations, in light of limited sample sizes and disparities between subgrid models \citep{Lagos}. Observations of lower redshift \acp{mqg}, such as those in the SSA22 protocluster at $z=3.09$ \citep{Umehata}, suggest alternative quenching mechanisms to the radiation pressure gradient scenario we propose. These authors claim that the halo retains its heat from its X-Ray \ac{agn}, which shuts off filamentary gas accretion and suppresses halo growth. At $z = 7.276$, however, \citet{Valentino2025} report the presence of an outflow with a substantial mass loading factor in the galaxy RUBIES-UDS-QG-z7, which the authors attribute to the existence of an undetected \ac{agn}. This may signify the rapid expulsion of star-forming gas: a ubiquitous feature of \textsc{Thesan} \acp{mqg}; yet the rapid quenching timescales are likely to be enhanced by the inclusion of a realistic feedback coupling strength.

\subsubsection{Need for further modelling}

The \textsc{Thesan} data shows that the earliest-forming \acp{mqg} are highly dependent on how \ac{agn} feedback is modelled, but these results correspond only to a small fraction of the \ac{mqg} population in the real universe. Given the amount of time after reionisation, there is a lot to be uncovered regarding the post-quenching timeline of these early \acp{mqg}, and the accumulation of later-quenching galaxies through various mechanisms. Modelling galaxy formation in the early universe is crucial for their initial formation, but in order to understand the complete picture of \ac{mqg} population statistics, the simulation must be extended past the epoch of reionisation, down to redshifts of $z\sim 3$. At these later times, the sample size of modelled galaxies should align with the growing anthology of \textit{JWST} studies.

Furthermore, these simulations still do not contain analogues for truly outlying observations, such as the massive galaxy with a formation redshift of $z\simeq 11$ \citep{Glazebrook2024}. It may transpire that with the extended dark matter power spectrum offered by high volume N-body simulations, we may capture primordial haloes which can host similar galaxies, and we may make projected estimates of the abundance of these exceptionally old \acp{mqg}. The number density, star formation history, structure, and morphology of extreme MQGs are likely dependent on the details of \ac{agn} feedback implementation and its coupling to the surrounding gas. Given the diversity of \ac{agn} feedback prescriptions across simulations and the theoretical uncertainties in high-redshift feedback efficiency, future simulations exploring different galaxy-\ac{agn} coevolution models \citep[e.g.][]{Mistral} will be valuable for understanding the range of possible evolutionary pathways for early massive galaxies.

Of course, alternative physical models may capture a broad population of \acp{mqg} with distinct properties to those in \textsc{Thesan}. The logistics of the \textsc{Thesan} model present opportunities to test separate quenching pathways. For instance, as the gas effectively remains above photoheating temperature thresholds in dense environments, this could suppress important gas cooling pathways, which could allow \acp{mqg} to manifest in greater abundance, or conversely, suppress their growth by means of feedback efficacy. New simulations with physically motivated stellar particle synthesis models may produce separate galaxy characteristics, and relationships with their haloes and environments over time, to what is achieved with the spatially and temporally varying radiation field in \textsc{Thesan}.

In addition to the importance of \ac{agn} feedback presented in this study, previous simulations have shown that the growth of \acp{smbh} in massive galaxies is regulated by supernova feedback. As these winds are not explicitly coupled to the gas in the \textsc{Thesan} model, it may be worthwhile to explore their role in future simulations. Strong supernova feedback can prevent the accumulation of cold gas in the galactic centre, or eject it entirely, preventing further accretion \citep{Dubois}. However, in massive haloes, the \ac{agn}-driven formation of a hot gas corona, combined with the deep graviational potential well, serves to contain the supernova-driven outflow, which causes a rapid increase in \ac{smbh} mass \citep{Bower}. At $z=5.5$, the masses of \ac{mqg} haloes only just reach the threshold for this regime, therefore this coronal gas retention is most likely to manifest shortly after reionisation. However, we have established that the \acp{mqg} are hosted by haloes with early-forming, deep potentials, which may contain the gas more easily. As gas particles in quiescent galaxies exhibit more concentrated rotation curves than their star-forming counterparts, this may already be the case.

\subsubsection{Interpreting MQG formation without enhanced star formation}

The newfound abundance of high-redshift galaxies has prompted speculation about elevated star formation efficiencies (SFE) during the epoch of reionisation \citep{Dekel, Harikane}. However, recent analyses have questioned this interpretation. \citet{Donnan2025}, using \textit{JWST} UV luminosity and stellar mass function data spanning $z=6-13$, find no evidence of enhanced SFE. Instead, they adopt a redshift-independent efficiency model that provides an excellent fit to the UV luminosity function at $z\geq 9$.

However, both \citet[][fig. 10]{Thesan} and \citet[][fig. 2]{Donnan2025} employ earlier stellar mass function measurements from \citet{Stefanon} at $z=6-8$, making it possible to directly compare performance against newer \textit{JWST}-era data. At $z=8$, \citet{Donnan2025}’s model aligns well with updated mass functions, but at $z=6-7$, it systematically overpredicts stellar masses relative to more recent estimates \citep{Harvey, Shuntov, Weibel}. \citet{Donnan2025} show that by contrast, the \textsc{Thesan} simulation better matches these newer observations, suggesting that the empirical SFE prescription may not fully capture the impact of late-stage processes in more developed galaxies, such as outflows and feedback, which cannot be encoded in efficiency models alone.

As \citet{Donnan2025} note, their model was optimised primarily for predicting UV luminosities at $z\geq 9$, and not for the detailed stellar mass buildup in galaxies at lower redshift. This opens the possibility that the observed decline in stellar mass at $z\sim 6-7$ originates from processes like \ac{agn} feedback and radiation-regulated accretion, particularly in dense environments. Our results support this hypothesis. We show that rapid \ac{smbh} growth in overdense haloes is critical to early quenching in the absence of mergers or unusually high SFEs.

Nevertheless, \citet{Donnan2025}'s model assumes a fixed halo mass function and a redshift-invariant initial stellar mass function. Both assumptions may be oversimplified at these epochs, where halo growth and stellar population assembly are strongly time-dependent, and not as predictable as they are at low redshifts. These limitations may make such models unable to characterise the most extreme galaxy populations during reionisation.

Finally, we note the emerging theoretical suggestion that \ac{smbh} growth may precede, rather than follow, significant stellar mass assembly in some early systems \citep{Silk2024, DSilva}. If supported observationally, this would require a re-evaluation of the standard galaxy-black hole coevolution paradigm, and point toward a broader diversity of high-redshift evolutionary pathways than that supported by current models.

\section{Conclusions}
\label{sec:conc}

In this work, we investigate the unique historical properties of \acfp{mqg} and their host haloes and environments in the \textsc{Thesan} cosmohydrodynamical simulation. A numerical issue in the simulation reduces \ac{agn} feedback efficiency by a factor of 25 in both thermal and kinetic modes while enhancing accretion rates \citep{ThesanAddendum}, creating a regime of suppressed feedback coupling. This advocates the reinterpretation of the \textsc{Thesan} results, not as a predictive model of galaxy evolution, but as a controlled experiment probing galaxy-black hole coevolution in such a regime; one which has observational analogues in the rapidly growing, overmassive black holes identified by \textit{JWST} at high redshift \citep{Maiolino, WWang, Prole}.

This configuration produces a population of galaxies which exceed a stellar mass of $10^{10} M_\odot$ and rapidly quench before $z=5.5$: a feat which simulations with standard feedback implementations fail to replicate. While this regime does not represent the originally intended \textsc{IllustrisTNG} calibration and would not reproduce its realism if evolved to lower redshifts, it provides insight into galaxy evolution under weakened \ac{agn} regulation; a regime that may have observational analogues, given the growing number of overmassive black holes and weakly-coupled \ac{agn} identified by \textit{JWST} at similar and higher redshifts \citep{Maiolino, WWang}.

We construct a statistically grounded framework for characterising the formation of \acp{mqg} according to their haloes and environments. By running a \acf{gmm} clustering algorithm on the \acl{pc} space encoding the key properties of the haloes' evolutionary histories, we isolate a distinct population of early-forming, massive haloes with merger-poor assembly, deep potentials, and dense environments. This statistical approach not only avoids arbitrary selection thresholds, but also quantifies the dominant physical variables which highlight the properties of haloes and environments which support this pathway of massive galaxy evolution under suppressed feedback, and differentiate \acp{mqg} from their high-mass star-forming counterparts, which may be of practical use for their modelling in N-body simulations. By further separating the \ac{mqg} population according to their quenching time, we determine how halo, environmental, and galaxy properties evolve over the course of quenching, providing new insight into the physics regulating early galaxy evolution.

Our findings can be summarised as follows:

\begin{itemize}
\item As shown in \cref{sec:fullcluster}, the smallest \ac{gmm} cluster that fully contains all \aclp{mqg} includes 90 haloes with the highest masses, radii, densities (measured within a 200 \ac{ckpc} aperture), and virial velocities, which serve as proxies for halo potential depth. The \ac{pca} decomposition shows that within this cluster, there is a mass-independent variance which separates recent and long quiescent galaxies, being correlated with their large scale environments, and against their largest merger ratios. This suggests that smooth accretion from megaparsec-scale cosmic structures can contribute to the growth timescale of massive galaxies; the timing of which separates early and late quenching galaxies.

\item In \cref{sec:incluster}, we find that all of these properties increase more rapidly for \acp{mqg} than for their star-forming counterparts. As a result, the quiescent galaxies emerge as the most massive and compact in terms of their stellar components and host some of the most massive \acp{agn} in the simulation. These deep potential wells appear to sustain \ac{agn} activity, which is likely the main driver of quenching. Mergers, by contrast, do not significantly influence either halo or galaxy evolution in these cases. However, \ac{agn} mass alone does not fully explain the quenching. These galaxies also have proportionally smaller gas reservoirs and more concentrated stellar mass, as reflected in their low stellar radii (\cref{fig:pcastats}) and early gas depletion (\cref{fig:ghe}).

\item We discuss in \cref{sec:incluster,sec:simvsim} that this relationship between halo sizes, small scale environment and star formation status is similar to that of the quiescent galaxies found at lower redshifts in the \textsc{IllustrisTNG} and \textsc{Colibre} simulations. However, they are distinct from the smaller haloes in underdense regions illustrated in the \textsc{Magneticum Pathfinder} model. With \textsc{Thesan} being closely related to \textsc{IllustrisTNG}, it suggests that environmental dependencies in \ac{mqg} formation may persist across different feedback implementations, though the specific \ac{agn} growth timescales and masses differ substantially from standard \textsc{IllustrisTNG} owing to the suppressed coupling in \textsc{Thesan}.

\item Haloes and large scale environments which harbour \aclp{mqg} continue to grow after quenching. In \cref{sec:incluster}, we show a clear hierarchy of halo mass, accretion gradient and small scale environment in relation to quenching time, which indicates that faster-growing haloes host massive galaxies which quench earlier. On 1 \ac{cMpc} scales, environmental densities around \acfp{rqg} are no different from massive star-forming galaxies; unlike \acfp{lqg}, whose environments remain relatively overdense. On scales of 3 \acp{cMpc}, \acp{lqg} also lose this enhancement. When tracking the distribution of neighbours over time using the radial skew parameter introduced in \citet{Chittenden}, we see the gravitation of large objects towards the target haloes, which are likely the source of accretion of matter which feeds the rapidly growing haloes, galaxies and \acp{agn}.

\item A crucial factor in the quenching of massive galaxies in \textsc{Thesan} is the growth of their central \acp{agn}, which reach among the highest black hole masses in the simulation. We show in \cref{fig:simcomp2} that the fiducial \textsc{Thesan} simulations exhibit faster \ac{agn} growth than their cross-matched counterparts using the \textsc{IllustrisTNG} model, resulting in an order-of-magnitude difference in black hole masses by $z=5.5$. This disparity arises from the suppressed \ac{agn} feedback coupling discussed above, which delays self-regulation and permits prolonged accretion \citep{ThesanAddendum}. Once the black holes exceed the $M_\mathrm{BH} \sim 10^{8.5},\mathrm{M_\odot}$ threshold, the kinetic feedback mode efficiently quenches star formation, demonstrating that the quenching mechanism itself remains robust even under artificially weak feedback prior to this transition. The coincidence of quenching with a sharp decline in Eddington ratio highlights the enduring importance of feedback mode transitions in regulating massive galaxy evolution. While the rapid black hole growth seen in \textsc{Thesan} is now recognised as arising from numerical issues rather than physical \ac{agn} models, it provides a valuable analogue for scenarios in which feedback is weakly coupled or obstructed by dense, optically thick environments at high redshift. In this sense, the \textsc{Thesan} \acp{mqg} offer a physically meaningful exploration of how inefficient self-regulation can still culminate in powerful kinetic feedback and early quenching, especially within the overdense environments typical of the epoch of reionisation where such conditions may be physically plausible.

\item We stress in \cref{sec:haloenv} that the relationship between the growth of large scale environments and the quenching of massive galaxies implies that the accretion of neighbouring haloes around \ac{mqg}-hosting haloes will continue after the end of reionisation. Consequently, the quiescent galaxies detected from $z=3$ to $z=5$ may be found in some of the densest large scale environments at these redshifts; potentially corroborated by examples given in \cref{sec:simvobs}. It would appear that while these galaxies remain quiescent, their haloes consume their surrounding neighbours and accumulate mass. The central \acp{agn} concurrently prevents further star formation, which constitutes a possible pathway for \acp{mqg} to become isolated while preserving their quiescent state throughout cosmic history. Since mergers are not a significant component of \ac{mqg} evolution, they will not be rejuvenated by the infall of smaller galaxies, which are likely to be stripped of their star-forming gas by ram pressure from the massive halo, further suppressing star formation. However, some observed quiescent galaxies exhibit potential evidence of past mergers, which could influence the morphology of these galaxies post-quenching.
\end{itemize}

The discovery of numerous massive quiescent galaxies in \textit{JWST} data offers a valuable opportunity to test and refine models of early galaxy evolution and \ac{agn} feedback at high redshift.

Comprehensive modelling of reionisation-era quenching could help resolve discrepancies between observed and simulated galaxy populations, yet the limited sample of simulated galaxies poses challenges for investigating their complex nature. Key questions remain unanswered, such as how some massive galaxies sustain star formation despite hosting massive \acp{agn} or residing in concentrated environments; conditions we associate with quiescent systems. Moreover, the recently identified \textit{JWST} galaxies with even earlier formation times remain unaccounted for in current models. The identification of significant haloes and environments in simulations could be pivotal for targeting regions in future deep-field surveys, and may offer deeper insights into these early galaxy populations.

While the suppressed \ac{agn} feedback in \textsc{Thesan} was initially unintended, it provides an exploration of galaxy evolution under a regime of weakened \ac{agn} regulation. Although there is no theoretical consensus that \ac{agn} feedback is systematically suppressed at high redshift, observational evidence from \textit{JWST} identifies systems with overmassive black holes at $4<z<11$ \citep{Maiolino} and cases where radiative feedback appears weakly coupled in individual objects \citep{WWang}. Whether and how commonly such conditions arise in the early universe remains an open question. Crucially, recent simulations show that permitting super-Eddington black hole accretion is sufficient to reproduce the observed number densities of massive quiescent galaxies at $z \geq 6$ \citep{Chaikin2026}, suggesting that the \textsc{Thesan} results may have a physical counterpart. In this sense, the \textsc{Thesan} data offer insights into how suppressed \ac{agn} self-regulation can culminate in powerful kinetic feedback and early quenching, offering valuable constraints on how feedback strength modulates the coevolution of galaxies and supermassive black holes during the epoch of reionisation.

Given that the conditions for forming \aclp{mqg} can be inferred from specific halo and environmental conditions and their historical properties, \acp{mqg} may be replicated in a greater abundance in future, by means of N-body simulation data, in order to reconcile them with the increasing number of detections in deep surveys. N-body simulations, unconstrained by computationally expensive baryonic physics, allow us to extend the \ac{mqg} population by applying machine learning models trained on \textsc{Thesan} to haloes with comparable masses and environments. Applying these predictors to high-volume simulations such as \textsc{Uchuu} \citep{Uchuu} could yield large-scale \ac{mqg} demographics and mock observables; yet doing so ultimately requires broader validation across multiple hydrodynamical models, to ensure the reliability of any extrapolation beyond \textsc{Thesan}.

Crucially, since \acp{agn} have minimal influence on the large-scale reionisation of the universe \citep{Yeh}, many reionisation simulations such as \textsc{Sphinx} \citep{Sphinx} and \textsc{Cosmic Dawn} \citep{Lewis} exclude \acp{agn} models to enhance computational efficiency. This even includes the recently announced \textsc{Thesan-Zoom} simulations\footnote{Further to the lack of AGN modelling, \textsc{Thesan-Zoom} is unsuitable for investigating MQGs as it resolves only fourteen selected haloes from the fiducial \textsc{Thesan} data. This small subset of haloes spans a halo mass range of $10^8 M_\odot$ to $10^{13} M_\odot$ with near loguniform separation, and so does not correspond to our MQG sample.} \citep{ThesanZoom}, where despite the success of their updated multiphase ISM and stellar feedback models in regulating star formation with improved accuracy, the lack of \ac{agn} feedback leads to centrally concentrated star formation bursts; which are rarely associated with \ac{agn}-dominated galaxies \citep{McClymont}.

Given our findings on the significant role of \acp{agn} in early galaxy quenching, their exclusion may overlook critical processes and impact the accuracy of these models, as well as other critical processes such as \ac{agn}-driven outflows, which are prevalent in recent ALMA observations \citep{Spilker}. Furthermore, it has been shown recently by \citet{DSilva} that the growth of supermassive black holes can outpace cosmic star formation as early as $z\sim 13.5$, which suggests that \ac{agn} assembly plays an integral role in early structure formation. The authors of \textsc{Thesan-Zoom} nonetheless acknowledge that incorporating \ac{agn} feedback is necessary for the comprehensive modelling of massive galaxies as they contribute significantly to star formation regulation, and intend to implement this in future work \citep{ThesanZoom, McClymont}. Our reduction of the parameter space to the environments capturing \acp{mqg} may be used in future to justify the selection of objects in future zoom simulations.

The unique population statistics of the high-redshift universe under different feedback implementations makes this a problem which benefits from exploring multiple AGN coupling scenarios. However, as we obtain a deeper survey of \aclp{mqg} with further details of their structures and environments, we may be able to use these galaxies to constrain AGN feedback processes and environmental effects operating during the first billion years of cosmic time.

\section*{Acknowledgements}

This research was generously funded under Australian Research Council Laureate Fellowship Grant FL180100060 and NASA Grant JWST-ERS-2565. HGC personally thanks Darren Croton, Philip Hopkins, Marion Farcy, Arianna Di Cintio, Enrico Garaldi and Xuejian Shen for valuable discussions; and the \textsc{Thesan} simulation team for access and support in relation to their public simulation data, as well as necessary revisions to this manuscript. We also thank the anonymous referee and journal editor for comments which improved the quality of this paper.

\vspace{-10pt}

\section*{Data Availability}
 
The primary author is presently willing to provide data relating to this work upon reasonable request.

\vspace{-10pt}



\bibliographystyle{mnras}
\bibliography{references} 

@article{IllustrisTNG,
  author =        {Nelson, Dylan and Springel, Volker and
                   Pillepich, Annalisa and Rodriguez-Gomez, Vicente and
                   Torrey, Paul and Genel, Shy and Vogelsberger, Mark and
                   Pakmor, Ruediger and Marinacci, Federico and
                   Weinberger, Rainer and Kelley, Luke and Lovell, Mark and
                   Diemer, Benedikt and Hernquist, Lars},
  journal =       {CompAC},
  month =         {05},
  pages =         {2},
  title =         {{The IllustrisTNG Simulations: Public Data Release}},
  volume =        {6},
  year =          {2019},
  doi =           {10.1186/s40668-019-0028-x},
}

@article{SimbaC,
   title={SIMBA-C: An updated chemical enrichment model for galactic chemical evolution in the SIMBA simulation},
   volume={525},
   ISSN={1365-2966},
   url={http://dx.doi.org/10.1093/mnras/stad2394},
   DOI={10.1093/mnras/stad2394},
   number={1},
   journal={MNRAS},
   publisher={Oxford University Press (OUP)},
   author={Hough, Renier T and Rennehan, Douglas and Kobayashi, Chiaki and Loubser, S Ilani and Davé, Romeel and Babul, Arif and Cui, Weiguang},
   year={2023},
   month={Aug},
   pages={1061}
}

@article{Szpila,
    author = {Jakub Szpila and Romeel Davé and Douglas Rennehan and Weiguang Cui and Renier T Hough},
    title = {The nature and evolution of early massive quenched galaxies in the Simba-C simulation},
    journal = {MNRAS},
    volume = {537},
    number = {2},
    pages = {1849},
    year = {2025},
    month = {Jan},
    doi = {10.1093/mnras/staf132},
    url = {https://doi.org/10.1093/mnras/staf132}
}

@article{Agarwal,
  author =        {Agarwal, Shankar and Dav{\'e}, Romeel and
                   Bassett, Bruce A},
  journal =       {MNRAS},
  month =         {May},
  number =        {3},
  pages =         {3410},
  publisher =     {Oxford University Press (OUP)},
  title =         {Painting galaxies into dark matter haloes using
                   machine learning},
  volume =        {478},
  year =          {2018},
  doi =           {10.1093/mnras/sty1169},
  issn =          {1365-2966},
  url =           {http://dx.doi.org/10.1093/mnras/sty1169},
}

@article{Lovell,
  author =        {Lovell, Christopher C and Wilkins, Stephen M and
                   Thomas, Peter A and Schaller, Matthieu and
                   Baugh, Carlton M and Fabbian, Giulio and
                   Bah{\'e}, Yannick},
  journal =       {MNRAS},
  month =         {Nov},
  number =        {4},
  pages =         {5046},
  publisher =     {Oxford University Press (OUP)},
  title =         {A machine learning approach to mapping baryons on to
                   dark matter haloes using the eagle and C-EAGLE
                   simulations},
  volume =        {509},
  year =          {2021},
  doi =           {10.1093/mnras/stab3221},
  issn =          {1365-2966},
  url =           {http://dx.doi.org/10.1093/mnras/stab3221},
}

@article{Adamo,
       author = {{Adamo}, Angela and {Atek}, Hakim and {Bagley}, Micaela B. and {Ba{\~n}ados}, Eduardo and {Barrow}, Kirk S.~S. and {Berg}, Danielle A. and {Bezanson}, Rachel and {Brada{\v{c}}}, Maru{\v{s}}a and {Brammer}, Gabriel and {Carnall}, Adam C. and {Chisholm}, John and {Coe}, Dan and {Dayal}, Pratika and {Eisenstein}, Daniel J. and {Eldridge}, Jan J. and {Ferrara}, Andrea and {Fujimoto}, Seiji and {de Graaff}, Anna and {Habouzit}, Melanie and {Hutchison}, Taylor A. and {Kartaltepe}, Jeyhan S. and {Kassin}, Susan A. and {Kriek}, Mariska and {Labb{\'e}}, Ivo and {Maiolino}, Roberto and {Marques-Chaves}, Rui and {Maseda}, Michael V. and {Mason}, Charlotte and {Matthee}, Jorryt and {McQuinn}, Kristen B.~W. and {Meynet}, Georges and {Naidu}, Rohan P. and {Oesch}, Pascal A. and {Pentericci}, Laura and {P{\'e}rez-Gonz{\'a}lez}, Pablo G. and {Rigby}, Jane R. and {Roberts-Borsani}, Guido and {Schaerer}, Daniel and {Shapley}, Alice E. and {Stark}, Daniel P. and {Stiavelli}, Massimo and {Strom}, Allison L. and {Vanzella}, Eros and {Wang}, Feige and {Wilkins}, Stephen M. and {Williams}, Christina C. and {Willott}, Chris J. and {Wylezalek}, Dominika and {Nota}, Antonella},
        title = {{The First Billion Years, According To JWST}},
      journal = {arXiv e-prints},
         year = 2024,
        month = {May},
          eid = {arXiv:2405.21054},
        pages = {arXiv:2405.21054},
          doi = {10.48550/arXiv.2405.21054},
archivePrefix = {arXiv},
       eprint = {2405.21054},
 primaryClass = {astro-ph.GA},
       adsurl = {https://ui.adsabs.harvard.edu/abs/2024arXiv240521054A}
}

@article{Pillepich2017,
  author =        {Pillepich, Annalisa and Springel, Volker and
                   Nelson, Dylan and Genel, Shy and Naiman, Jill and
                   Pakmor, Rüdiger and Hernquist, Lars and Torrey, Paul and
                   Vogelsberger, Mark and Weinberger, Rainer and
                   Marinacci, Federico},
  journal =       {MNRAS},
  month =         {Oct},
  number =        {3},
  pages =         {4077},
  publisher =     {Oxford University Press (OUP)},
  title =         {Simulating galaxy formation with the IllustrisTNG
                   model},
  volume =        {473},
  year =          {2017},
  doi =           {10.1093/mnras/stx2656},
  issn =          {1365-2966},
  url =           {http://dx.doi.org/10.1093/mnras/stx2656},
}

@article{Arepo,
    doi = {10.3847/1538-4365/ab908c},
  
    url = {https://doi.org/10.3847%2F1538-4365%2Fab908c},
  
    year = 2020,
    month = {Jun},
  
    publisher = {AAS},
  
    volume = {248},
  
    number = {2},
  
    pages = {32},
  
    author = {Rainer Weinberger and Volker Springel and Rüdiger Pakmor},
  
    title = {The Arepo Public Code Release},
  
    journal = {ApJSS}
}

@article{ELU,
  author =        {{Clevert}, Djork-Arn{\'e} and {Unterthiner}, Thomas and
                   {Hochreiter}, Sepp},
  journal =       {arXiv e-Print:1511.07289},
  month =         Nov,
  title =         {{Fast and Accurate Deep Network Learning by
                   Exponential Linear Units (ELUs)}},
  year =          {2015},
  doi =           {10.48550/ARXIV.1511.07289},
}

@article{Chaikin2026,
  author = {Chaikin, Evgenii and Schaye, Joop and Hu{\v s}ko, Filip and Lacey, Cedric G. and Ploeckinger, Sylvia and Schaller, Matthieu},
  title = {The importance of super-{E}ddington black hole accretion for the emergence of massive quiescent galaxies at high redshift},
  journal = {MNRAS},
  year = {2026},
  doi = {10.48550/arXiv.2601.15207}
}

@ARTICLE{ChandroGomez,
       author = {{Chandro-Gómez}, {\'A}ngel and {Lagos}, Claudia del P. and {Power}, Chris and {Baker}, Willian M. and {Ben{\'\i}tez-Llambay}, Alejandro and {Chaikin}, Evgenii and {Chittenden}, Harry G. and {Correa}, Camila and {Frenk}, Carlos S. and {Hu{\v{s}}ko}, Filip and {McGibbon}, Robert J. and {Nanayakkara}, Themiya and {Ploeckinger}, Sylvia and {Richings}, Alexander J. and {Schaller}, Matthieu and {Schaye}, Joop and {Trayford}, James W.},
        title = "{Unveiling the properties and origin of massive quenched galaxies at $z\ge2$ in the COLIBRE hydrodynamical simulations}",
      journal = {arXiv e-prints},
         year = 2025,
        month = dec,
          eid = {arXiv:2512.16208},
        pages = {arXiv:2512.16208},
          doi = {10.48550/arXiv.2512.16208},
archivePrefix = {arXiv},
       eprint = {2512.16208},
 primaryClass = {astro-ph.GA},
       adsurl = {https://ui.adsabs.harvard.edu/abs/2025arXiv251216208C}
}

@ARTICLE{Colibre,
       author = {{Schaye}, Joop and {Chaikin}, Evgenii and {Schaller}, Matthieu and {Ploeckinger}, Sylvia and {Hu{\v{s}}ko}, Filip and {McGibbon}, Rob and {Trayford}, James W. and {Ben{\'\i}tez-Llambay}, Alejandro and {Correa}, Camila and {Frenk}, Carlos S. and {Richings}, Alexander J. and {Forouhar Moreno}, Victor J. and {Bah{\'e}}, Yannick M. and {Borrow}, Josh and {Durrant}, Anna and {Gebek}, Andrea and {Helly}, John C. and {Jenkins}, Adrian and {Lacey}, Cedric G. and {Ludlow}, Aaron and {Nobels}, Folkert S.~J.},
        title = "{The COLIBRE project: cosmological hydrodynamical simulations of galaxy formation and evolution}",
      journal = {MNRAS},
         year = 2025,
          doi = {10.48550/arXiv.2508.21126}
}

@ARTICLE{Colibre2,
       author = {{Chaikin}, Evgenii and {Schaye}, Joop and {Schaller}, Matthieu and {Ploeckinger}, Sylvia and {Bah{\'e}}, Yannick M. and {Ben{\'\i}tez-Llambay}, Alejandro and {Correa}, Camila and {Forouhar Moreno}, Victor J. and {Frenk}, Carlos S. and {Hu{\v{s}}ko}, Filip and {Kugel}, Roi and {McGibbon}, Robert and {Richings}, Alexander J. and {Trayford}, James W. and {Borrow}, Josh and {Crain}, Robert A. and {Helly}, John C. and {Lacey}, Cedric G. and {Ludlow}, Aaron and {Nobels}, Folkert S.~J.},
        title = "{COLIBRE: calibrating subgrid feedback in cosmological simulations that include a cold gas phase}",
      journal = {MNRAS},
         year = 2025,
          doi = {10.48550/arXiv.2509.04067}
}

@article{Glazebrook2024,
   title={A massive galaxy that formed its stars at z = 11},
   volume={628},
   ISSN={1476-4687},
   url={http://dx.doi.org/10.1038/s41586-024-07191-9},
   DOI={10.1038/s41586-024-07191-9},
   number={8007},
   journal={Nature},
   publisher={Springer Science and Business Media LLC},
   author={Glazebrook, Karl and Nanayakkara, Themiya and Schreiber, Corentin and Lagos, Claudia and Kawinwanichakij, Lalitwadee and Jacobs, Colin and Chittenden, Harry and Brammer, Gabriel and Kacprzak, Glenn G. and Labbé, Ivo and Marchesini, Danilo and Marsan, Z. Cemile and Oesch, Pascal A. and Papovich, Casey and Remus, Rhea-Silvia and Tran, Kim-Vy H. and Esdaile, James and Chandro-Gomez, Angel},
   year={2024},
   month={Feb},
   pages={277}}

@article{Marinacci,
    author = {Marinacci, Federico and Vogelsberger, Mark and Pakmor, Rüdiger and Torrey, Paul and Springel, Volker and Hernquist, Lars and Nelson, Dylan and Weinberger, Rainer and Pillepich, Annalisa and Naiman, Jill and Genel, Shy},
    title = {{First results from the IllustrisTNG simulations: radio haloes and magnetic fields}},
    journal = {MNRAS},
    volume = {480},
    number = {4},
    pages = {5113},
    year = {2018},
    month = {08},
    issn = {0035-8711},
    doi = {10.1093/mnras/sty2206},
    url = {https://doi.org/10.1093/mnras/sty2206},
}

@ARTICLE{Prole,
       author = {{Prole}, Lewis R. and {Regan}, John A. and {Mehta}, Daxal and {Pakmor}, Rudiger and {Koudmani}, Sophie and {Bourne}, Martin A. and {Glover}, Simon C.~O. and {Wise}, John H. and {Klessen}, Ralf S. and {Tremmel}, Michael and {Sijacki}, Debora and {Beckmann}, Ricarda S. and {Haehnelt}, Martin G. and {Brennan}, John and {van de Bor}, Pelle and {Clark}, Paul C.},
        title = "{The SEEDZ Simulations: Methodology and First Results on Massive Black Hole Seeding and Early Galaxy Growth}",
      journal = {arXiv e-prints},
         year = 2025,
        month = {Nov},
          eid = {arXiv:2511.09640},
        pages = {arXiv:2511.09640},
          doi = {10.48550/arXiv.2511.09640},
archivePrefix = {arXiv},
       eprint = {2511.09640},
 primaryClass = {astro-ph.GA},
       adsurl = {https://ui.adsabs.harvard.edu/abs/2025arXiv251109640P}
}

@article{Bennett,
    author = {Jake S. Bennett and Debora Sijacki and Tiago Costa and Nicolas Laporte and Callum Witten},
    title = {The growth of the gargantuan black holes powering high-redshift quasars and their impact on the formation of early galaxies and protoclusters},
    DOI= "10.1093/mnras/stad3179",
    url= "https://doi.org/10.1093/mnras/stad3179",
    journal = {MNRAS},
    year = 2024,
    volume = 527,
    pages = 1033
}

@ARTICLE{Maiolino,
       author = {{Maiolino}, Roberto and {Scholtz}, Jan and {Curtis-Lake}, Emma and {Carniani}, Stefano and {Baker}, William and {de Graaff}, Anna and {Tacchella}, Sandro and {{\"U}bler}, Hannah and {D'Eugenio}, Francesco and {Witstok}, Joris and {Curti}, Mirko and {Arribas}, Santiago and {Bunker}, Andrew J. and {Charlot}, St{\'e}phane and {Chevallard}, Jacopo and {Eisenstein}, Daniel J. and {Egami}, Eiichi and {Ji}, Zhiyuan and {Jones}, Gareth C. and {Lyu}, Jianwei and {Rawle}, Tim and {Robertson}, Brant and {Rujopakarn}, Wiphu and {Perna}, Michele and {Sun}, Fengwu and {Venturi}, Giacomo and {Williams}, Christina C. and {Willott}, Chris},
        title = "{JADES: The diverse population of infant black holes at 4 < z < 11: Merging, tiny, poor, but mighty}",
      journal = {\aap},
         year = 2024,
        month = {Nov},
       volume = {691},
          eid = {A145},
        pages = {A145},
          doi = {10.1051/0004-6361/202347640},
archivePrefix = {arXiv},
       eprint = {2308.01230},
 primaryClass = {astro-ph.GA},
       adsurl = {https://ui.adsabs.harvard.edu/abs/2024A&A...691A.145M}
}

@ARTICLE{Gaspari1,
       author = {{Gaspari}, M. and {Ruszkowski}, M. and {Oh}, S. Peng},
        title = "{Chaotic cold accretion on to black holes}",
      journal = {\mnras},
         year = 2013,
        month = {Jul},
       volume = {432},
       number = {4},
        pages = {3401-3422},
          doi = {10.1093/mnras/stt692},
archivePrefix = {arXiv},
       eprint = {1301.3130},
 primaryClass = {astro-ph.CO},
       adsurl = {https://ui.adsabs.harvard.edu/abs/2013MNRAS.432.3401G}
}

@article{DESI,
  author =        {{DESI Collaboration}},
  journal =       {arXiv e-Print: 1611.00036},
  month =         {Oct},
  title =         {{The DESI Experiment Part I: Science, Targeting, and
                   Survey Design}},
  year =          {2016},
  doi =           {10.48550/ARXIV.1611.00036}
}

@article{Wilkins,
    doi = {10.33232/001c.132374},
    url = {http://dx.doi.org/10.33232/001c.132374},
    year = 2025,
    volume={8},
    pages={29},
    author = {Stephen M. Wilkins and Jussi K. Kuusisto and Dimitrios Irodotou and Shihong Liao and Christopher C. Lovell and Sonja Soininen and Sabrina C. Berger and Sophie L. Newman and William J. Roper and Louise T. C. Seeyave and Peter A. Thomas and Aswin P. Vijayan},
    title = {{First Light and Reionization Epoch Simulations (FLARES) -- XV: The physical properties of super-massive black holes and their impact on galaxies in the early universe}},
    journal = {Open Journal Of Astrophysics}
}

@article{Harvey,
   title={EPOCHS. IV. SED Modeling Assumptions and Their Impact on the Stellar Mass Function at 6.5 < z < 13.5 Using PEARLS and Public JWST Observations},
   volume={978},
   ISSN={1538-4357},
   url={http://dx.doi.org/10.3847/1538-4357/ad8c29},
   DOI={10.3847/1538-4357/ad8c29},
   number={1},
   journal={\apj},
   publisher={American Astronomical Society},
   author={Harvey, Thomas and Conselice, Christopher J. and Adams, Nathan J. and Austin, Duncan and Juodžbalis, Ignas and Trussler, James and Li, Qiong and Ormerod, Katherine and Ferreira, Leonardo and Lovell, Christopher C. and Duan, Qiao and Westcott, Lewi and Harris, Honor and Bhatawdekar, Rachana and Coe, Dan and Cohen, Seth H. and Caruana, Joseph and Cheng, Cheng and Driver, Simon P. and Frye, Brenda and Furtak, Lukas J. and Grogin, Norman A. and Hathi, Nimish P. and Holwerda, Benne W. and Jansen, Rolf A. and Koekemoer, Anton M. and Marshall, Madeline A. and Nonino, Mario and Vijayan, Aswin P. and Wilkins, Stephen M. and Windhorst, Rogier and Willmer, Christopher N. A. and Yan, Haojing and Zitrin, Adi},
   year={2024},
   month={Dec},
   pages={89}
}

@article{Belli,
   title={Star formation shut down by multiphase gas outflow in a galaxy at a redshift of 2.45},
   volume={630},
   ISSN={1476-4687},
   url={http://dx.doi.org/10.1038/s41586-024-07412-1},
   DOI={10.1038/s41586-024-07412-1},
   number={8015},
   journal={Nature},
   publisher={Springer Science and Business Media LLC},
   author={Sirio Belli and Minjung Park and Rebecca L. Davies and J. Trevor Mendel and Benjamin D. Johnson and Charlie Conroy and Chloë Benton and Letizia Bugiani and Razieh Emami and Joel Leja and Yijia Li and Gabriel Maheson and Elijah P. Mathews and Rohan P. Naidu and Erica J. Nelson and Sandro Tacchella and Bryan A. Terrazas and Rainer Weinberger},
   year={2024},
   month={Apr},
   pages={54}
}

@article{Lyu,
    doi = {10.1051/0004-6361/202451067},
    url = {https://doi.org/10.1051/0004-6361/202451067},
    year = 2025,
    month = {Jan},
    pages = {A313},
    author = {{Lyu}, Yipeng and {Magnelli}, Benjamin and {Elbaz}, David and {P{\'e}rez-Gonz{\'a}lez}, Pablo G. and {Correa}, Camila and {Daddi}, Emanuele and {G{\'o}mez-Guijarro}, Carlos and {Dunlop}, James S. and {Grogin}, Norman A. and {Koekemoer}, Anton M. and {McLeod}, Derek J. and {Lu}, Shiying},
    title = {PRIMER: JWST/MIRI reveals the evolution of star-forming structures in galaxies at z ≤ 2.5},
    volume = {693},
    journal = {A\&A}
}

@article{Wright,
    doi = {10.3847/2041-8213/ad2b6d},
    url = {https://doi.org/10.3847/2041-8213/ad2b6d},
    year = 2024,
    month = {Mar},
    pages = {L10},
    author = {Lillian Wright and Katherine E. Whitaker and John R. Weaver and Sam E. Cutler and Bingjie Wang and Adam Carnall and Katherine A. Suess and Rachel Bezanson and Erica Nelson and Tim B. Miller and Kei Ito and Francesco Valentino},
    title = {Remarkably Compact Quiescent Candidates at 3 < z < 5 in JWST-CEERS},
    volume = {964},
    journal = {ApJL}
}

@ARTICLE{Stefanon,
       author = {{Stefanon}, Mauro and {Bouwens}, Rychard J. and {Labb{\'e}}, Ivo and {Illingworth}, Garth D. and {Gonzalez}, Valentino and {Oesch}, Pascal A.},
        title = {Galaxy Stellar Mass Functions from z   10 to z   6 using the Deepest Spitzer/Infrared Array Camera Data: No Significant Evolution in the Stellar-to-halo Mass Ratio of Galaxies in the First Gigayear of Cosmic Time},
      journal = {\apj},
         year = 2021,
        month = {Nov},
       volume = {922},
       number = {1},
          eid = {29},
        pages = {29},
          doi = {10.3847/1538-4357/ac1bb6},
archivePrefix = {arXiv},
       eprint = {2103.16571},
 primaryClass = {astro-ph.GA},
       adsurl = {https://ui.adsabs.harvard.edu/abs/2021ApJ...922...29S}
}

@article{Bower,
   title={The dark nemesis of galaxy formation: why hot haloes trigger black hole growth and bring star formation to an end},
   volume={465},
   ISSN={1365-2966},
   url={http://dx.doi.org/10.1093/mnras/stw2735},
   DOI={10.1093/mnras/stw2735},
   number={1},
   journal={\mnras},
   publisher={Oxford University Press (OUP)},
   author={Bower, Richard G. and Schaye, Joop and Frenk, Carlos S. and Theuns, Tom and Schaller, Matthieu and Crain, Robert A. and McAlpine, Stuart},
   year={2017},
   month={Feb},
   pages={32}
}

@ARTICLE{Hopkins,
       author = {{Hopkins}, Philip F. and {Grudic}, Michael Y. and {Su}, Kung-Yi and {Wellons}, Sarah and {Angles-Alcazar}, Daniel and {Steinwandel}, Ulrich P. and {Guszejnov}, David and {Murray}, Norman and {Faucher-Giguere}, Claude-Andre and {Quataert}, Eliot and {Keres}, Dusan},
        title = "{FORGE'd in FIRE: Resolving the End of Star Formation and Structure of AGN Accretion Disks from Cosmological Initial Conditions}",
      journal = {The Open Journal of Astrophysics},
         year = 2024,
        month = mar,
       volume = {7},
          eid = {18},
        pages = {18},
          doi = {10.21105/astro.2309.13115},
archivePrefix = {arXiv},
       eprint = {2309.13115},
 primaryClass = {astro-ph.GA},
       adsurl = {https://ui.adsabs.harvard.edu/abs/2024OJAp....7E..18H}
}

@article{Nelson2018,
  author =        {Nelson, Dylan and Pillepich, Annalisa and
                   Springel, Volker and Weinberger, Rainer and
                   Hernquist, Lars and Pakmor, Rüdiger and Genel, Shy and
                   Torrey, Paul and Vogelsberger, Mark and
                   Kauffmann, Guinevere and Marinacci, Federico and
                   Naiman, Jill},
  journal =       {MNRAS},
  month =         {Nov},
  number =        {1},
  pages =         {624},
  publisher =     {Oxford University Press (OUP)},
  title =         {First results from the IllustrisTNG simulations: the
                   galaxy colour bimodality},
  volume =        {475},
  year =          {2017},
  doi =           {10.1093/mnras/stx3040},
  issn =          {1365-2966},
  url =           {http://dx.doi.org/10.1093/mnras/stx3040},
}

@article{Naiman,
    doi = {10.1093/mnras/sty618},
  
    url = {https://doi.org/10.1093%2Fmnras%2Fsty618},
  
    year = 2018,
    month = {Mar},
  
    publisher = {Oxford University Press ({OUP})},
  
    volume = {477},
  
    number = {1},
  
    pages = {1206},
  
    author = {Jill P Naiman and Annalisa Pillepich and Volker Springel and Enrico Ramirez-Ruiz and Paul Torrey and Mark Vogelsberger and Rüdiger Pakmor and Dylan Nelson and Federico Marinacci and Lars Hernquist and Rainer Weinberger and Shy Genel},
  
    title = {First results from the {IllustrisTNG} simulations: a tale of two elements {\textendash} chemical evolution of magnesium and europium},
  
    journal = {MNRAS}
}

@article{Baker,
       author = {William M. Baker and Seunghwan Lim and Francesco D'Eugenio and Roberto Maiolino and Zhiyuan Ji and Santiago Arribas and Andrew J. Bunker and Stefano Carniani and Stephane Charlot and Anna de Graaff and Kevin Hainline and Tobias J. Looser and Jianwei Lyu and Pierluigi Rinaldi and Brant Robertson and Matthieu Schaller and Joop Schaye and Jan Scholtz and Hannah Ubler and Christina C. Williams and Christopher N. A. Willmer and Chris Willott and Yongda Zhu},
        title = {The abundance and nature of high-redshift quiescent galaxies from JADES spectroscopy and the FLAMINGO simulations},
      journal = {MNRAS},
         year = 2025,
        month = {Mar},
        volume={539},
        number={1},
        pages={557},
          doi = {10.1093/mnras/staf475}
}

@book{Kendall,
  title={Kendall's Advanced Theory of Statistics},
  author={Kendall, M.G. and Stuart, A. and Ord, J.K. and Arnold, S.F. and O'Hagan, A.},
  number={v. 2, pt. 1},
  isbn={9780340807521},
  lccn={94188490},
  series={Kendall's Advanced Theory of Statistics},
  url={https://books.google.bs/books?id=JY5TnwEACAAJ},
  year={1999},
  publisher={Edward Arnold}
}

@BOOK{MurtaghHeck,
       author = {{Murtagh}, Fionn and {Heck}, Andre},
        title = "{Multivariate Data Analysis}",
         year = 1987,
       volume = {131},
       series={2},
       publisher={Springer},
          doi = {10.1007/978-94-009-3789-5},
       adsurl = {https://ui.adsabs.harvard.edu/abs/1987ASSL..131.....M}
}

@article{Ubler,
    author = {Übler, Hannah and Maiolino, Roberto and Pérez-González, Pablo G and D’Eugenio, Francesco and Perna, Michele and Curti, Mirko and Arribas, Santiago and Bunker, Andrew and Carniani, Stefano and Charlot, Stéphane and Rodríguez Del Pino, Bruno and Baker, William and Böker, Torsten and Cresci, Giovanni and Dunlop, James and Grogin, Norman A and Jones, Gareth C and Kumari, Nimisha and Lamperti, Isabella and Laporte, Nicolas and Marshall, Madeline A and Mazzolari, Giovanni and Parlanti, Eleonora and Rawle, Tim and Scholtz, Jan and Venturi, Giacomo and Witstok, Joris},
    title = {GA-NIFS: JWST discovers an offset AGN 740 million years after the big bang},
    journal = {MNRAS},
    volume = {531},
    number = {1},
    pages = {355},
    year = {2024},
    month = {May},
    issn = {0035-8711},
    doi = {10.1093/mnras/stae943},
    url = {https://doi.org/10.1093/mnras/stae943},
    eprint = {https://academic.oup.com/mnras/article-pdf/531/1/355/57604632/stae943.pdf},
}

@article{Popesso,
    author = {P. Popesso and I. Marini and K. Dolag and G. Lamer and B. Csizi and V. Biffi and A. Robothan and M. Bravo and A. Biviano and S. Vladutesku-Zopp and L. Lovisari and S. Ettori and M. Angelinelli and S. Driver and V. Toptun and A. Dev and D. Mazengo and A. Merloni and Y. Zhang and J. Comparat and G. Ponti and T. Mroczkowski and E. Bulbul},
    title = {Average X-ray properties of galaxy groups. From Milky Way-like halos to massive clusters},
    DOI= "10.48550/arXiv.2411.17120",
    url= "https://doi.org/10.48550/arXiv.2411.17120",
    volume = {704},
          eid = {A278},
        pages = {A278},
    journal = {A\&A},
    year = 2025
}

@article{Forrest2025,
    author = {{Forrest}, Ben and {Muzzin}, Adam and {Marchesini}, Danilo and {Pan}, Richard and {Ozden}, Nehir and {Antwi-Danso}, Jacqueline and {Chang}, Wenjun and {Cooper}, M.~C. and {Edward}, Adit H. and {Gomez}, Percy and {Kimmig}, Lucas and {Lemaux}, Brian C. and {McConachie}, Ian and {Noble}, Allison and {Remus}, Rhea-Silvia and {Urbano Stawinski}, Stephanie M. and {Wilson}, Gillian and {Wisz}, M.~E.},
    title = {A massive, evolved slow-rotating galaxy in the early Universe},
    DOI= "10.48550/arXiv.2508.10987",
    url= "https://doi.org/10.48550/arXiv.2508.10987",
    journal = {Nat. Astr.},
    year = 2025
}

@article{EncyclopediaMagneticum,
    author = {Klaus Dolag and Rhea-Silvia Remus and Lucas M. Valenzuela and Lucas C. Kimmig and Benjamin Seidel and Silvio Fortuné and Johannes Stoiber and Anna Ivleva and Tadziu Hoffmann and Veronica Biffi and Ilaria Marini and Paola Popesso and Stephan Vladutescu-Zopp},
    title = {Encyclopedia Magneticum: Scaling Relations from Cosmic Dawn to Present Day},
    DOI= "10.48550/arXiv.2504.01061",
    url= "https://doi.org/10.48550/arXiv.2504.01061",
    journal = {A\&A},
    year = 2025
}

@article{Springel,
    doi = {10.1093/mnras/stx3304},
  
    url = {https://doi.org/10.1093%2Fmnras%2Fstx3304},
  
    year = 2017,
    month = {Dec},
  
    publisher = {Oxford University Press ({OUP})},
  
    volume = {475},
  
    number = {1},
  
    pages = {676},
  
    author = {Volker Springel and Rüdiger Pakmor and Annalisa Pillepich and Rainer Weinberger and Dylan Nelson and Lars Hernquist and Mark Vogelsberger and Shy Genel and Paul Torrey and Federico Marinacci and Jill Naiman},
  
    title = {First results from the {IllustrisTNG} simulations: matter and galaxy clustering},
  
    journal = {MNRAS}
}

@article{Wu,
    doi = {10.3847/1538-4357/ad7bb3},
    url = {https://doi.org/10.3847/1538-4357/ad7bb3},
    year = 2024,
    month = {Feb},
    pages = {37},
    author = {John F. Wu and Christian Kragh Jespersen and Risa H. Wechsler},
    number = {1},
    title = {How the Galaxy-Halo Connection Depends on Large-Scale Environment},
    volume = {976},
    journal = {ApJ}
}

@article{PFWu,
    doi = {10.3847/1538-4357/ad98ef},
    url = {https://doi.org/10.3847/1538-4357/ad98ef},
    year = 2025,
    month = {Jan},
    pages = {131},
    author = {Po-Feng Wu},
    number = {2},
    title = {Ejective feedback as a quenching mechanism in the 1.5 billion years of the universe: detection of neutral gas outflow in a z=4 recently quenched galaxy},
    volume = {978},
    journal = {ApJ}
}

@article{RemusKimmig,
    doi = {10.3847/1538-4357/ad8b4b},
    url = {https://iopscience.iop.org/article/10.3847/1538-4357/ad8b4b},
    year = 2025,
    month = {Mar},
    volume = {982},
    number = {1},
    pages = {30},
    author = {Rhea-Silvia Remus and Lucas C. Kimmig},
    title = {Relight the Candle: What happens to High Redshift Massive Quenched Galaxies},
    journal = {ApJ}
}

@article{Kakimoto,
    doi = {10.3847/1538-4357/ad1ff1},
    url = {https://doi.org/10.3847/1538-4357/ad1ff1},
    year = 2024,
    month = {Mar},
    pages = {49},
    author = {Takumi Kakimoto and Masayuki Tanaka and Masato Onodera and Rhythm Shimakawa and Po-Feng Wu and Katriona M. L. Gould and Kei Ito and Shuowen Jin and Mariko Kubo and Tomoko L. Suzuki and Sune Toft and Francesco Valentino and Kiyoto Yabe},
    number = {1},
    title = {A massive quiescent galaxy in a group environment at z=4.53},
    volume = {963},
    journal = {ApJ}
}

@article{Dekel,
    author = {Avishai Dekel and Kartick C Sarkar and Yuval Birnboim and Nir Mandelker and Zhaozhou Li},
    title = {Efficient formation of massive galaxies at cosmic dawn by feedback-free starbursts},
    journal = {MNRAS},
    volume = {523},
    number = {3},
    pages = {3201},
    year = {2023},
    month = {May},
    issn = {0035-8711},
    doi = {10.1093/mnras/stad1557},
    url = {https://doi.org/10.1093/mnras/stad1557}
}

@article{Harikane,
doi = {10.3847/1538-4365/acaaa9},
url = {https://doi.org/10.3847/1538-4365/acaaa9},
year = {2023},
month = {Feb},
publisher = {The American Astronomical Society},
volume = {265},
number = {1},
pages = {5},
author = {Harikane, Yuichi and Ouchi, Masami and Oguri, Masamune and Ono, Yoshiaki and Nakajima, Kimihiko and Isobe, Yuki and Umeda, Hiroya and Mawatari, Ken and Zhang, Yechi},
title = {A Comprehensive Study of Galaxies at z ∼ 9–16 Found in the Early JWST Data: Ultraviolet Luminosity Functions and Cosmic Star Formation History at the Pre-reionization Epoch},
journal = {ApJSS}
}

@article{Park,
    doi = {10.3847/1538-4357/ad7e15},
    url = {https://doi.org/10.3847/1538-4357/ad7e15},
    year = 2024,
    month = {Nov},
    pages = {72},
    author = {Minjung Park and Sirio Belli and Charlie Conroy and Benjamin D. Johnson and Rebecca L. Davies and Joel Leja and Sandro Tacchella and J. Trevor Mendel and Chloë Benton and Letizia Bugiani and Razieh Emami and Amirhossein Khoram and Yijia Li and Gabriel Maheson and Elijah P. Mathews and Rohan P. Naidu and Erica J. Nelson and Bryan A. Terrazas and Rainer Weinberger},
    number = {1},
    title = {Widespread rapid quenching at cosmic noon revealed by JWST deep spectroscopy},
    volume = {976},
    journal = {ApJ}
}

@ARTICLE{Pacucci2,
       author = {{Pacucci}, Fabio and {Volonteri}, Marta and {Ferrara}, Andrea},
      journal = {\mnras},
         year = 2015,
        month = {Sep},
       volume = {452},
       number = {2},
        pages = {1922},
          doi = {10.1093/mnras/stv1465}
}

@article{Chittenden,
  author =        {{Chittenden}, Harry George and {Tojeiro}, Rita},
  journal =       {MNRAS},
  title =         {{Modelling the galaxy-halo connection with semi-recurrent neural networks}},
  year =          {2023},
  month =         {Feb},
  volume =        {518},
  number =        {4},
  pages =         {5670},
  url =           {https://dx.doi.org/10.1093/mnras/stac3498},
  doi =           {10.1093/mnras/stac3498}
}

@article{Chittenden2,
  author =        {{Chittenden}, Harry George and {Behera}, Jayashree and {Tojeiro}, Rita},
  journal =       {MNRAS},
  year =          {2025},
  month =         {Jul},
  volume =        {541},
  number =        {2},
  pages =         {1682},
  title =         {{Evaluating the galaxy formation histories predicted by a neural network in pure dark matter simulations}},
  doi =           {10.1093/mnras/staf1086},
  url =           {https://dx.doi.org/10.1093/mnras/staf1086}
}

@article{Behera,
  author =        {{Behera}, Jayashree and {Tojeiro}, Rita and {Chittenden}, Harry George},
  journal =       {MNRAS},
  title =         {{Optimised neural network predictions of galaxy formation histories using semi-stochastic corrections}},
  volume =        {540},
  number =        {4},
  pages =         {3753},
  year =          {2025},
  month =         {Jun},
  doi =           {10.1093/mnras/staf920},
  url =           {https://dx.doi.org/10.1093/mnras/staf920}
}

@article{Lagos,
  author =        {Claudia Del P Lagos and Francesco Valentino and Ruby J. Wright and Anna {De Graaff} and Karl Glazebrook and Gabriella {De Lucia} and Aaron S. G. Robotham and Themiya Nanayakkara and Angel Chandro-Gomez and Matías Bravo and Carlton M. Baugh and Katherine E. Harborne and Michaela Hirschmann and Fabio Fontanot and Lizhi Xie and Harry Chittenden},
  journal =       {MNRAS},
  title =         {{The diverse star formation histories of early massive, quenched galaxies in modern galaxy formation simulations}},
  year =          {2025},
  volume =        {536},
  number =        {3},
  pages =         {2324},
  doi =           {10.1093/mnras/stae2626},
  url =           {https://doi.org/10.1093/mnras/stae2626}
}

@article{Zier,
       author = {Oliver Zier and Rahul Kannan and Aaron Smith and Ewald Puchwein and Mark Vogelsberger and Josh Borrow and Enrico Garaldi and Laura Keating and William McClymont and Xuejian Shen and Lars Hernquist},
      journal = {MNRAS},
         year = 2025,
        month = nov,
       volume = {544},
       number = {1},
        pages = {391},
          doi = {10.1093/mnras/staf1052},
archivePrefix = {arXiv},
       eprint = {2503.02927},
 primaryClass = {astro-ph.GA},
       adsurl = {https://ui.adsabs.harvard.edu/abs/2025MNRAS.544..391Z}
}

@article{GMM,
      title={Deep Gaussian Mixture Models}, 
      author={Cinzia Viroli and Geoffrey J. McLachlan},
      year={2017},
      journal={arXiv e-Print 1711.06929},
      url={https://arxiv.org/abs/1711.06929},
      doi={10.48550/arXiv.1711.06929}
}

@article{Russell,
      title={Cosmic Stillness: High Quiescent Galaxy Fractions Across Upper Mass Scales In The Early Universe to z = 7 With JWST},
      author={Tobias A. Russell and Neva Dobric and Nathan J. Adams and Christopher J. Conselice and Duncan Austin and Thomas Harvey and James Trussler and Leonardo Ferreira and Lewi Westcott and Honor Harris and Rogier A. Windhorst and Dan Coe and Seth H. Cohen and Simon P. Driver and Brenda Frye and Norman A. Grogin and Nimish P. Hathi and Rolf A. Jansen and Anton M. Koekemoer and Madeline A. Marshall and Rafael Ortiz and Nor Pirzkal and Aaron Robotham and Russell E. Ryan and Jake Summers and Jordan C. J. {D'Silva} and Christopher N. A. Willmer and Haojing Yan},
      year={2025},
      journal={MNRAS},
      volume = {544},
       number = {4},
        pages = {4482},
          doi = {10.1093/mnras/staf1916}
}

@article{DSilva,
      title={Self-Consistent JWST Census of Star Formation and AGN activity at z=5.5-13.5},
      author={Jordan C. J. {D'Silva} and Simon P. Driver and Claudia D. P. Lagos and Aaron S. G. Robotham and Nathan J. Adams and Christopher J. Conselice and Brenda Frye and Nimish P. Hathi and Thomas Harvey and Rafael Ortiz and Massimo Ricotti and Clayton Robertson and Ross M. Silver and Stephen M. Wilkins and Christopher N. A. Willmer and Rogier A. Windhorst and Seth H. Cohen and Rolf A. Jansen and Jake Summers and Anton M. Koekemoer and Dan Coe and Norman A. Grogin and Madeline A. Marshall and Mario Nonino and Nor Pirzkal and Russell E. Ryan and Haojing Yan},
      year={2025},
      journal={ApJ},
      volume = {990},
       number = {1},
          eid = {44},
        pages = {44},
          doi = {10.3847/1538-4357/adf19e}
}

@article{Spilker,
      title={Direct Evidence for AGN Feedback from Fast Molecular Outflows in Reionization-Era Quasars},
      author={Justin S. Spilker and Jaclyn B. Champagne and Xiaohui Fan and Seiji Fujimoto and Paul P. {Van Der Werf} and Jinyi Yang and Minghao Yue},
      year={2025},
      journal={ApJ},
      volume={982},
      number={2},
      pages={72},
      month={Mar},
      url={https://iopscience.iop.org/article/10.3847/1538-4357/adb750},
      doi={10.3847/1538-4357/adb750}
}

@article{Conaboy,
      title={The connection between high-redshift galaxies and Lyman α transmission in the Sherwood-Relics simulations of patchy reionisation},
      author={Luke Conaboy and James S. Bolton and Laura C. Keating and Martin G. Haehnelt and Girish Kulkarni and Ewald Puchwein},
      year={2025},
      volume = {539},
    number = {3},
    pages = {2790},
      journal={MNRAS},
      month={Apr},
      url={https://doi.org/10.1093/mnras/staf648},
      doi={10.1093/mnras/staf648}
}

@article{Donnan2025,
    doi = {10.1093/mnras/staf641},
    url = {https://doi.org/10.1093/mnras/staf641},
    year = 2025,
    month = {Apr},
    volume = {539},
    number = {3},
    pages = {2409},
    author = {C. T. Donnan and J. S. Dunlop and R. J. McLure and D. J. McLeod and F. Cullen},
    title = {No evidence (yet) for increased star-formation efficiency at early times},
    journal = {MNRAS}
}

@article{Shuntov,
    doi = {10.1051/0004-6361/202452570},
    url = {https://doi.org/10.1051/0004-6361/202452570},
    year = 2025,
    month = {Jan},
    author = {M. Shuntov and O. Ilbert and S. Toft and R. C. Arango-Toro and H. B. Akins and C. M. Casey and M. Franco and S. Harish and J. S. Kartaltepe and A. M. Koekemoer and H. J. McCracken and L. Paquereau and C. Laigle and M. Bethermin and Y. Dubois and N. E. Drakos and A. Faisst and G. Gozaliasl and S. Gillman and C. C. Hayward and M. Hirschmann and M. Huertas-Company and C. K. Jespersen and S. Jin and V. Kokorev and E. Lambrides and D. Le Borgne and D. Liu and G. Magdis and R. Massey and C. J. R. McPartland and W. Mercier and J. E. McCleary and J. McKinney and P. A. Oesch and J. D. Rhodes and R. M. Rich and B. E. Robertson and D. Sanders and M. Trebitsch and L. Tresse and F. Valentino and A. P. Vijayan and J. R. Weaver and A. Weibel and S. M. Wilkins},
    title = {COSMOS-Web: stellar mass assembly in relation to dark matter halos across 0.2<z<12 of cosmic history},
    volume={695},
    pages={A20},
    journal = {A\&A}
}

@article{Valentino,
   title={Quiescent Galaxies 1.5 Billion Years after the Big Bang and Their Progenitors},
   volume={889},
   ISSN={1538-4357},
   url={http://dx.doi.org/10.3847/1538-4357/ab64dc},
   DOI={10.3847/1538-4357/ab64dc},
   number={2},
   journal={ApJ},
   publisher={American Astronomical Society},
   author={Valentino, Francesco and Tanaka, Masayuki and Davidzon, Iary and Toft, Sune and Gómez-Guijarro, Carlos and Stockmann, Mikkel and Onodera, Masato and Brammer, Gabriel and Ceverino, Daniel and Faisst, Andreas L. and Gallazzi, Anna and Hayward, Christopher C. and Ilbert, Olivier and Kubo, Mariko and Magdis, Georgios E. and Selsing, Jonatan and Shimakawa, Rhythm and Sparre, Martin and Steinhardt, Charles and Yabe, Kiyoto and Zabl, Johannes},
   year={2020},
   month={Jan},
   pages={93}}

@article{Weinberger,
    doi = {10.1093/mnras/stw2944},
  
    url = {https://doi.org/10.1093%2Fmnras%2Fstw2944},
  
    year = 2016,
    month = {Nov},
  
    publisher = {Oxford University Press ({OUP})},
  
    volume = {465},
  
    number = {3},
  
    pages = {3291},
  
    author = {Rainer Weinberger and Volker Springel and Lars Hernquist and Annalisa Pillepich and Federico Marinacci and Rüdiger Pakmor and Dylan Nelson and Shy Genel and Mark Vogelsberger and Jill Naiman and Paul Torrey},
  
    title = {Simulating galaxy formation with black hole driven thermal and kinetic feedback},
  
    journal = {MNRAS}
}

@article{Nanayakkara,
   title={A population of faint, old, and massive quiescent galaxies at 3 < z < 4 revealed by JWST NIRSpec Spectroscopy},
   volume={14},
   pages={3724},
   url={http://dx.doi.org/10.1038/s41598-024-52585-4},
   DOI={10.1038/s41598-024-52585-4},
   number={1},
   journal={Scientific Reports},
   publisher={Springer Science and Business Media LLC},
   author={Nanayakkara, Themiya and Glazebrook, Karl and Jacobs, Colin and Kawinwanichakij, Lalitwadee and Schreiber, Corentin and Brammer, Gabriel and Esdaile, James and Kacprzak, Glenn G. and Labbé, Ivo and Lagos, Claudia and Marchesini, Danilo and Marsan, Z. Cemile and Oesch, Pascal A. and Papovich, Casey and Remus, Rhea-Silvia and Tran, Kim-Vy H.},
   year={2024},
   month={Feb}}

@article{Nanayakkara2,
   title={The formation histories of massive and quiescent galaxies in the 3 < z < 4.5 Universe},
   url={https://doi.org/10.3847/1538-4357/ada6ac},
   DOI={10.3847/1538-4357/ada6ac},
   journal={ApJ},
   author={Themiya Nanayakkara and Karl Glazebrook and Corentin Schreiber and Harry Chittenden and Gabriel Brammer and James Esdaile and Colin Jacobs and Glenn G. Kacprzak and Lalitwadee Kawinwanichakij and Lucas C. Kimmig and Ivo Labbe and Claudia Lagos and Danilo Marchesini and M. Martìnez-Marìn and Z. Cemile Marsan and Pascal A. Oesch and Casey Papovich and Rhea-Silvia Remus and Kim-Vy H. Tran},
   year={2025},
   volume={981},
   pages={78},
   number={1},
   month={Oct}}

@ARTICLE{Looser,
       author = {Tobias J. Looser and Francesco D'Eugenio and Roberto Maiolino and Joris Witstok and Lester Sandles and Emma Curtis-Lake and Jacopo Chevallard and Sandro Tacchella and Benjamin D. Johnson and William M. Baker and Katherine A. Suess and Stefano Carniani and Pierre Ferruit and Santiago Arribas and Nina Bonaventura and Andrew J. Bunker and Alex J. Cameron and Stephane Charlot and Mirko Curti and Anna de Graaff and Michael V. Maseda and Tim Rawle and Hans-Walter Rix and Bruno Rodriguez Del Pino and Renske Smit and Hannah Übler and Chris Willott and Stacey Alberts and Eiichi Egami and Daniel J. Eisenstein and Ryan Endsley and Ryan Hausen and Marcia Rieke and Brant Robertson and Irene Shivaei and Christina C. Williams and Kristan Boyett and Zuyi Chen and Zhiyuan Ji and Gareth C. Jones and Nimisha Kumari and Erica Nelson and Michele Perna and Aayush Saxena and Jan Scholtz},
        title = "{A recently quenched galaxy 700 million years after the Big Bang}",
      journal = {Nature},
         year = 2024,
        month = {May},
       volume = {629},
       number = {8010},
        pages = {53},
          doi = {10.1038/s41586-024-07227-0},
archivePrefix = {arXiv},
       eprint = {2302.14155},
 primaryClass = {astro-ph.GA},
       adsurl = {https://ui.adsabs.harvard.edu/abs/2024Natur.629...53L}
}

@article{Nipoti,
      title={Evolution of massive quiescent galaxies via envelope accretion},
      author={Carlo Nipoti},
      year={2025},
      journal={A\&A},
      volume = {697},
      pages={A74},
      doi={10.1051/0004-6361/202553930}
}

@article{Valentino2025,
      title={Gas outflows in two recently quenched galaxies at z = 4 and 7},
      author={F. Valentino and K. E. Heintz and G. Brammer and K. Ito and V. Kokorev and K. E. Whitaker and A. Gallazzi and A. {De Graaff} and A. Weibel and B. L. Frye and P. S. Kamieneski and S. Jin and D. Ceverino and A. Faisst and M. Farcy and S. Fujimoto and S. Gillman and R. Gottumukkala and M. Hamadouche and K. C. Harrington and M. Hirschmann and C. K. Jespersen and T. Kakimoto and M. Kubo and C. D. P. Lagos and M. Lee and G. E. Magdis and A. W. S. Man and M. Onodera and F. Rizzo and R. Shimakawa and D. J. Setton and M. Tanaka and S. Toft and P.-F. Wu and P. Zhu},
      year={2025},
      journal={A\&A},
      month = jul,
       volume = {699},
          eid = {A358},
        pages = {A358},
          doi = {10.1051/0004-6361/202553908}
}

@article{Ito,
      title={A merging pair of massive quiescent galaxies at z=3.44 in the Cosmic Vine},
      author={K. Ito and F. Valentino and M. Farcy and G. {De Lucia} and C. D. P. Lagos and M. Hirschmann and G. Brammer and A. {De Graaff} and D. Blánquez-Sesé and D. Ceverino and A. L. Faisst and F. Fontanot and S. Gillman and M. L. Hamadouche and K. E. Heintz and S. Jin and C. K. Jespersen and M. Kubo and M. Lee and G. Magdis and A. W. S. Man and M. Onodera and F. Rizzo and R. Shimakawa and M. Tanaka and S. Toft and K. E. Whitaker and L. Xie and P. Zhu},
      year={2025},
      journal={A\&A},
      volume={697},
      pages={A111},
      doi={10.1051/0004-6361/202453211}
}

@article{Weibel,
      title={RUBIES Reveals a Massive Quiescent Galaxy at z=7.3},
      author={Andrea Weibel and Anna de Graaff and David J. Setton and Tim B. Miller and Pascal A. Oesch and Gabriel Brammer and Claudia D. P. Lagos and Katherine E. Whitaker and Christina C. Williams and Josephine F.W. Baggen and Rachel Bezanson and Leindert A. Boogaard and Nikko J. Cleri and Jenny E. Greene and Michaela Hirschmann and Raphael E. Hviding and Adarsh Kuruvanthodi and Ivo Labbé and Joel Leja and Michael V. Maseda and Jorryt Matthee and Ian McConachie and Rohan P. Naidu and Guido Roberts-Borsani and Daniel Schaerer and Katherine A. Suess and Francesco Valentino and Pieter van Dokkum and Bingjie Wang},
      year={2025},
      journal={ApJ},
      volume={983},
      pages={11},
      number={1},
      doi={10.3847/1538-4357/adab7a}
}

@article{Valentino2023,
doi = {10.3847/1538-4357/acbefa},
url = {https://dx.doi.org/10.3847/1538-4357/acbefa},
year = {2023},
month = {Apr},
publisher = {The American Astronomical Society},
volume = {947},
number = {1},
pages = {20},
author = {Valentino, Francesco and Brammer, Gabriel and Gould, Katriona M. L. and Kokorev, Vasily and Fujimoto, Seiji and Jespersen, Christian Kragh and Vijayan, Aswin P. and Weaver, John R. and Ito, Kei and Tanaka, Masayuki and Ilbert, Olivier and Magdis, Georgios E. and Whitaker, Katherine E. and Faisst, Andreas L. and Gallazzi, Anna and Gillman, Steven and Giménez-Arteaga, Clara and Gómez-Guijarro, Carlos and Kubo, Mariko and Heintz, Kasper E. and Hirschmann, Michaela and Oesch, Pascal and Onodera, Masato and Rizzo, Francesca and Lee, Minju and Strait, Victoria and Toft, Sune},
title = {An Atlas of Color-selected Quiescent Galaxies at z > 3 in Public JWST Fields},
journal = {ApJ}
}

@article{Lewis,
   title={The short ionizing photon mean free path at z=6 in Cosmic Dawn III, a new fully coupled radiation-hydrodynamical simulation of the Epoch of Reionization},
   volume={516},
   ISSN={1365-2966},
   url={http://dx.doi.org/10.1093/mnras/stac2383},
   DOI={10.1093/mnras/stac2383},
   number={3},
   journal={MNRAS},
   publisher={Oxford University Press (OUP)},
   author={Lewis, Joseph S W and Ocvirk, Pierre and Sorce, Jenny G and Dubois, Yohan and Aubert, Dominique and Conaboy, Luke and Shapiro, Paul R and Dawoodbhoy, Taha and Teyssier, Romain and Yepes, Gustavo and Gottlöber, Stefan and Rasera, Yann and Ahn, Kyungjin and Iliev, Ilian T and Park, Hyunbae and Thélie, Émilie},
   year={2022},
   month={Aug}, pages={3389} }

@article{Shi,
  author =        {Shi, Jingjing and Wang, Huiyuan and Mo, Houjun and
                   Vogelsberger, Mark and Ho, Luis C. and Du, Min and
                   Nelson, Dylan and Pillepich, Annalisa and
                   Hernquist, Lars},
  journal =       {ApJ},
  month =         {Apr},
  number =        {2},
  pages =         {139},
  publisher =     {American Astronomical Society},
  title =         {The Formation History of Subhalos and the Evolution
                   of Satellite Galaxies},
  volume =        {893},
  year =          {2020},
  doi =           {10.3847/1538-4357/ab8464},
  issn =          {1538-4357},
  url =           {http://dx.doi.org/10.3847/1538-4357/ab8464},
}

@article{FlaresVIII,
    author = {Christopher C Lovell and Will Roper and Aswin P Vijayan and Louise Seeyave and Dimitrios Irodotou and Stephen M Wilkins and Christopher J Conselice and Flaminia Fortuni and Jussi K Kuusisto and Emiliano Merlin and Paola Santini and Peter Thomas},
    title = {First light and reionisation epoch simulations (FLARES) – VIII. The emergence of passive galaxies at z > 5},
    journal = {MNRAS},
    volume = {525},
    number = {4},
    pages = {5520},
    year = {2023},
    month = {Aug},
    issn = {0035-8711},
    doi = {10.1093/mnras/stad2550},
    url = {https://doi.org/10.1093/mnras/stad2550},
    eprint = {https://academic.oup.com/mnras/article-pdf/525/4/5520/51606230/stad2550.pdf}
}

@article{Kurinchi-Vendhan,
   title={On the origin of star formation quenching in massive galaxies at z>3 in the cosmological simulations IllustrisTNG},
   volume={534},
   ISSN={1365-2966},
   url={http://dx.doi.org/10.1093/mnras/stae2297},
   DOI={10.1093/mnras/stae2297},
   number={4},
   journal={MNRAS},
   publisher={Oxford University Press (OUP)},
   author={Kurinchi-Vendhan, Shalini and Farcy, Marion and Hirschmann, Michaela and Valentino, Francesco},
   year={2024},
   month={Oct},
   pages={3974}
}

@article{Thesan,
    doi = {10.1093/mnras/stab3710},
  
    url = {https://doi.org/10.1093%2Fmnras%2Fstab3710},
  
    year = 2021,
    month = {Dec},
  
    publisher = {Oxford University Press ({OUP})},
  
    volume = {511},
  
    number = {3},
  
    pages = {4005},
  
    author = {Rahul Kannan and E Garaldi and A Smith and R Pakmor and V Springel and M Vogelsberger and L Hernquist},
  
    title = {Introducing the THESAN project: radiation-magnetohydrodynamic simulations of the epoch of reionization},
  
    journal = {MNRAS}
}

@article{Xu,
    doi = {10.1093/mnras/stad789},
  
    url = {https://doi.org/10.1093%2Fmnras%2Fstad789},
  
    year = 2023,
    month = {Mar},
  
    publisher = {Oxford University Press ({OUP})},
  
    volume = {521},
  
    number = {3},
  
    pages = {4356},
  
    author = {Clara Xu and Aaron Smith and Josh Borrow and Enrico Garaldi and Rahul Kannan and Mark Vogelsberger and Rüdiger Pakmor and Volker Springel and Lars Hernquist},
  
    title = {The THESAN project: Lyman-alpha emitter luminosity function calibration},
  
    journal = {MNRAS}
}

@article{Yeh,
    doi = {10.1093/mnras/stad210},
  
    url = {https://doi.org/10.1093%2Fmnras%2Fstad210},
  
    year = 2023,
    month = {Jan},
  
    publisher = {Oxford University Press ({OUP})},
  
    volume = {520},
  
    number = {2},
  
    pages = {2757},
  
    author = {Jessica Y.-C. Yeh and Aaron Smith and Rahul Kannan and Enrico Garaldi and Mark Vogelsberger and Josh Borrow and Rüdiger Pakmor and Volker Springel and Lars Hernquist},
  
    title = {The THESAN project: ionizing escape fractions of reionization-era galaxies},
  
    journal = {MNRAS}
}

@article{Alberts,
  author =        {Stacey Alberts and Christina C. Williams and Jakob M. Helton and Katherine A. Suess and Zhiyuan Ji and Irene Shivaei and Jianwei Lyu and George Rieke and William M. Baker and Nina Bonaventura and Andrew J. Bunker and Stefano Carniani and Stephane Charlot and Emma Curtis-Lake and Francesco D'Eugenio and Daniel J. Eisenstein and Anna {De Graaff} and Kevin N. Hainline and Ryan Hausen and Benjamin D. Johnson and Roberto Maiolino and Eleonora Parlanti and Marcia J. Rieke and Brant E. Robertson and Yang Sun and Sandro Tacchella and Christopher N. A. Willmer and Chris J. Willott},
  journal =       {ApJ},
  month =         Oct,
  volume = {975},
  number = {1},
  pages = {85},
  title =         {To high redshift and low mass: exploring the emergence of quenched galaxies and their environments at 3<z<6 in the ultra-deep JADES MIRI F770W parallel},
  year =          {2024},
  doi =           {10.3847/1538-4357/ad66cc},
}

@article{Smith,
    doi = {10.1093/mnras/stac713},
  
    url = {https://doi.org/10.1093%2Fmnras%2Fstac713},
  
    year = 2022,
    month = {Mar},
  
    publisher = {Oxford University Press ({OUP})},
  
    volume = {512},
  
    number = {3},
  
    pages = {3243},
  
    author = {A Smith and R Kannan and E Garaldi and M Vogelsberger and R Pakmor and V Springel and L Hernquist},
  
    title = {The THESAN project: Lyman-alpha emission and transmission during the Epoch of Reionization},
  
    journal = {MNRAS}
}

@article{XShen,
      title={THESAN-HR: Galaxies in the Epoch of Reionization in warm dark matter, fuzzy dark matter and interacting dark matter}, 
      author={Xuejian Shen and Josh Borrow and Mark Vogelsberger and Enrico Garaldi and Aaron Smith and Rahul Kannan and Sandro Tacchella and Jesús Zavala and Lars Hernquist and Jessica Y.-C. Yeh and Chunyuan Zheng},
      doi={10.1093/mnras/stad3397},
      year={2024},
      volume={527},
      number={2},
      pages={2835},
      month={Jan},
      journal={MNRAS}
}

@article{XShen2,
      title={The THESAN project: galaxy sizes during the epoch of reionization},
      author={Xuejian Shen and Mark Vogelsberger and Josh Borrow and Yongao Hu and Evan Erickson and Rahul Kannan and Aaron Smith and Enrico Garaldi and Lars Hernquist and Takahiro Morishita and Sandro Tacchella and Oliver Zier and Guochao Sun and Anna-Christina Eilers and Hui Wang},
      doi={10.1093/mnras/stae2156},
      year={2024},
      volume={534},
      number={2},
      pages={1433},
      month={Oct},
      journal={MNRAS}
}

@article{Neyer,
   title={The THESAN project: connecting ionized bubble sizes to their local environments during the Epoch of Reionization},
   volume={531},
   ISSN={1365-2966},
   url={http://dx.doi.org/10.1093/mnras/stae1325},
   DOI={10.1093/mnras/stae1325},
   number={3},
   journal={MNRAS},
   publisher={Oxford University Press (OUP)},
   author={Neyer, Meredith and Smith, Aaron and Kannan, Rahul and Vogelsberger, Mark and Garaldi, Enrico and Galárraga-Espinosa, Daniela and Borrow, Josh and Hernquist, Lars and Pakmor, Rüdiger and Springel, Volker},
   year={2024},
   month={May},
   pages={2943}
}

@article{Garaldi,
    doi = {10.1093/mnras/stac257},
  
    url = {https://doi.org/10.1093%2Fmnras%2Fstac257},
  
    year = 2022,
    month = {Feb},
  
    publisher = {Oxford University Press ({OUP})},
  
    volume = {512},
  
    number = {4},
  
    pages = {4909},
  
    author = {E Garaldi and R Kannan and A Smith and V Springel and R Pakmor and M Vogelsberger and L Hernquist},
  
    title = {The THESAN project: properties of the intergalactic medium and its connection to reionization-era galaxies},
  
    journal = {MNRAS}
}

@article{Garaldi2,
    author = {Garaldi, Enrico and Kannan, Rahul and Smith, Aaron and Borrow, Josh and Vogelsberger, Mark and Pakmor, Rüdiger and Springel, Volker and Hernquist, Lars and Galárraga-Espinosa, Daniela and Yeh, Jessica Y.-C. and Shen, Xuejian and Xu, Clara and Neyer, Meredith and Spina, Benedetta and Almualla, Mouza and Zhao, Yu},
    title = {The thesan project: public data release of radiation-hydrodynamic simulations matching reionization-era JWST observations},
    journal = {MNRAS},
    volume = {530},
    number = {4},
    pages = {3765},
    year = {2024},
    month = {Mar},
    issn = {0035-8711},
    doi = {10.1093/mnras/stae839},
    url = {https://doi.org/10.1093/mnras/stae839},
    eprint = {https://academic.oup.com/mnras/article-pdf/530/4/3765/57420532/stae839.pdf},
}

@ARTICLE{Sanati,
       author = {{Sanati}, Mahsa and {Devriendt}, Julien and {SergioMartin-Alvarez} and {Slyz}, Adrianne and {Tan}, Jonathan C.},
      journal = {MNRAS},
         year = 2025,
        month = {Dec},
          volume = {544},
       number = {4},
        pages = {4317-4335},
          doi = {10.1093/mnras/staf2000}
}

@article{Bassini,
    author = {Bassini, Luigi and Feldmann, Robert and Gensior, Jindra and Hayward, Christopher C and Faucher-Giguère, Claude-André and Cenci, Elia and Liang, Lichen and Bernardini, Mauro},
    title = {The inefficiency of stellar feedback in driving galactic outflows in massive galaxies at high redshift},
    journal = {MNRAS},
    volume = {525},
    number = {4},
    pages = {5388-5405},
    year = {2023},
    month = {Sep},
    issn = {0035-8711},
    doi = {10.1093/mnras/stad2617},
    url = {https://doi.org/10.1093/mnras/stad2617}
}

@ARTICLE{YShi,
       author = {{Shi}, Yanlong and {Kremer}, Kyle and {Hopkins}, Philip F.},
      journal = {A\&A},
         year = 2024,
        month = {Nov},
       volume = {691},
          eid = {A24},
        pages = {A24},
          doi = {10.1051/0004-6361/202450964},
archivePrefix = {arXiv},
       eprint = {2405.12164},
 primaryClass = {astro-ph.GA},
       adsurl = {https://ui.adsabs.harvard.edu/abs/2024A&A...691A..24S}
}

@ARTICLE{WWang,
       author = {{Wang}, Wuji and {Wylezalek}, Dominika and {De Breuck}, Carlos and {Vernet}, Joel and {Rupke}, David S.~N. and {Zakamska}, Nadia L. and {Vayner}, Andrey and {Lehnert}, Matthew D. and {Nesvadba}, Nicole P.~H. and {Stern}, Daniel},
      journal = {A\&A},
         year = 2024,
        month = {Mar},
       volume = {683},
          eid = {A169},
        pages = {A169},
          doi = {10.1051/0004-6361/202348531},
archivePrefix = {arXiv},
       eprint = {2401.02479},
 primaryClass = {astro-ph.GA},
       adsurl = {https://ui.adsabs.harvard.edu/abs/2024A&A...683A.169W}
}

@article{NanayakkaraPASA,
       author = {{Nanayakkara}, Themiya and {Esdaile}, James and {Glazebrook}, Karl and {Espejo Salcedo}, Juan M. and {Durre}, Mark and {Jacobs}, Colin},
        title = {{Massive High-Redshift Quiescent Galaxies With JWST}},
      journal = {PASA},
         year = 2022,
        month = {Jan},
       volume = {39},
        pages = {e002},
          doi = {10.1017/pasa.2021.61}
}

@article{Marshall,
    author = {Marshall, Madeline A and Wyithe, J Stuart B and Windhorst, Rogier A and Matteo, Tiziana Di and Ni, Yueying and Wilkins, Stephen and Croft, Rupert A C and Mechtley, Mira},
    title = {Observing the host galaxies of high-redshift quasars with JWST: predictions from the BlueTides simulation},
    journal = {MNRAS},
    volume = {506},
    number = {1},
    pages = {1209},
    year = {2021},
    month = {Jun},
    issn = {0035-8711},
    doi = {10.1093/mnras/stab1763}
}

@article{Umehata,
    author = {Hideki Umehata and Mariko Kubo and Kouichiro Nakanishi},
    title = {ADF22-WEB: Detection of a molecular gas reservoir in a massive quiescent galaxy located in a $z\approx3$ proto-cluster core},
    journal = {ApJ},
    year = {2025},
    volume = {985},
       number = {1},
          eid = {L8},
        pages = {L8},
          doi = {10.3847/2041-8213/add1d4}
}

@article{Bera,
   title={Studying cosmic dawn using redshifted HI 21-cm signal: A brief review},
   volume={44},
   pages={10},
   url={http://dx.doi.org/10.1007/s12036-022-09904-w},
   DOI={10.1007/s12036-022-09904-w},
   number={1},
   journal={Journal of Astrophysics and Astronomy},
   publisher={Springer Science and Business Media LLC},
   author={Bera, Ankita and Ghara, Raghunath and Chatterjee, Atrideb and Datta, Kanan K. and Samui, Saumyadip},
   year={2023},
   month={Feb}
}

@article{Williams,
   title={ALMA Measures Rapidly Depleted Molecular Gas Reservoirs in Massive Quiescent Galaxies at z ∼ 1.5},
   volume={908},
   ISSN={1538-4357},
   url={http://dx.doi.org/10.3847/1538-4357/abcbf6},
   DOI={10.3847/1538-4357/abcbf6},
   number={1},
   journal={ApJ},
   publisher={American Astronomical Society},
   author={Williams, Christina C. and Spilker, Justin S. and Whitaker, Katherine E. and Davé, Romeel and Woodrum, Charity and Brammer, Gabriel and Bezanson, Rachel and Narayanan, Desika and Weiner, Benjamin},
   year={2021},
   month={Feb},
   pages={54}
}

@article{Galform,
    doi = {10.1093/mnras/stw1888},
  
    url = {https://doi.org/10.1093%2Fmnras%2Fstw1888},
  
    year = 2016,
    month = {Aug},
  
    publisher = {Oxford University Press ({OUP})},
  
    volume = {462},
  
    number = {4},
  
    pages = {3854},
  
    author = {Cedric G. Lacey and Carlton M. Baugh and Carlos S. Frenk and Andrew J. Benson and Richard G. Bower and Shaun Cole and Violeta Gonzalez-Perez and John C. Helly and Claudia D. P. Lagos and Peter D. Mitchell},
  
    title = {A unified multiwavelength model of galaxy formation},
  
    journal = {MNRAS}
}

@article{McGibbonKhochfar,
  author =        {Robert J. McGibbon and Sadegh Khochfar},
  journal =       {MNRAS},
  month =         {May},
  number =        {4},
  pages =         {5423},
  publisher =     {Oxford University Press (OUP)},
  title =         {{Multi-Epoch Machine Learning 1: Unravelling Nature Versus Nurture For Galaxy Formation}},
  volume =        {513},
  year =          {2022},
  doi =           {10.1093/mnras/stac1269},
  url =           {https://doi.org/10.1093/mnras/stac1269},
}

@article{McGibbonKhochfar2,
   title={Multi-epoch machine learning 2: identifying physical drivers of galaxy properties in simulations},
   volume={523},
   ISSN={1365-2966},
   url={http://dx.doi.org/10.1093/mnras/stad1811},
   DOI={10.1093/mnras/stad1811},
   number={4},
   journal={MNRAS},
   publisher={Oxford University Press (OUP)},
   author={Robert J. McGibbon and Sadegh Khochfar},
   year={2023},
   month={Jun},
   pages={5583}}

@article{Kimmig,
    doi = {10.3847/1538-4357/ad9472},
    volume = {979},
    url = {https://doi.org/10.3847/1538-4357/ad9472},
    number = {1},
    year = 2025,
    month = {Jan},
    pages = {15},
    publisher = {American Astronomical Society},
    author = {Lucas C. Kimmig and Rhea-Silvia Remus and Benjamin Seidel and Lucas M. Valenzuela and Klaus Dolag and Andreas Burkert},
    title = {Blowing out the Candle: How to Quench Galaxies at High Redshift -- an Ensemble of Rapid Starbursts, AGN Feedback and Environment},
    journal = {ApJ}
}

@ARTICLE{Lustig,
       author = {{Lustig}, Peter and {Strazzullo}, Veronica and {Remus}, Rhea-Silvia and {D'Eugenio}, Chiara and {Daddi}, Emanuele and {Burkert}, Andreas and {De Lucia}, Gabriella and {Delvecchio}, Ivan and {Dolag}, Klaus and {Fontanot}, Fabio and {Gobat}, Raphael and {Mohr}, Joseph J. and {Onodera}, Masato and {Pannella}, Maurilio and {Pillepich}, Annalisa},
        title = "{Massive quiescent galaxies at z=3: A comparison of selection, stellar population, and structural properties with simulation predictions}",
      journal = {MNRAS},
         year = 2023,
        month = {Feb},
       volume = {518},
       number = {4},
        pages = {5953},
          doi = {10.1093/mnras/stac3450},
archivePrefix = {arXiv},
       eprint = {2201.09068},
 primaryClass = {astro-ph.GA},
       adsurl = {https://ui.adsabs.harvard.edu/abs/2023MNRAS.518.5953L}
}

@ARTICLE{Steinborn,
       author = {{Steinborn}, Lisa K. and {Dolag}, Klaus and {Hirschmann}, Michaela and {Prieto}, M. Almudena and {Remus}, Rhea-Silvia},
        title = "{A refined sub-grid model for black hole accretion and AGN feedback in large cosmological simulations}",
      journal = {MNRAS},
         year = 2015,
        month = {Apr},
       volume = {448},
       number = {2},
        pages = {1504},
          doi = {10.1093/mnras/stv072},
archivePrefix = {arXiv},
       eprint = {1409.3221},
 primaryClass = {astro-ph.GA},
       adsurl = {https://ui.adsabs.harvard.edu/abs/2015MNRAS.448.1504S}
}

@ARTICLE{Bondi,
       author = {Hermann Bondi},
        title = "{On spherically symmetrical accretion}",
      journal = {\mnras},
         year = 1952,
        month = {Jan},
       volume = {112},
        pages = {195},
          doi = {10.1093/mnras/112.2.195},
       adsurl = {https://ui.adsabs.harvard.edu/abs/1952MNRAS.112..195B}
}

@article{Kawinwanichakij,
      title={Stellar Mass-Size Relation and Morphology of Massive Quiescent Galaxies at $3 < z < 4$ with JWST},
      author={Lalitwadee Kawinwanichakij and Karl Glazebrook and Themiya Nanayakkara and Glenn G. Kacprzak and Harry George Chittenden and Colin Jacobs and Ángel Chandro-Gómez and Claudia Lagos and Danilo Marchesini and Monserrat Martínez-Marín and Pascal A. Oesch and Rhea-Silvia Remus},
      year = 2026,
      journal={ApJ},
        month = jan,
       volume = {997},
       number = {1},
          eid = {29},
        pages = {29},
          doi = {10.3847/1538-4357/ae0a18}
}

@article{Labbe,
    doi = {10.3847/1538-4357/ad3551},
    url = {https://doi.org/10.3847/1538-4357/ad3551},
    year = 2025,
    month = {Jan},
    pages = {92},
    author = {Ivo Labbé and Jenny E. Greene and Rachel Bezanson and Seiji Fujimoto and Lukas J. Furtak and Andy D. Goulding and Jorryt Matthee and Rohan P. Naidu and Pascal A. Oesch and Hakim Atek and Gabriel Brammer and Iryna Chemerynska and Dan Coe and Sam E. Cutler and Pratika Dayal and Robert Feldmann and Marijn Franx and Karl Glazebrook and Joel Leja and Danilo Marchesini and Michael Maseda and Themiya Nanayakkara and Erica J. Nelson and Richard Pan and Casey Papovich and Sedona H. Price and Katherine A. Suess and Bingjie Wang and Katherine E. Whitaker and Christina C. Williams and Adi Zitrin},
    number = {1},
    title = {UNCOVER: Candidate Red Active Galactic Nuclei at 3<z<7 with JWST and ALMA},
    volume = {978},
    journal = {ApJ}
}

@article{Dubois,
   title={Black hole evolution – I. Supernova-regulated black hole growth},
   volume={452},
   ISSN={1365-2966},
   url={http://dx.doi.org/10.1093/mnras/stv1416},
   DOI={10.1093/mnras/stv1416},
   number={2},
   journal={\mnras},
   publisher={Oxford University Press (OUP)},
   author={Dubois, Yohan and Volonteri, Marta and Silk, Joseph and Devriendt, Julien and Slyz, Adrianne and Teyssier, Romain},
   year={2015},
   month={Jul},
   pages={1502}
}

@ARTICLE{deGraaff,
       author = {Anna {De Graaff} and David J. Setton and Gabriel Brammer and Sam Cutler and Katherine A. Suess and Ivo Labbe and Joel Leja and Andrea Weibel and Michael V. Maseda and Katherine E. Whitaker and Rachel Bezanson and Leindert A. Boogaard and Nikko J. Cleri and Gabriella De Lucia and Marijn Franx and Jenny E. Greene and Michaela Hirschmann and Jorryt Matthee and Ian McConachie and Rohan P. Naidu and Pascal A. Oesch and Sedona H. Price and Hans-Walter Rix and Francesco Valentino and Bingjie Wang and Christina C. Williams},
        title = "{Efficient formation of a massive quiescent galaxy at redshift 4.9}",
      journal = {Nat. Astr.},
         year = 2024,
        month = apr,
      volume = {9},
        pages = {280},
          doi = {10.1038/s41550-024-02424-3}
}

@ARTICLE{deGraaff2,
       author = {{De Graaff}, Anna and {Rix}, Hans-Walter and {Carniani}, Stefano and {Suess}, Katherine A. and {Charlot}, St{\'e}phane and {Curtis-Lake}, Emma and {Arribas}, Santiago and {Baker}, William M. and {Boyett}, Kristan and {Bunker}, Andrew J. and {Cameron}, Alex J. and {Chevallard}, Jacopo and {Curti}, Mirko and {Eisenstein}, Daniel J. and {Franx}, Marijn and {Hainline}, Kevin and {Hausen}, Ryan and {Ji}, Zhiyuan and {Johnson}, Benjamin D. and {Jones}, Gareth C. and {Maiolino}, Roberto and {Maseda}, Michael V. and {Nelson}, Erica and {Parlanti}, Eleonora and {Rawle}, Tim and {Robertson}, Brant and {Tacchella}, Sandro and {{\"U}bler}, Hannah and {Williams}, Christina C. and {Willmer}, Christopher N.~A. and {Willott}, Chris},
        title = "{Ionised gas kinematics and dynamical masses of z {\ensuremath{\gtrsim}} 6 galaxies from JADES/NIRSpec high-resolution spectroscopy}",
      journal = {\aap},
         year = 2024,
        month = apr,
       volume = {684},
          eid = {A87},
        pages = {A87},
          doi = {10.1051/0004-6361/202347755},
archivePrefix = {arXiv},
       eprint = {2308.09742},
 primaryClass = {astro-ph.GA},
       adsurl = {https://ui.adsabs.harvard.edu/abs/2024A&A...684A..87D}
}

@article{ThesanAddendum,
    author = {Garaldi, Enrico and Kannan, Rahul and Smith, Aaron and Borrow, Josh and Vogelsberger, Mark and Pakmor, Rüdiger and Springel, Volker and Hernquist, Lars and Galárraga-Espinosa, Daniela and Yeh, Jessica Y-C and Shen, Xuejian and Xu, Clara and Neyer, Meredith and Spina, Benedetta and Almualla, Mouza and Zhao, Yu},
    journal = {MNRAS},
    volume = {545},
    number = {4},
    pages = {staf2196},
    year = {2025},
    month = {Dec},
    doi = {10.1093/mnras/staf2196}
}

@article{Sphinx,
  author =        {Joakim Rosdahl and Harley Katz and J{\'{e}}r{\'{e}}my Blaizot and Taysun Kimm and L{\'{e}}o Michel-Dansac and Thibault Garel and Martin Haehnelt and Pierre Ocvirk and Romain Teyssier},
  journal =       {MNRAS},
  month =         {Jun},
  number =        {1},
  pages =         {994},
  publisher =     {Oxford University Press (OUP)},
  title =         {{The SPHINX Cosmological Simulations Of The First Billion Years: The Impact Of Binary Stars On Reionisation}},
  volume =        {479},
  year =          {2018},
  doi =           {10.1093/mnras/sty1655},
  url =           {http://dx.doi.org/10.1093/mnras/sty1655},
}

@article{Chadayammuri,
   title={Painting baryons on to N-body simulations of galaxy clusters with image-to-image deep learning},
   volume={526},
   ISSN={1365-2966},
   url={http://dx.doi.org/10.1093/mnras/stad2596},
   DOI={10.1093/mnras/stad2596},
   number={2},
   journal={MNRAS},
   publisher={Oxford University Press (OUP)},
   author={Urmila Chadayammuri and Michelle Ntampaka and John ZuHone and {\`A}kos Bogdàn and Ralph Kraft},
   year={2023},
   month={Sep},
   pages={2812}}

@article{Cuevas,
   title={Not hydro: using neural networks to estimate galaxy properties on a dark-matter-only simulation},
   volume={524},
   ISSN={1365-2966},
   url={http://dx.doi.org/10.1093/mnras/stad2112},
   DOI={10.1093/mnras/stad2112},
   number={3},
   journal={MNRAS},
   publisher={Oxford University Press (OUP)},
   author={{Hernández Cuevas}, Cristian A and González, Roberto E and Padilla, Nelson D},
   year={2023},
   month={Jul},
   pages={4653}}

@ARTICLE{Chworowsky,
       author = {{Chworowsky}, Katherine and {Finkelstein}, Steven L. and {Spilker}, Justin S. and {Leung}, Gene C.~K. and {Bagley}, Micaela B. and {Casey}, Caitlin M. and {Gronwall}, Caryl and {Jogee}, Shardha and {Larson}, Rebecca L. and {Papovich}, Casey and {Somerville}, Rachel S. and {Stevans}, Matthew and {Wold}, Isak G.~B. and {Yung}, L.~Y. Aaron},
        title = "{ALMA 1.1 mm Observations of a Conservative Sample of High-redshift Massive Quiescent Galaxies in SHELA}",
      journal = {\apj},
         year = 2023,
        month = {Jul},
       volume = {951},
       number = {1},
          eid = {49},
        pages = {49},
          doi = {10.3847/1538-4357/acd1e3},
archivePrefix = {arXiv},
       eprint = {2305.06309},
 primaryClass = {astro-ph.GA},
       adsurl = {https://ui.adsabs.harvard.edu/abs/2023ApJ...951...49C}
}

@article{Ortiz,
   title={The VANDELS Survey: Star formation and quenching in two over-densities at 3<z<4},
   url={https://doi.org/10.1051/0004-6361/202449535},
   DOI={10.1051/0004-6361/202449535},
   journal={A\&A},
   publisher={EDP Sciences},
   author={M. {Espinoza Ortiz} and L. Guaita and R. Demarco and A. Calabró and L. Pentericci and M. Castellano and M. {Celeste Artale} and N. P. Hathi and Anton M. Koekemoer and F. Mannucci and P. Hibon and D. J. McLeod and A. Gargiulo and E. Pompei},
   year={2024},
   pages = {A42},
   volume = {692},
   month={Oct}
}

@article{Jin,
   title={Cosmic Vine: A z = 3.44 large-scale structure hosting massive quiescent galaxies},
   volume={683},
   ISSN={1432-0746},
   url={http://dx.doi.org/10.1051/0004-6361/202348540},
   DOI={10.1051/0004-6361/202348540},
   journal={A\&A},
   publisher={EDP Sciences},
   author={Jin, Shuowen and Sillassen, Nikolaj B. and Magdis, Georgios E. and Brinch, Malte and Shuntov, Marko and Brammer, Gabriel and Gobat, Raphael and Valentino, Francesco and Carnall, Adam C. and Lee, Minju and Vijayan, Aswin P. and Gillman, Steven and Kokorev, Vasily and Le Bail, Aurélien and Greve, Thomas R. and Gullberg, Bitten and Gould, Katriona M. L. and Toft, Sune},
   year={2024},
   month={Feb},
   pages={L4}
}

@article{SKLearn,
  author =        {Fabian Pedregosa and Ga{{\"e}}l Varoquaux and
                   Alexandre Gramfort and Vincent Michel and
                   Bertrand Thirion and Olivier Grisel and
                   Mathieu Blondel and Peter Prettenhofer and Ron Weiss and
                   Vincent Dubourg and Jake Vanderplas and
                   Alexandre Passos and David Cournapeau and
                   Matthieu Brucher and Matthieu Perrot and
                   {{\'E}}douard Duchesnay},
  journal =       {Journal of Machine Learning Research},
  number =        {85},
  pages =         {2825},
  title =         {Scikit-learn: Machine Learning in Python},
  volume =        {12},
  year =          {2011},
  doi =           {10.48550/ARXIV.1201.0490},
}

@article{Mistral,
    author = {Marion Farcy and Michaela Hirschmann and Rachel S. Somerville and Ena Choi and Sophie Koudmani and Thorsten Naab and Rainer Weinberger and Jake S. Bennett and Aklant K. Bhowmick and Hyunseop Choi and Lars Hernquist and Julie Hlavacek-Larrondo and Bryan A. Terrazas and Francesco Valentino},
    title = {MISTRAL: a model for radiatively efficient AGN winds in cosmological simulations},
    journal = {MNRAS},
    year = {2025},
    volume = {543},
       number = {2},
        pages = {967},
          doi = {10.1093/mnras/staf1464}
}

@article{Uchuu,
    doi = {10.1093/mnras/stab1755},
  
    url = {https://doi.org/10.1093%2Fmnras%2Fstab1755},
  
    year = 2021,
    month = {Jun},
  
    publisher = {Oxford University Press ({OUP})},
  
    volume = {506},
  
    number = {3},
  
    pages = {4210},
  
    author = {Tomoaki Ishiyama and Francisco Prada and Anatoly A Klypin and Manodeep Sinha and R Benton Metcalf and Eric Jullo and Bruno Altieri and Sof{\'{\i}}a A Cora and Darren Croton and Sylvain de~la~Torre and David E Mill{\'{a}}n-Calero and Taira Oogi and Jos{\'{e}} Ruedas and Cristian A Vega-Mart{\'{\i}}nez},
  
    title = {{The Uchuu Simulations: Data Release 1 and Dark Matter Halo Concentrations}},
  
    journal = {MNRAS}
}

@article{Turner,
    author = {Crispin Turner and Sandro Tacchella and Francesco D’Eugenio and Stefano Carniani and Mirko Curti and Karl Glazebrook and Benjamin D. Johnson and Seunghwan Lim and Tobias Looser and Roberto Maiolino and Themiya Nanayakkara and Jenny Wan},
    title = {Age-dating early quiescent galaxies: high star formation efficiency, but consistent with direct, higher-redshift observations},
    journal = {MNRAS},
    volume = {537},
    number = {2},
    pages = {1826},
    year = {2025},
    month = {Jan},
    doi = {10.1093/mnras/staf128}
}

@article{ArepoRT,
   title={AREPO-RT: Radiation hydrodynamics on a moving mesh},
   volume={485},
   ISSN={1365-2966},
   url={http://dx.doi.org/10.1093/mnras/stz287},
   DOI={10.1093/mnras/stz287},
   number={1},
   journal={MNRAS},
   publisher={Oxford University Press (OUP)},
   author={Rahul Kannan and Mark Vogelsberger and Federico Marinacci and Ryan McKinnon and Rüdiger Pakmor and Volker Springel},
   year={2019},
   month={Jan},
   pages={117}}

@ARTICLE{DelPino,
       author = {{Rodr{\'\i}guez Del Pino}, B. and {Perna}, M. and {Arribas}, S. and {D'Eugenio}, F. and {Lamperti}, I. and {P{\'e}rez-Gonz{\'a}lez}, P.~G. and {{\"U}bler}, H. and {Bunker}, A. and {Carniani}, S. and {Charlot}, S. and {Maiolino}, R. and {Willott}, C.~J. and {B{\"o}ker}, T. and {Chevallard}, J. and {Cresci}, G. and {Curti}, M. and {Jones}, G.~C. and {Parlanti}, E. and {Scholtz}, J. and {Venturi}, G.},
        title = "{GA-NIFS: Co-evolution within a highly star-forming galaxy group at z {\ensuremath{\sim}} 3.7 witnessed by JWST/NIRSpec IFS}",
      journal = {\aap},
         year = 2024,
        month = {Apr},
       volume = {684},
          eid = {A187},
        pages = {A187},
          doi = {10.1051/0004-6361/202348057},
archivePrefix = {arXiv},
       eprint = {2309.14431},
 primaryClass = {astro-ph.GA},
       adsurl = {https://ui.adsabs.harvard.edu/abs/2024A&A...684A.187R}
}

@article{Carnallmnras,
   title={A surprising abundance of massive quiescent galaxies at 3 < z < 5 in the first data from JWST CEERS},
   volume={520},
   ISSN={1365-2966},
   url={http://dx.doi.org/10.1093/mnras/stad369},
   DOI={10.1093/mnras/stad369},
   number={3},
   journal={MNRAS},
   publisher={Oxford University Press (OUP)},
   author={A. C. Carnall and D. J. McLeod and R. J. McLure and J. S. Dunlop and R. Begley and F. Cullen and C. T. Donnan and M. L. Hamadouche and S. M. Jewell and E. W. Jones and C. L. Pollock and V. Wild},
   year={2023},
   month={Feb},
   pages={3974}}

@article{Carnallmnras2,
   title={The JWST EXCELS survey: Too much, too young, too fast? Ultra-massive quiescent galaxies at 3 < z < 5},
   url={https://doi.org/10.1093/mnras/stae2092},
   DOI={10.1093/mnras/stae2092},
   journal={MNRAS},
   volume={534},
   number={1},
   publisher={Oxford University Press (OUP)},
   author={A. C. Carnall and F. Cullen and R. J. McLure and D. J. McLeod and R. Begley and C. T. Donnan and J. S. Dunlop and A. E. Shapley and K. Rowlands and O. Almaini and K. Z. Arellano-Córdova and L. Barrufet and A. Cimatti and R. S. Ellis and N. A. Grogin and M. L. Hamadouche and G. D. Illingworth and A. M. Koekemoer and H.-H. Leung and C. C. Lovell and P. G. Pérez-González and P. Santini and T. M. Stanton and V. Wild},
   year={2024},
   pages={325},
   month={Oct}}

@article{Carnallnature,
   title={A massive quiescent galaxy at redshift 4.658},
   volume={619},
   ISSN={1476-4687},
   url={http://dx.doi.org/10.1038/s41586-023-06158-6},
   DOI={10.1038/s41586-023-06158-6},
   number={7971},
   journal={Nature},
   publisher={Springer Science and Business Media LLC},
   author={A. C. Carnall and R. J. McLure and J. S. Dunlop and D. J. McLeod and V. Wild and F. Cullen and D. Magee and R. Begley and A. Cimatti and C. T. Donnan and M. L. Hamadouche and S. M. Jewell and S. Walker},
   year={2023},
   month={May},
   pages={716}}

@ARTICLE{Faucher-Giguere,
       author = {{Faucher-Gigu{\`e}re}, Claude-Andr{\'e} and {Lidz}, Adam and {Hernquist}, Lars and {Zaldarriaga}, Matias},
        title = "{Evolution of the Intergalactic Opacity: Implications for the Ionizing Background, Cosmic Star Formation, and Quasar Activity}",
      journal = {\apj},
         year = 2008,
        month = {Nov},
       volume = {688},
       number = {1},
        pages = {85},
          doi = {10.1086/592289},
archivePrefix = {arXiv},
       eprint = {0807.4177},
 primaryClass = {astro-ph},
       adsurl = {https://ui.adsabs.harvard.edu/abs/2008ApJ...688...85F}
}

@article{Montero-Dorta,
  author =        {Montero-Dorta, Antonio D and Chaves-Montero, Jonás and
                   Artale, M Celeste and Favole, Ginevra},
  journal =       {MNRAS},
  month =         {Sep},
  number =        {1},
  pages =         {940},
  publisher =     {Oxford University Press (OUP)},
  title =         {On the influence of halo mass accretion history on
                   galaxy properties and assembly bias},
  volume =        {508},
  year =          {2021},
  doi =           {10.1093/mnras/stab2556},
  issn =          {1365-2966},
  url =           {http://dx.doi.org/10.1093/mnras/stab2556},
}

@article{Flares,
    doi = {10.1093/mnras/staa3360},
  
    url = {https://doi.org/10.1093%2Fmnras%2Fstaa3360},
  
    year = 2020,
    month = {Oct},
  
    publisher = {Oxford University Press ({OUP})},
  
    volume = {500},
  
    number = {2},
  
    pages = {2127},
  
    author = {Christopher C Lovell and Aswin P Vijayan and Peter A Thomas and Stephen M Wilkins and David J Barnes and Dimitrios Irodotou and Will Roper},
  
    title = {{First Light And Reionization Epoch Simulations (FLARES) - I. Environmental dependence of high-redshift galaxy evolution}},
  
    journal = {MNRAS}
}

@article{Disperse,
  author =        {Sousbie, T.},
  journal =       {MNRAS},
  month =         {Apr},
  number =        {1},
  pages =         {350},
  publisher =     {Oxford University Press (OUP)},
  title =         {The persistent cosmic web and its filamentary
                   structure - I. Theory and implementation},
  volume =        {414},
  year =          {2011},
  doi =           {10.1111/j.1365-2966.2011.18394.x},
  issn =          {0035-8711},
  url =           {http://dx.doi.org/10.1111/j.1365-2966.2011.18394.x},
}

@article{Eagle,
    author = {Schaye, Joop and Crain, Robert A. and Bower, Richard G. and Furlong, Michelle and Schaller, Matthieu and Theuns, Tom and Dalla Vecchia, Claudio and Frenk, Carlos S. and McCarthy, I. G. and Helly, John C. and Jenkins, Adrian and Rosas-Guevara, Y. M. and White, Simon D. M. and Baes, Maarten and Booth, C. M. and Camps, Peter and Navarro, Julio F. and Qu, Yan and Rahmati, Alireza and Sawala, Till and Thomas, Peter A. and Trayford, James},
    title = {{The EAGLE Project: Simulating the Evolution and Assembly of Galaxies and their Environments}},
    journal = {MNRAS},
    volume = {446},
    number = {1},
    pages = {521},
    year = {2014},
    month = {11},
    issn = {0035-8711},
    doi = {10.1093/mnras/stu2058},
    url = {https://doi.org/10.1093/mnras/stu2058}
}

@ARTICLE{Gaea,
       author = {{De Lucia}, Gabriella and {Fontanot}, Fabio and {Xie}, Lizhi and {Hirschmann}, Michaela},
        title = "{Tracing the quenching journey across cosmic time}",
      journal = {\aap},
         year = 2024,
        month = {Jul},
       volume = {687},
          eid = {A68},
        pages = {A68},
          doi = {10.1051/0004-6361/202349045},
archivePrefix = {arXiv},
       eprint = {2401.06211},
 primaryClass = {astro-ph.GA},
       adsurl = {https://ui.adsabs.harvard.edu/abs/2024A&A...687A..68D}
}

@article{Shark,
    author = {Lagos, Claudia Del P and Tobar, Rodrigo J and Robotham, Aaron S G and Obreschkow, Danail and Mitchell, Peter D and Power, Chris and Elahi, Pascal J},
    title = {Shark: introducing an open source, free, and flexible semi-analytic model of galaxy formation},
    journal = {MNRAS},
    volume = {481},
    number = {3},
    pages = {3573},
    year = {2018},
    doi = {10.1093/mnras/sty2440},
    URL = {http://dx.doi.org/10.1093/mnras/sty2440},
    eprint = {/oup/backfile/content_public/journal/mnras/481/3/10.1093_mnras_sty2440/1/sty2440.pdf}
}

@article{Shark2,
    author = {Lagos, Claudia Del P and Bravo, Matías and Tobar, Rodrigo and Obreschkow, Danail and Power, Chris and Robotham, Aaron S G and Proctor, Katy L and Hansen, Samuel and Chandro-Gómez, Ángel and Carrivick, Julian},
    title = {Quenching massive galaxies across cosmic time with the semi-analytic model shark v2.0},
    journal = {MNRAS},
    volume = {531},
    number = {3},
    pages = {3551},
    year = {2024},
    month = {May},
    issn = {0035-8711},
    doi = {10.1093/mnras/stae1024},
    url = {https://doi.org/10.1093/mnras/stae1024},
    eprint = {https://academic.oup.com/mnras/article-pdf/531/3/3551/58211224/stae1024.pdf},
}

@article{Silk2024,
doi = {10.3847/2041-8213/ad1bf0},
url = {https://dx.doi.org/10.3847/2041-8213/ad1bf0},
year = {2024},
month = {Jan},
publisher = {The American Astronomical Society},
volume = {961},
number = {2},
pages = {L39},
author = {Silk, Joseph and Begelman, Mitchell C. and Norman, Colin and Nusser, Adi and Wyse, Rosemary F. G.},
title = {Which Came First: Supermassive Black Holes or Galaxies? Insights from JWST},
journal = {ApJL}
}

@article{Grispy,
  author =        {{Chalela}, Martin and {Sillero}, Emanuel and
                   {Pereyra}, Luis and {Alejandro Garc{\'\i}a}, Mario and
                   {Cabral}, Juan B. and {Lares}, Marcelo and
                   {Merch{\'a}n}, Manuel},
  journal =       {Astronomy \& Computing},
  month =         Dec,
  volume =        {34},
  pages =         {100443},
  title =         {{GriSPy: A Python Package for Fixed-Radius Nearest
                   Neighbors Search}},
  year =          {2021},
  doi =           {https://doi.org/10.1016/j.ascom.2020.100443},
}

@article{Zou,
   title={The Cosmic Evolution of the Supermassive Black Hole Population: A Hybrid Observed Accretion and Simulated Mergers Approach},
   volume={976},
   url={http://dx.doi.org/10.3847/1538-4357/ad815d},
   DOI={10.3847/1538-4357/ad815d},
   number={1},
   journal={\apj},
   publisher={American Astronomical Society},
   author={Fan Zou and W. N. Brandt and Elena Gallo and Bin Luo and Qingling Ni and Yongquan Xue and and Zhibo Yu},
   year={2024},
   month={Nov},
   pages={6}
}

@article{Pacucci,
   title={JWST CEERS and JADES Active Galaxies at z = 4-7 Violate the Local Mbh-Ms Relation at 3σ: Implications for Low-mass Black Holes and Seeding Models},
   volume={957},
   ISSN={2041-8213},
   url={http://dx.doi.org/10.3847/2041-8213/ad0158},
   DOI={10.3847/2041-8213/ad0158},
   number={1},
   journal={\apjl},
   publisher={American Astronomical Society},
   author={Pacucci, Fabio and Nguyen, Bao and Carniani, Stefano and Maiolino, Roberto and Fan, Xiaohui},
   year={2023},
   month={Oct},
   pages={L3}
}

@article{Li_2022,
   title={On the Connection between Supermassive Black Holes and Galaxy Growth in the Reionization Epoch},
   volume={931},
   ISSN={2041-8213},
   url={http://dx.doi.org/10.3847/2041-8213/ac6de8},
   DOI={10.3847/2041-8213/ac6de8},
   number={1},
   journal={\apjl},
   publisher={American Astronomical Society},
   author={Li, Junyao and Silverman, John D. and Izumi, Takuma and He, Wanqiu and Akiyama, Masayuki and Inayoshi, Kohei and Matsuoka, Yoshiki and Onoue, Masafusa and Toba, Yoshiki},
   year={2022},
   month={May},
   pages={L11}
}

@ARTICLE{Suh,
       author = {{Suh}, Hyewon and {Civano}, Francesca and {Trakhtenbrot}, Benny and {Shankar}, Francesco and {Hasinger}, G{\"u}nther and {Sanders}, David B. and {Allevato}, Viola},
        title = "{No Significant Evolution of Relations between Black Hole Mass and Galaxy Total Stellar Mass Up to z {\ensuremath{\sim}} 2.5}",
      journal = {\apj},
         year = 2020,
        month = {Jan},
       volume = {889},
       number = {1},
          eid = {32},
        pages = {32},
          doi = {10.3847/1538-4357/ab5f5f},
archivePrefix = {arXiv},
       eprint = {1912.02824},
 primaryClass = {astro-ph.GA},
       adsurl = {https://ui.adsabs.harvard.edu/abs/2020ApJ...889...32S}
}

@article{Mezcua,
   title={Overmassive Black Holes at Cosmic Noon: Linking the Local and the High-redshift Universe},
   volume={966},
   ISSN={2041-8213},
   url={http://dx.doi.org/10.3847/2041-8213/ad3c2a},
   DOI={10.3847/2041-8213/ad3c2a},
   number={2},
   journal={ApJL},
   publisher={American Astronomical Society},
   author={Mezcua, Mar and Pacucci, Fabio and Suh, Hyewon and Siudek, Malgorzata and Natarajan, Priyamvada},
   year={2024},
   month={May},
   pages={L30}
}

@article{ThesanZoom,
      title={Introducing the THESAN-ZOOM project: radiation-hydrodynamic simulations of high-redshift galaxies with a multi-phase interstellar medium},
      author={Rahul Kannan and Ewald Puchwein and Aaron Smith and Josh Borrow and Enrico Garaldi and Laura Keating and Mark Vogelsberger and Oliver Zier and William McClymont and Xuejian Shen and Filip Popovic and Sandro Tacchella and Lars Hernquist and Volker Springel},
      journal = {The Open Journal of Astrophysics},
         year = 2025,
        month = oct,
       volume = {8},
          eid = {153},
        pages = {153},
          doi = {10.33232/001c.145804}
}

@article{McClymont,
      title={The THESAN-ZOOM project: Burst, quench, repeat -- unveiling the evolution of high-redshift galaxies along the star-forming main sequence},
      author={William McClymont and Sandro Tacchella and Aaron Smith and Rahul Kannan and Ewald Puchwein and Josh Borrow and Enrico Garaldi and Laura Keating and Mark Vogelsberger and Oliver Zier and Xuejian Shen and Filip Popovic and Charlotte Simmonds},
      year={2025},
      journal={MNRAS},
      volume = {544},
       number = {1},
        pages = {513},
      doi={10.48550/arXiv.2503.00106}
}

@article{Planck,
    doi = {10.1051/0004-6361/201525830},
  
    url = {https://doi.org/10.1051%2F0004-6361%2F201525830},
  
    year = 2016,
    month = {Sep},
  
    publisher = {{EDP Sciences}},
  
    volume = {594},
  
    pages = {A13},
  
    author = {{Planck Collaboration}},
  
    title = {{Planck 2015 Results. XIII. Cosmological Parameters}},
  
    journal = {A\&A}
}

@book{Jolliffe,
  author={Ian Jolliffe},
  title={{Principal Component Analysis}},
  edition={$2^\text{nd}$},
  publisher={Springer-Verlag Publications},
  address={New York, USA},
  year={2002}
}

@Article{JoKim,
  author        = {Jo, Yongseok and Kim, Ji-Hoon},
  title         = {Machine-assisted Semi-Simulation Model (MSSM): Estimating Galactic Baryonic Properties from Their Dark Matter Using A Machine Trained on Hydrodynamic Simulations},
  journal       = {MNRAS},
  year          = {2019},
  volume        = {489},
  number        = {3},
  pages         = {3565},
  month         = {Aug},
  issn          = {1365-2966},
  __markedentry = {[hgc-swin:6]},
  doi           = {10.1093/mnras/stz2304},
  publisher     = {Oxford University Press (OUP)},
  url           = {http://dx.doi.org/10.1093/mnras/stz2304},
}





\bsp	
\label{lastpage}
\end{document}